\documentclass[useAMS,usenatbib]{mn2e}

\usepackage{graphicx,amssymb,natbib,cite,aas_macros,amsmath}		
\usepackage{graphics,times,caption}
\usepackage{longtable,lscape,rotating}
\def\etal{et al}	
\def\swift{\textit{Swift}}
\def\fermi{\textit{Fermi}}

\title[The AMI 15.7 GHz GRB Catalogue]{The Arcminute Microkelvin Imager Catalogue of Gamma-ray Burst afterglows at 15.7 GHz}
\author[G. E. Anderson et al.]{G. E. Anderson,$^{1,2}$\thanks{E-mail: gemma.anderson@curtin.edu.au} 
T. D. Staley,$^1$
A. J. van der Horst,$^{3,4}$ 
R. P. Fender,$^1$
A. Rowlinson,$^{5,6}$
\newauthor 
K. P. Mooley,$^1$ 
J. W. Broderick,$^{1,5}$
R. A. M. J. Wijers,$^6$
C. Rumsey,$^7$  
and D. J. Titterington$^7$ 	\\
$^1$Department of Physics, Astrophysics, University of Oxford, Denys Wilkinson Building, Oxford, OX1 3RH, UK\\
$^2$International Centre for Radio Astronomy Research, Curtin University, GPO Box U1987, Perth, WA 6845, Australia\\
$^3$Department of Physics, The George Washington University, 725 21st Street NW, Washington, DC 20052, USA\\
$^4$Astronomy, Physics, and Statistics Institute of Sciences (APSIS), The George Washington University, Washington, DC 20052, USA\\
$^5$ASTRON, The Netherlands Institute for Radio Astronomy, Postbus 2, NL-7990 AA Dwingeloo, the Netherlands\\
$^6$Anton Pannekoek Institute for Astronomy, University of Amsterdam, Postbus 94249, NL-1090 GE Amsterdam, the Netherlands\\
$^7$Astrophysics Group, Cavendish Laboratory, 19 J J Thomson Avenue, Cambridge CB3 0HE, UK\\
}

\begin{document}

\date{Accepted 2017 September 13. Received 2017 August 29; in original form 2017 July 02}

\pagerange{\pageref{firstpage}--\pageref{lastpage}} \pubyear{2017}

\maketitle

\label{firstpage}

\begin{abstract}

We present the Arcminute Microkelvin Imager (AMI) Large Array catalogue of 139 gamma-ray bursts (GRBs). AMI observes at a central frequency of 15.7\,GHz and is equipped with a fully automated rapid-response mode, which enables the telescope to respond to high-energy transients detected by \swift. On receiving a transient alert, AMI can be on-target within two minutes, scheduling later start times if the source is below the horizon. Further AMI observations are manually scheduled for several days following the trigger. The AMI GRB programme probes the early-time ($<1$\,day) radio properties of GRBs, and has obtained some of the earliest radio detections (GRB 130427A at 0.36 and GRB 130907A at 0.51\,days post-burst). As all \swift\ GRBs visible to AMI are observed, this catalogue provides the first representative sample of GRB radio properties, unbiased by multi-wavelength selection criteria. We report the detection of six GRB radio afterglows that were not previously detected by other radio telescopes, increasing the rate of radio detections by 50\% over an 18-month period. The AMI catalogue implies a \swift\ GRB radio detection rate of $\gtrsim15\%$, down to $\sim0.2$\,mJy/beam. However, scaling this by the fraction of GRBs AMI would have detected in the Chandra \& Frail sample (all radio-observed GRBs between $1997-2011$), it is possible $\sim44-56\%$ of \swift\ GRBs are radio bright, down to $\sim0.1-0.15$\,mJy/beam. This increase from the Chandra \& Frail rate ($\sim30$\%) is likely due to the AMI rapid-response mode, which allows observations to begin while the reverse-shock is contributing to the radio afterglow. 

\end{abstract}


\section{Introduction}

The first detection of a gamma-ray burst (GRB) at radio wavelengths \citep[GRB 970508;][]{galama98,frail00b}, bought about a new era of transient astrophysics that has lead to over twenty years of discovery \citep{frail03,chandra12,ugarte12}. With the launch of \swift\ in 2004, the improved ability to localise GRBs to within $4'$ \citep{krimm13} using the Burst Alert Telescope \citep[BAT;][]{barthelmy05}, with more precise localisation being provided by the \swift\ X-ray Telescope \citep[XRT;][]{burrows00} and the \swift\ Ultraviolet/Optical Telescope \citep[UVOT;][]{roming05}, has allowed for the rapid identification of hundreds of multi-wavelength counterparts. However, despite such a rich dataset, a conclusive picture of the radio afterglow properties of GRBs is yet to emerge. 

The study of GRBs at radio wavelengths is important because the radio domain provides a unique probe of the associated jet and its interaction with the surrounding circumstellar environment. Such observations, particularly when the jet has decelerated to non-relativistic speeds, also allow us to investigate the total energy budget from these events \citep{frail01}. The standard model for GRBs is the internal-external shock scenario \citep{rees92,piran99}. This model suggests that along with the forward-shock created by the blast wave at the front of the relativistic jet propagating into the circumstellar medium (observed as the classical afterglow), there is a reverse-shock that propagates back into the relativistic ejecta causing a much faster flash of emission \citep{sari99s}. The transient radio emission associated with the reverse-shock occurs within a few hours to a few days post-burst \citep{kulkarni99}, and therefore requires a rapid observing response in order to be detected. 

In an attempt to understand the radio properties of GRBs, \citet{chandra12} conducted a complete investigation of all historical events observed in the radio domain. These included both of the main GRB populations \citep{kouveliotou93}: long-duration GRBs \citep[likely produced by massive stellar collapse where the gamma-ray emission lasts for more than 2\,s;][]{woosley93,kulkarni98,woosley06} and short-duration GRBs \citep[likely caused by the coalescence of two neutron stars or a neutron star and black hole, which lasts for less than 2\,s;][]{lattimer76,eichler89,narayan92}. Only 30\% of their sample had a detectable radio afterglow, with the radio emission peaking within a very narrow flux range. This led them to conclude that the low percentage of detections was likely due to the sensitivity of radio telescopes rather than there being two distinct GRB populations: radio-bright and radio-faint. \citet{ghirlanda13} and \citet{burlon15} then conducted simulations to demonstrate that potentially all \swift\ GRBs will be detectable at radio frequencies with phase 1 of the Square Kilometre Array (SKA), specifically SKA1-MID in Band 5 ($\sim9$\,GHz)\footnote{see SKA baseline documents http://skatelescope.org/key-documents/} between $2-10$ days post-burst, as well as with the recently upgraded Karl G. Jansky Very Large Array (VLA)\footnote{https://science.nrao.edu/facilities/vla} and MeerKAT \citep[the South African SKA precursor telescope;][]{jonas09}. In fact, SKA1-MID will be so sensitive it could detect the radio counterparts from GRBs with gamma-ray emission up to five times fainter than those currently detected with \swift-BAT \citep[note that these simulations do not account for radio emission produced by the reverse-shock, only considering contributions from the forward-shock;][]{burlon15}. However, a study conducted by \citet{hancock13}, which involved visibility stacking of VLA GRB radio observations, suggested the low radio detection rate may be due to there being separate radio-bright and radio-faint GRB populations, and that $\leq70$\% are likely to be truly radio bright. 

While \citet{chandra12} provide a very comprehensive study of all radio-observed GRBs up until 2011 January, their sample may not be representative of the entire GRB population. Due to the limited number of radio telescopes, the amount of available radio observing time is at a premium, so choosing the best GRBs to follow-up is often based on existing knowledge of the event to ensure the greatest chance of a radio afterglow detection. Such criteria usually include a bright optical or X-ray counterpart, its close proximity to the Milky Way, or the suspicion of the GRB being an optically dark burst \citep{vanderhorst09,vanderhorst15}. Additionally, while a vast quantity of early-time (within minutes of the burst) optical and X-ray data on GRBs have been collected \citep[see][and references therein]{gehrels09}, once again, the rarity of radio telescopes has led to fewer experiments designed to obtain similar early-time observations at radio wavelengths. We clearly require a programme capable of targeting the early-time radio properties of GRBs ($<1$ day post-burst), which are specifically sensitive to the reverse-shock contributions to the radio afterglow. The programme design would also need to provide a radio detection rate more representative of the GRB sample (i.e. not informed by multi-wavelength properties). 
 
In-order to probe the early-time radio properties of GRBs, one solution is to implement a rapid-response observing system, which enables telescopes to trigger on transient alerts, such as Swift-detected GRBs. Such a system automatically repoints the
telescope, allowing it to begin observing the transient within minutes of its detection. While uncommon, radio telescopes capable of responding to external triggers have existed for at least 20 years. The first triggering programmes were specifically designed to probe for prompt, coherent radio emission associated with GRBs, with timescales on the order of milliseconds. For example, the Cambridge Low Frequency Synthesis Telescope performed triggered observations of Burst And Transient Source Experiment (BATSE) GRBs, placing limits on prompt emission on the order of 10's Jy at 151~MHz \citep{green95,dessenne96}. In fact, with the discovery of fast radio bursts \citep[FRBs;][]{lorimer07}, prompt radio emission associated with GRBs became one of the top progenitor candidates. \citet{bannister12} used a 12\,m radio dish at 1.4 GHz to trigger on nine \swift\ GRBs, possibly detecting a single, highly dispersed short duration radio pulse at $6$ times the root-mean square (RMS) noise ($\sigma_s$) from two GRBs. While work by \citet{zhang14} supports a possible link between GRBs and FRBs (particularly short GRBs), triggered observations performed by \citet{palaniswamy14} on five \swift\ GRBs using a 26m radio dish at 2.3~GHz, failed to detect prompt radio emission above $6\sigma_s$, discouraging an association. The Murchison Widefield Array \citep[MWA;][]{tingay13} also triggers on \swift\ GRBs, and a recent search for prompt radio emission associated with the short GRB 150424A placed 3\,Jy flux limits on 4\,s, 2 minute and 30 minute timescales between $80-133$\,MHz \citep{kaplan15}. In each case these experiments were conducted at low frequencies ($\leq2.3$\,GHz) with the specific task of searching for coherent radio emission associated with GRBs. None of these programmes probed for early-time incoherent, synchrotron emission signatures from the forward- or reverse-shock afterglows, which are likely to be bright and evolving on daily timescales at higher radio frequencies ($\gtrsim5$\,GHz).  

Clearly there is a need for a longer running radio programme capable of performing rapid-response and long-term monitoring of GRBs to probe associated incoherent radio emission. Over the last 5 years, we have been running a robotised follow-up programme that automatically triggers the Large Array (LA) interferometer of the Arcminute Microkelvin Imager \citep[AMI;][]{zwart08} on \swift-BAT detected GRBs. This programme is called the AMI-LA Rapid Response Mode (ALARRM), which is currently the longest running GRB rapid-response follow-up project in the radio domain \citep{staley13}. Following a \swift\ trigger, AMI-LA (henceforth referred to as AMI) is capable of being on-target and beginning observations within 2 minutes post-burst. This programme is therefore capable of statistically constraining the radio properties of GRBs within the first few hours to day post-burst, making it sensitive to radio reverse-shock emission. Continued AMI monitoring is then manually scheduled throughout the following weeks and months, allowing us to obtain a global view of the radio properties of both long and short \swift-detected GRBs from a representative sample that have not been informed by multi-wavelength observations. 

Early AMI-ALARRM results include observations of GRB 130427A, obtaining one of the earliest published radio detections of a long GRB at 0.36 days post-burst, allowing us to follow the rise and decline of the reverse-shock flare in the radio band \citep{anderson14}. Since 2014 May, the ALARRM programme has been expanded to also trigger on non-GRB \swift\ transients. Consequently, AMI triggering on the gamma-ray superflare from the rapidly rotating M-dwarf DG CVn, detecting the associated giant radio flare, which represents one of the earliest radio transient detections resulting from a high-energy trigger \citep{fender15}. Using the ALARRM mode, AMI was also the first radio telescope to begin observing V404 Cyg just 2 hours after the \swift\ detection of its 2015 June 15 outburst, following 26 years of quiescence \citep{mooley15}.

In this paper we present the AMI GRB catalogue from the first three years of ALARRM triggering and AMI follow-up. This catalogue includes 871 radio flux densities and limits at 15.7 GHz for 139 GRBs, 132 of which were detected by \swift-BAT, with comprehensive and systematic temporal coverage spanning $<2$ minutes up to several months post-burst. A description of the AMI/ALARRM observing strategy, data reduction and GRB radio counterpart identification can be found in Section 2 with the complete AMI GRB catalogue presented in Section 3. In Section 4 we discuss individual GRBs that were detected with AMI. These include new radio GRBs, which were first detected in the radio band by AMI, and radio-detected GRBs (initially identified by other radio telescopes). We also briefly describe those GRBs for which we have possible AMI detections and those that appear coincident with a steady radio source. Known radio-detected GRBs that were not detected with AMI are also examined. In Section 5 we discuss the overall statistical properties and implications of the AMI GRB catalogue, which represent the first systematic radio survey of \swift\ detected GRBs. This includes discussions on the early time radio properties of GRBs ($<1$ hr post-burst), the radio GRB detection rate, and the radio brightness temperatures, minimum Lorentz factors and luminosities. Our summary and conclusions can be found in Section 6.

\section{Observing strategy and data analysis}

\subsection{AMI Strategy and Observations}\label{strat}

The radio observations of GRBs presented in this paper were obtained using AMI, which is a radio interferometer consisting of eight 12.8~m diameter dishes with baselines between 18--110~m. As all the observations were conducted prior to 2015 June, the effective frequency range was 13.9--17.5~GHz using channels 3--7, each with a bandwidth of 0.72 GHz, with channels 1, 2 and 8 being disregarded due to their susceptibility to radio frequency interference (RFI). AMI measures a single polarisation ($I + Q$) and has a flux RMS noise sensitivity of 3.3 mJy s$^{-1/2}$ for five frequency channels. At the central operating frequency of 15.7 GHz, AMI has a primary beam of 5.5 arcmin and a $\approx30$ arcsec resolution \citep{zwart08}. During each AMI observation a bright unresolved source within a few degrees of the target of interest is visited for 1 minute in every 11 minutes to provide phase and amplitude calibration \citep[for further details on the absolute flux calibration of AMI see][]{franzen11}.

The first stage of the ALARRM observing strategy involves AMI receiving a GRB alert from \swift-BAT, which triggers a fully automated AMI observation of the event, now with response times within 2~mins post-burst. The \swift-BAT trigger is broadcast via a VOEvent, which is a standard format for distributing information regarding astronomical transient alerts such as the source position, classification and fluxes.\footnote{http://wiki.ivoa.net/bin/view/IVOA/IvoaVOEvent} The VOEvent alerts are then parsed by the ``4 Pi Sky" VOEvent broker \citep{staley16}, which triggers a rapid-response AMI observation of the transient. If the GRB is above the declination cut-off but below the horizon, the software will automatically update the AMI schedule to begin observations when the source has risen above the horizon. As the \swift-BAT position is only accurate to within $1-4$ arcmins \citep{gehrels04} it is likely that the true GRB position will be off-centre, but still contained within the AMI primary beam. Follow-up AMI observations designed to detect late-time radio emission are then manually scheduled using updated positions supplied by \swift-XRT or \swift-UVOT. 

The early results of the ALARRM programme were first described by \citet{staley13}. Of the 11 GRBs reported, only GRB 120326A was detected by AMI, with all non-detections listed as upper-limits. At this stage of the programme, AMI triggered on all \swift-BAT detected GRBs with a declination $\delta>-10^{\circ}$. If the GRB was above the horizon then AMI was capable of being on-target within 5 minutes post-burst. The resulting triggered observation and subsequent manually scheduled observations were one hour in duration and followed a logarithmic follow-up schedule \citep[for specific details on the automation of the AMI telescope and the original trigger policy please see][]{staley13}.

By the end of one year of operation the only radio counterparts detected were from GRB 120326A and GRB 130427A, with only one spectacular early detection coming from the latter \citep{anderson14}. The average sensitivity of these observations were $0.1$ mJy/beam, with progressively worse RMS noise levels for lower declinations. The ALARRM triggered observations also tended to have worse RMS noise levels as these often occurred closer to the horizon and therefore suffered more severely from terrestrial RFI. With the detection of only two out of 68 AMI observed GRBs, nine of which were detected in the radio band by other instruments, we decided to adjust the ALARRM strategy. 

The updated ALARRM strategy, which was implemented in 2013 August and ran until the old correlator was shut down in mid 2015, was aimed at obtaining a larger proportion of GRB radio counterpart detections with a smaller number of triggered and manually scheduled observations. ALARRM observations were restricted to all \swift\ triggers that had a declination $\delta \geq 15^{\circ}$ to decrease the amount of RFI due to low elevation angles. The \swift\ triggered AMI observations were extended to 2\,hrs in order to obtain a more sensitive observation, but short enough to not significantly disrupt the calibrator observing schedule, which is crucial for telescope operations. Additional software changes were implemented to decrease the reaction time of the telescope, allowing us to be observing the target $<2$~mins post-burst. 

The duration of the manually scheduled follow-up observations were also increased to improve the likelihood of a radio detection. The recent investigation of the entire sample of radio detected GRBs before 2011 April by \citet[][]{chandra12} demonstrated that the majority of GRBs detected in the radio band at 8.5\,GHz had a peak flux between $0.1-0.2$ mJy/beam at 5 to 10 days post burst \citep[see Figure 4 of][]{chandra12}. A 4~hr AMI observation is therefore required to reach an RMS noise of $\sim0.03-0.04$ mJy/beam that will allow the reliable detection of $>0.1-0.2$ mJy/beam sources. However, it is worth noting that since GRB relativistic blast waves generate synchrotron radiation as they expand into the circumstellar (wind generated) medium \citep{granot02}, we expect the forward shock of the afterglow to peak more brightly at 15.7\,GHz and at earlier times than the peaks recorded by \citet{chandra12}. We therefore require a higher monitoring cadence at early-times (within 5 days post-burst) to detect similar radio peaks. As the range of radio peaks observed by \citet{chandra12} will be brighter at 15.7\,GHz, the RMS achieved by a 4\,hr AMI observation will be sufficient for detecting events similar to those seen in their sample. The follow-up observations are manually scheduled to occur near transit approximately 24 hrs, 3, 7, and 10 days post-burst, with this temporal spacing designed to catch the peak of the forward- or reverse-shock at 15.7 GHz at a range of redshifts \citep[z $\lesssim 5$; for example see Figure 22 and 23 of][]{chandra12}. In the event that a GRB radio counterpart was detected, the AMI observing cadence was increased to a 4~hr observation every one or two days. As part of the AMI GRB observing programme we also obtained manually scheduled observations of GRBs that were detected with the \fermi\ Large Area Telescope \citep[LAT;][]{atwood09}, the \fermi\ Gamma-ray Burst Monitor \citep[GBM;][]{meegan09} and the \textit{International Gamma-Ray Astrophysics Laboratory} \citep[\textit{INTEGRAL};][]{winkler03}, whose positions had been more precisely localised through the identification of X-ray and/or optical counterparts, usually by the \swift-XRT, \swift-UVOT, or one of the ground based GRB follow-up programmes. 

\subsection{Pipeline reduction and analysis}

The development of fully automated reduction pipelines, capable of calibrating, imaging and analysing radio data are crucial for the preparation of the SKA (and its pathfinders) in-order to minimise the human effort required for processing the projected vast data volumes. With this in mind, we constructed a pipeline specifically designed to deal with multi-epoch radio observations of transients that was built upon mature radio astronomy software packages using Python as the interface. The resulting software package {\sc AMIsurvey} \citep{staley15b} utilises dedicated Python libraries that were built to allow use of the AMI-{\sc REDUCE} software suite \citep{dickinson04} and the Common Astronomy Software Applications package \citep[{\sc CASA};][]{jaeger08}. 

The calibration stage of {\sc AMIsurvey} calls on the Python library {\sc drive-ami} \citep[first introduced by][]{staley13} built upon AMI {\sc REDUCE}. AMI {\sc REDUCE} is designed to take the raw AMI dataset and automatically flag for interference, shadowing, and hardware errors, apply phase and amplitude calibration, and Fourier transform the data. These processes are automated by a {\sc REDUCE} script that also searches for, and applies, adaptive amplitude flagging to known sources of interference. The {\sc REDUCE} script applied to the AMI GRB observations used a more relaxed adaptive flagging than what is usually applied to AMI data as we found that standard amplitude cut-offs potentially attenuated the measured point source fluxes by $\sim10$\%. It is therefore possible that the measured fluxes reported in this catalogue are slightly overestimated due to potentially unflagged interference. AMI {\sc REDUCE} then outputs the data as \textit{uv}-FITS files that are suitable for imaging in standard radio analysis software. 

The imaging stage of {\sc AMIsurvey} is conduced using {\sc chimenea}, which is built upon {\sc CASA} and is specifically designed to clean and image multi-epoch radio transient observations \citep{staley15a,staley15c}. {\sc chimenea} first takes a list of \textit{uv}-FITS files, converts them to {\sc CASA} measurement sets and concatenates all epochs with the same pointing centre. The {\sc CASA} {\sc clean} algorithm is then used to invert, deconvolve, and restore the concatenated data, creating a deep image. {\sc chimenea} then uses source finding algorithms developed for the LOFAR Transient Key Science Project, \footnote{http://docs.transientskp.org\\ https://github.com/transientskp/tkp} specifically the LOFAR Transients Pipeline \citep[{\sc TraP};][]{swinbank15} and the Transients Project source extraction \& measurement code ({\sc PySE}; Carbone et al. submitted).  {\sc chimenea} identifies sources in the deep image down to a flux significance of 4 times the RMS ($4\sigma_s$ level), the positions of which are used to create a clean mask to be applied during future cleaning steps, ensuring model components are only placed at known source locations. Additional clean apertures can also be applied by the user at the location of other sources, such as at the known position of a GRB. The final clean mask is applied to each individual radio epoch, along with the concatenated data, which then undergo an iterative cleaning process. This re-cleaning action continues down to a predefined flux threshold that is usually 3 times the RMS noise. This iterative cleaning process is necessary as the sidelobes from bright field sources reduce with each clean attempt, lowering the background RMS noise and therefore allowing for a deeper clean. The $3\sigma_s$ threshold then prevents over-cleaning, which can cause artificial changes in source fluxes and image background noise levels (also known as ``clean bias"). The final cleaned images from both the individual epochs and concatenated data (deep image) are output in FITS format. Note that the final images were not primary beam corrected as this correction severely distorts the Gaussian shape of the point sources in the AMI data. The primary beam correction is applied manually to the measured fluxes and upper-limits in the final catalogue. 

\subsection{Radio counterpart detection and identification}

All the individual AMI epochs and deep concatenated images (created by only concatenating those epochs with the same pointing center) were searched for GRB radio afterglow detections using {\sc PySE}. We searched for all sources with a $>4\sigma_s$ flux significance that were within 3 times the positional uncertainty ($\sigma_p$) of the best known \swift\ position of the GRB (see details below). These sources then became candidate afterglow detections. For those GRBs where no radio counterpart candidate was detected, we used {\sc PySE} to perform a forced fit at the best known \swift\ position, assuming a point source with a size and shape fixed to that of the restoring beam. These algorithms report the position of any detected radio sources, plus the flux and significance of all detections and forced fits, along with the statistical errors \citep[based on][]{condon98}.

In order to determine the false source detection rate in our AMI observations at different levels of significance, we ran {\sc PySE} on the deep concatenated image of each GRB to obtain a true and deep catalogue of all the radio sources in each field. All those sources with a flux significance $>3\sigma_s$ that were detected in the individual epochs by {\sc PySE} within the primary beam, but were not detected in the deep concatenated image of that particular GRB, were considered false detections (further investigations as to whether these sources are transients is beyond the scope of this paper). A radio source in an individual epoch was considered to be the same as a source in the deep concatenated image if the angular distance between them was less than 3 times the $1\sigma_p$ position error (this excludes all $>3\sigma_s$ detected sources within 3 times the positional error of the \swift\ GRB position to avoid contamination from a radio counterpart below the detection threshold defined below). After visually inspecting the data and removing all false sources that appeared to be artefacts from nearby bright sources, our analysis found 12 possible false sources with a flux significance between $3 < \sigma_s < 4$, and 6 possible false sources within a flux significance between $4 < \sigma_s < 5$, with none above $5\sigma_s$ in 871 AMI observations. If we assume that these false sources came from random Gaussian noise fluctuations then we can calculate the probability of a false source to occur at the position of a GRB in the 871 observations. In most cases the position error of a GRB is within the AMI synthesised beam ($<30"$), of which $\sim100$ fill the AMI primary beam. Therefore in 87100 samples (100 beams in 871 observations) there is a 0.014\% and 0.007\% chance of a false source being detected at the position of the GRB with a flux significance between $3 < \sigma_s < 4$ and $\sigma_s >4$, respectively. Given it is twice as likely for a false source to be detected with a flux significance between $3 < \sigma_s < 4$ than with a flux significance $\sigma_s >4$, we have chosen to only report radio detections at the position of the GRB with a flux significance of $\sigma_s >4$. 

To identify those GRBs with a candidate radio afterglows, we ran a source matching algorithm designed to pair any AMI detected radio source with a GRB provided the angular distance between the two positions were within 3 times the total $1\sigma_p$ position error, which is the \swift\ GRB position error added in quadrature to the radio source position error calculated by {\sc PySE}. In some sets of AMI observations the angular distance between an AMI detected radio source and GRB was $\leq3\sigma_p$ at some epochs but $>3\sigma_p$ at others. We therefore define an AMI radio source to be the same source at multiple epochs if their positions agree within $\lesssim1.5\sigma_p$ of the source position in the deep concatenated image. After a coincident radio source was identified using the above method, the individual AMI observations were visually inspected to ensure it was not diffuse emission or an artefact from a nearby bright source. Once verified, all GRBs with a coincident radio source with a $>4\sigma_s$ significant detection are considered highly likely to be associated. 

\subsubsection{Radio counterpart selection criteria}\label{criteria}

Identifying a coincident radio source as a GRB radio afterglow needed to be assessed on a case by case basis. Many of the afterglow candidates detected had a $\leq5\sigma_s$ flux significance and therefore large errors. As a result, it was difficult to test for variability in the radio light curves by deriving a reduced $\chi^{2}$ to a weighted mean fit \citep{gaensler00}. We therefore based our identification of a coincident radio source as a \textit{confirmed} AMI-detected GRB radio afterglow using the following criteria:
\begin{enumerate}
\item The radio source must have been detected in at least two epochs with a flux significance $>4\sigma_s$. This is due to our detecting several false sources in the AMI observations with a significance between $4<\sigma_s<5$ (as demonstrated in Section 2.3), which suggests a small chance of a single candidate afterglow detection not being real.
\item There must be an early- or late-time non-detections in the AMI observations with RMS noise levels that would have detect the radio source at its brightest measured flux with a significance $>5\sigma_s$.
\end{enumerate}
Those GRBs with a coincident radio source that was only detected in one epoch, but are consistent with our second criterion, are considered \textit{possible} AMI-detected GRB radio afterglow. 

There were also several GRBs for which no coincident radio source was detected in a single epoch, but where one was detected in the deep concatenated image. In such cases it is possible that the deep concatenated image was sensitive enough to detect the radio afterglow. To investigate this possibility we divided the individual epochs of the GRB into two even groups, the early epochs and the later epochs, and created two new concatenated images. A radio afterglow would be expected to have different brightnesses in these two concatenated images, thus confirming its variable nature. If the source is detected in both of the concatenated images but with the flux measurements differing by $\geq4 \sigma_{s}$; or if the source is only detected in one concatenated image with the non-detection in the second concatenated image obeying criterion (ii), then the source will be considered variable and therefore classified as a \textit{confirmed} AMI-detected GRB radio afterglow.

\section{The AMI GRB Catalogue}

The AMI GRB Catalogue presented in Tables~\ref{tab:1} and \ref{tab:2} is the complete list of GRBs observed with AMI as part of the ALARRM programme, from the first triggered observation of GRB 120305A, up until the last observation of GRB 150413A taken on 2015 May 9. Table~\ref{tab:1} only reports those GRBs that have a \textit{confirmed} or \textit{possible} radio afterglow detection in the AMI data as defined in Section~\ref{criteria}, all of which are long GRBs. Table~\ref{tab:2} lists all other AMI observed GRBs including those reporting coincident/serendipitous steady sources. The full catalogue includes the AMI 15.7~GHz fluxes and upper limits of 139 GRBs from 871 AMI observations, totalling $90.2$ days of observing time. Of the 139 GRBs in this catalogue, 132 were detected with \swift, which resulted in a triggered rapid-response AMI observation. Of the 132 \swift-detection events, 12 were short GRBs. Another five \fermi- and two \textit{INTEGRAL-}detected long GRBs were also observed with AMI after their positions were localised through optical and/or X-ray afterglow detections, which were consequently reported on the GRB Coordinates Network (GCN).\footnote{http://gcn.gsfc.nasa.gov/} In Tables~\ref{tab:1} and \ref{tab:2}, all GRB observations from GRB 130907A and GRB 130813A onwards, respectively, follow the updated ALARRM strategy (see Section~\ref{strat}) and are thus used in the detection statistics in Section~\ref{grbstat}.

The AMI GRB Catalogue Tables~\ref{tab:1} and \ref{tab:2} include a column indicating whether the GRB has been reported to have radio afterglow, either through the GCN or in the literature. Table~\ref{tab:2} also specifies whether the GRB was long or short, where we classify it as a short GRB if its $T_{90}$ (time that the cumulative counts increase from 5\% to 95\% above background, therefore including 90\% of the total GRB counts) is $<2$\,s \citep{kouveliotou93}. Note that only long-duration GRBs were detected with AMI. The Telescope column indicates the name of the \swift\ telescope that provided the most localised position of the GRB. This position was used as the reference GRB position when searching for radio afterglows in the AMI data, and is also the position at which a forced fit was applied in the case that a coincident radio source was not detected. The preferred position is from the \swift-UVOT, which has a 90\% error radius better than 1 arcsecond \citep{roming05}, followed by the \swift-XRT position, with a 90\% error radius better than 5 arcseconds \citep{burrows08}. If neither of these positions were available, which could be due to a faint counterpart or an observing constraint, we tried to find a counterpart position reported on the GCN. If none were reported, we used the position from \swift-BAT, which has a 90\% error radius between $1-4$ arcmin.

Tables~\ref{tab:1} and \ref{tab:2} then list the basic information of each AMI observation, including the start date of the observations in both Gregorian and MJD formats, the duration of the observation in hours, the time post-burst that the observation commenced in days, and the observational RMS noise. The reported peak fluxes for the individual epochs and concatenated deep images (labelled ``concat" in the Tables) are those output by {\sc PySE}, which are then primary beam corrected. The peak fluxes measured from detections of radio sources coincident with GRBs are indicated by asterisks in the Tables. Otherwise the reported peak fluxes are obtained by performing forced Gaussian fits at the position of either the radio counterpart seen in another AMI epoch, or at the best known \swift\ position of the GRB. The flux errors are the errors output by {\sc PySE}, added in quadrature to the AMI 5\% calibration error \citep{perrott13} and are also primary beam corrected. The significance of the peak fluxes reported by {\sc PySE} are also included in these Tables.


\onecolumn

\begin{center}
\begin{table}

\caption{The AMI 15.7 GHz GRB catalogue: GRBs that have been detected or possibly detected with AMI}
\begin{tabular}{lccccccccc} \label{tab:1}
\\
\hline
GRB$^{\mathrm{~a}}$ & Radio$^{\mathrm{~b}}$ & Telescope$^{\mathrm{~c}}$ & Date$^{\mathrm{~d}}$ & Start$^{\mathrm{~e}}$ & Integration$^{\mathrm{~f}}$ & Days$^{\mathrm{~g}}$ & Peak Flux$^{\mathrm{~h}}$ & Sig$^{\mathrm{~i}}$ & RMS$^{\mathrm{~j}}$ \\
 & Flag & & (yyyy-mm-dd) & (MJD) & (hours) & (post-burst) & (mJy/beam) & & (mJy/beam) \\
\hline
120320A &  P  &  XRT        &  2012-03-21 &  56007.09  & 1.0  & 0.60          &  * $0.38  \pm  0.09$   & 4.0   &  0.08   \\  
120320A &  P  &  XRT        &  2012-04-05 &  56022.06  & 1.0  & 15.56         &    $0.15  \pm  0.28$   & 9.0   &  0.09   \\  
120320A &  P  &  XRT        &  Concat     &            &      &               &    $0.15  \pm  0.28$   & 9.0   &  0.06   \\  \hline
120326A &  R  &  UVOT       &  2012-03-26 &  56012.36  & 1.0  & 0.31          &    $-1.20 \pm  8.88$   & 2.6   &  0.14   \\  
120326A &  R  &  UVOT       &  2012-04-02 &  56019.21  & 1.0  & 7.15          &  * $0.86  \pm  0.1 $   & 8.4   &  0.08   \\  
120326A &  R  &  UVOT       &  2012-04-16 &  56033.30  & 1.0  & 21.25         &  * $0.37  \pm  0.09$   & 4.0   &  0.08   \\  
120326A &  R  &  UVOT       &  2012-04-28 &  56045.29  & 1.0  & 33.24         &    $0.42  \pm  0.14$   & 3.3   &  0.14   \\  
120326A &  R  &  UVOT       &  Concat     &            &      &               &  * $0.59  \pm  0.08$   & 7.8   &  0.05   \\  \hline
130427A &  R  &  XRT        &  2013-04-27 &  56409.66  & 1.0  & 0.34          &  * $3.58  \pm  0.25$   & 21.6  &  0.15   \\  
130427A &  R  &  XRT        &  2013-04-27 &  56409.94  & 1.0  & 0.62          &  * $4.54  \pm  0.25$   & 43.5  &  0.08   \\  
130427A &  R  &  XRT        &  2013-04-28 &  56410.82  & 3.0  & 1.49          &  * $1.83  \pm  0.12$   & 22.9  &  0.08   \\  
130427A &  R  &  XRT        &  2013-04-29 &  56411.78  & 1.0  & 2.46          &  * $1.18  \pm  0.09$   & 15.3  &  0.07   \\  
130427A &  R  &  XRT        &  2013-04-29 &  56411.95  & 1.0  & 2.62          &  * $0.87  \pm  0.13$   & 6.9   &  0.1    \\  
130427A &  R  &  XRT        &  2013-04-30 &  56412.85  & 1.5  & 3.52          &  * $0.61  \pm  0.07$   & 9.2   &  0.06   \\  
130427A &  R  &  XRT        &  2013-05-01 &  56413.84  & 3.0  & 4.52          &  * $0.60  \pm  0.06$   & 10.9  &  0.05   \\  
130427A &  R  &  XRT        &  2013-05-02 &  56414.79  & 2.8  & 5.47          &  * $0.55  \pm  0.06$   & 9.2   &  0.05   \\  
130427A &  R  &  XRT        &  2013-05-04 &  56416.79  & 2.8  & 7.47          &  * $0.46  \pm  0.07$   & 7.2   &  0.06   \\  
130427A &  R  &  XRT        &  2013-05-07 &  56419.90  & 3.0  & 10.58         &  * $0.46  \pm  0.06$   & 8.0   &  0.06   \\  
130427A &  R  &  XRT        &  2013-05-10 &  56422.86  & 2.0  & 13.53         &    $0.21  \pm  0.07$   & 3.3   &  0.08   \\  
130427A &  R  &  XRT        &  2013-05-21 &  56433.72  & 2.0  & 24.40         &  * $0.26  \pm  0.06$   & 4.3   &  0.05   \\  
130427A &  R  &  XRT        &  2013-06-05 &  56448.70  & 3.0  & 39.38         &  * $0.24  \pm  0.05$   & 4.2   &  0.05   \\  
130427A &  R  &  XRT        &  2013-06-25 &  56468.56  & 4.0  & 59.24         &    $0.11  \pm  0.04$   & 2.7   &  0.05   \\  
130427A &  R  &  XRT        &  Concat     &            &      &               &  * $0.78  \pm  0.05$   & 25.2  &  0.02   \\  \hline
130625A &  P  &  XRT        &  2013-06-25 &  56468.29  & 1.0  & 0.00 (4.32)   &    $-0.11 \pm  0.07$   & 1.6   &  0.08   \\  
130625A &  P  &  XRT        &  2013-06-26 &  56469.34  & 1.0  & 1.05          &    $-0.18 \pm  0.42$   & 2.0   &  0.09   \\  
130625A &  P  &  XRT        &  2013-06-30 &  56473.25  & 1.0  & 4.96          &    $0.16  \pm  0.3 $   & 1.9   &  0.11   \\  
130625A &  P  &  XRT        &  2013-07-03 &  56476.21  & 1.0  & 7.92          &    $-0.35 \pm  0.47$   & 2.6   &  0.45   \\  
130625A &  P  &  XRT        &  2013-07-15 &  56488.19  & 1.0  & 19.90         &    $0.15  \pm  0.08$   & 2.7   &  0.11   \\  
130625A &  P  &  XRT        &  2013-07-22 &  56495.27  & 1.0  & 26.98         &  * $0.59  \pm  0.14$   & 4.4   &  0.13   \\  
130625A &  P  &  XRT        &  Concat     &            &      &               &    $0.08  \pm  0.04$   & 3.1   &  0.04   \\  \hline
130702A$^{\dagger}$ &  R  &  UVOT       &  2013-07-05 &  56478.80  & 2.0  & 3.79          &  * $1.54  \pm  0.15$   & 10.8  &  0.12   \\  
130702A$^{\dagger}$ &  R  &  UVOT       &  2013-07-06 &  56479.73  & 1.0  & 4.72          &  * $1.56  \pm  0.16$   & 11.1  &  0.13   \\  
130702A$^{\dagger}$ &  R  &  UVOT       &  2013-07-09 &  56482.79  & 1.5  & 7.78          &  * $0.73  \pm  0.13$   & 5.9   &  0.1    \\  
130702A$^{\dagger}$ &  R  &  UVOT       &  2013-07-10 &  56483.78  & 1.5  & 8.77          &  * $0.89  \pm  0.1 $   & 10.1  &  0.07   \\  
130702A$^{\dagger}$ &  R  &  UVOT       &  2013-07-12 &  56485.78  & 1.7  & 10.78         &  * $0.69  \pm  0.12$   & 5.6   &  0.1    \\  
130702A$^{\dagger}$ &  R  &  UVOT       &  2013-07-14 &  56487.77  & 1.5  & 12.77         &  * $0.57  \pm  0.09$   & 6.6   &  0.08   \\  
130702A$^{\dagger}$ &  R  &  UVOT       &  2013-07-17 &  56490.80  & 2.5  & 15.79         &  * $0.50  \pm  0.12$   & 4.3   &  0.1    \\  
130702A$^{\dagger}$ &  R  &  UVOT       &  2013-07-22 &  56495.74  & 2.5  & 20.74         &    $0.47  \pm  0.13$   & 3.7   &  0.12   \\  
130702A$^{\dagger}$ &  R  &  UVOT       &  2013-07-26 &  56499.87  & 2.0  & 24.86         &  * $0.59  \pm  0.11$   & 5.3   &  0.09   \\  
130702A$^{\dagger}$ &  R  &  UVOT       &  2013-07-28 &  56501.79  & 2.5  & 26.78         &  * $0.45  \pm  0.09$   & 4.8   &  0.08   \\  
130702A$^{\dagger}$ &  R  &  UVOT       &  2013-07-31 &  56504.64  & 2.5  & 29.64         &    $0.01  \pm  0.01$   & 1.9   &  0.15   \\  
130702A$^{\dagger}$ &  R  &  UVOT       &  Concat     &            &      &               &  * $0.70  \pm  0.06$   & 15.9  &  0.03   \\  \hline
130907A &  R  &  UVOT       &  2013-09-08 &  56543.41  & 2.0  & 0.51          &  * $1.04  \pm  0.13$   & 8.6   &  0.1    \\  
130907A &  R  &  UVOT       &  2013-09-09 &  56544.61  & 3.5  & 1.71          &  * $0.75  \pm  0.12$   & 6.1   &  0.12   \\  
130907A &  R  &  UVOT       &  2013-09-11 &  56546.51  & 2.4  & 3.61          &    $0.36  \pm  0.1 $   & 3.2   &  0.09   \\  
130907A &  R  &  UVOT       &  2013-09-11 &  56546.63  & 4.0  & 3.73          &    $0.14  \pm  0.07$   & 2.9   &  0.08   \\  
130907A &  R  &  UVOT       &  2013-09-14 &  56549.60  & 2.1  & 6.70          &    $0.21  \pm  0.07$   & 3.1   &  0.07   \\  
130907A &  R  &  UVOT       &  2013-09-16 &  56551.54  & 3.0  & 8.64          &    $0.06  \pm  0.04$   & 1.7   &  0.06   \\  
130907A &  R  &  UVOT       &  Concat     &            &      &               &  * $0.46  \pm  0.05$   & 10.2  &  0.03   \\  \hline
140209A &  P  &  XRT        &  2014-02-10 &  56698.68  & 4.0  & 1.36          &  * $0.43  \pm  0.1 $   & 4.2   &  0.09   \\  
140209A &  P  &  XRT        &  2014-02-11 &  56699.82  & 4.0  & 2.51          &    $-0.07 \pm  0.11$   & 1.6   &  0.04   \\  
140209A &  P  &  XRT        &  2014-02-13 &  56701.70  & 4.0  & 4.39          &    $0.18  \pm  0.06$   & 2.5   &  0.04   \\  
140209A &  P  &  XRT        &  2014-02-18 &  56706.73  & 4.0  & 9.42          &    $0.06  \pm  0.04$   & 2.5   &  0.05   \\  
140209A &  P  &  XRT        &  Concat     &            &      &               &    $0.18  \pm  0.05$   & 2.6   &  0.02   \\  \hline
\end{tabular}
\end{table}
\begin{table}
\contcaption{}
\begin{tabular}{lccccccccc} 
\\
\hline
GRB$^{\mathrm{~a}}$ & Radio$^{\mathrm{~b}}$ & Telescope$^{\mathrm{~c}}$ & Date$^{\mathrm{~d}}$ & Start$^{\mathrm{~e}}$ & Integration$^{\mathrm{~f}}$ & Days$^{\mathrm{~g}}$ & Peak Flux$^{\mathrm{~h}}$ & Sig$^{\mathrm{~i}}$ & RMS$^{\mathrm{~j}}$ \\
 & Flag & & (yyyy-mm-dd) & (MJD) & (hours) & (post-burst) & (mJy/beam) & & (mJy/beam) \\
\hline
140304A &  R  &  XRT        &  2014-03-04 &  56720.56  & 2.0  & 0.00 (4.32)   &    $-1.17 \pm  24.4$   & 2.8   &  0.05   \\  
140304A &  R  &  XRT        &  2014-03-05 &  56721.58  & 1.1  & 1.02          &    $-0.47 \pm  2.6 $   & 1.9   &  0.07   \\  
140304A &  R  &  XRT        &  2014-03-05 &  56721.71  & 3.2  & 1.16          &    $0.24  \pm  0.07$   & 3.8   &  0.05   \\  
140304A &  R  &  XRT        &  2014-03-06 &  56722.63  & 4.5  & 2.07          &    $0.07  \pm  0.03$   & 2.6   &  0.04   \\  
140304A &  R  &  XRT        &  2014-03-07 &  56723.47  & 4.0  & 2.91          &  * $0.38  \pm  0.05$   & 9.1   &  0.04   \\  
140304A &  R  &  XRT        &  2014-03-08 &  56724.60  & 4.0  & 4.04          &  * $0.18  \pm  0.04$   & 4.3   &  0.04   \\  
140304A &  R  &  XRT        &  2014-03-09 &  56725.48  & 4.0  & 4.92          &  * $0.24  \pm  0.04$   & 6.2   &  0.04   \\  
140304A &  R  &  XRT        &  2014-03-10 &  56726.63  & 4.0  & 6.07          &  * $0.23  \pm  0.04$   & 5.2   &  0.04   \\  
140304A &  R  &  XRT        &  2014-03-11 &  56727.62  & 4.0  & 7.06          &    $0.12  \pm  0.04$   & 3.4   &  0.04   \\  
140304A &  R  &  XRT        &  2014-03-12 &  56728.57  & 3.5  & 8.01          &    $0.16  \pm  0.05$   & 3.1   &  0.04   \\  
140304A &  R  &  XRT        &  2014-03-13 &  56729.45  & 6.0  & 8.90          &    $0.08  \pm  0.03$   & 4.2   &  0.04   \\  
140304A &  R  &  XRT        &  2014-03-14 &  56730.61  & 4.0  & 10.05         &    $0.05  \pm  0.03$   & 3.5   &  0.04   \\  
140304A &  R  &  XRT        &  2014-03-17 &  56733.47  & 4.0  & 12.92         &    $0.04  \pm  0.02$   & 2.5   &  0.04   \\  
140304A &  R  &  XRT        &  2014-03-21 &  56737.56  & 3.5  & 17.00         &    $-0.23 \pm  1.03$   & 2.4   &  0.05   \\  
140304A &  R  &  XRT        &  2014-03-23 &  56739.44  & 4.0  & 18.88         &    $0.06  \pm  0.03$   & 2.7   &  0.04   \\  
140304A &  R  &  XRT        &  2014-03-29 &  56745.43  & 4.0  & 24.88         &    $1.17  \pm  26.8$6  & 1.9   &  0.05   \\  
140304A &  R  &  XRT        &  2014-04-01 &  56748.48  & 4.0  & 27.92         &    $-3.3  \pm  166.$5  & 2.8   &  0.06   \\  
140304A &  R  &  XRT        &  Concat     &            &      &               &  * $0.16  \pm  0.02$   & 10.7  &  0.01   \\  \hline
140305A &  A  &  BAT        &  2014-03-05 &  56721.63  & 2.0  & 0.00 (4.32)   &    $0.12  \pm  0.17$   & 4.0   &  0.11   \\  
140305A &  A  &  BAT        &  2014-03-06 &  56722.45  & 4.0  & 0.83          &    $-0.11 \pm  0.25$   & 3.3   &  0.04   \\  
140305A &  A  &  BAT        &  2014-03-08 &  56724.43  & 4.0  & 2.80          &  * $0.29  \pm  0.06$   & 5.0   &  0.04   \\  
140305A &  A  &  BAT        &  2014-03-10 &  56726.49  & 3.0  & 4.87          &    $0.14  \pm  0.07$   & 6.4   &  0.05   \\  
140305A &  A  &  BAT        &  2014-03-12 &  56728.39  & 4.0  & 6.76          &  * $0.34  \pm  0.06$   & 5.6   &  0.04   \\  
140305A &  A  &  BAT        &  2014-03-15 &  56731.42  & 4.0  & 9.79          &  * $0.42  \pm  0.06$   & 6.9   &  0.04   \\  
140305A &  A  &  BAT        &  2014-03-18 &  56734.26  & 5.0  & 12.64         &  * $0.34  \pm  0.07$   & 4.6   &  0.04   \\  
140305A &  A  &  BAT        &  2014-03-21 &  56737.41  & 3.5  & 15.78         &  * $0.39  \pm  0.07$   & 5.4   &  0.04   \\  
140305A &  A  &  BAT        &  2014-03-27 &  56743.28  & 4.0  & 21.66         &  * $0.28  \pm  0.06$   & 4.7   &  0.04   \\  
140305A &  A  &  BAT        &  2014-03-30 &  56746.38  & 4.0  & 24.76         &    $-0.09 \pm  0.15$   & 2.7   &  0.05   \\  
140305A &  A  &  BAT        &  2014-04-01 &  56748.30  & 4.0  & 26.68         &    $0.19  \pm  0.08$   & 8.9   &  0.06   \\  
140305A &  A  &  BAT        &  Concat     &            &      &               &  * $0.24  \pm  0.03$   & 9.7   &  0.01   \\  \hline
140318A &  P  &  XRT        &  2014-03-18 &  56734.02  & 2.0  & 0.01          &    $-0.02 \pm  0.02$   & 1.8   &  0.06   \\  
140318A &  P  &  XRT        &  2014-03-18 &  56734.94  & 4.0  & 0.93          &    $0.06  \pm  0.04$   & 2.5   &  0.04   \\  
140318A &  P  &  XRT        &  2014-03-25 &  56741.98  & 4.0  & 7.97          &  * $0.28  \pm  0.05$   & 5.0   &  0.04   \\  
140318A &  P  &  XRT        &  2014-03-28 &  56744.96  & 4.0  & 10.95         &    $0.03  \pm  0.02$   & 2.8   &  0.04   \\  
140318A &  P  &  XRT        &  Concat     &            &      &               &  * $0.15  \pm  0.03$   & 4.9   &  0.02   \\  \hline
140320C$^{\ddagger}$ &  P  &  XRT        &  2014-03-21 &  56737.71  & 3.5  & 1.16          &    $-0.02 \pm  0.03$   & 2.7   &  0.04   \\  
140320C$^{\ddagger}$ &  P  &  XRT        &  2014-03-22 &  56738.69  & 4.0  & 2.13          &  * $0.14  \pm  0.03$   & 4.1   &  0.04   \\  
140320C$^{\ddagger}$ &  P  &  XRT        &  2014-03-24 &  56740.72  & 4.0  & 4.16          &    $-0.02 \pm  0.02$   & 2.2   &  0.04   \\  
140320C$^{\ddagger}$ &  P  &  XRT        &  2014-03-28 &  56744.78  & 4.0  & 8.23          &    $-0.0  \pm  0.0 $   & 1.6   &  0.04   \\  
140320C$^{\ddagger}$ &  P  &  XRT        &  2014-03-30 &  56746.83  & 4.0  & 10.28         &    $0.03  \pm  0.02$   & 2.1   &  0.04   \\  
140320C$^{\ddagger}$ &  P  &  XRT        &  Concat     &            &      &               &    $0.03  \pm  0.01$   & 2.3   &  0.02   \\  \hline
140607A &  P  &  BAT        &  2014-06-07 &  56815.72  & 1.1  & 0.00 (4.32)   &    $0.36  \pm  0.59$   & 3.3   &  0.22   \\  
140607A &  P  &  BAT        &  2014-06-09 &  56817.47  & 4.0  & 1.75          &  * $0.59  \pm  0.12$   & 4.8   &  0.08   \\  
140607A &  P  &  BAT        &  2014-06-11 &  56819.48  & 4.0  & 3.76          &    $0.49  \pm  0.3 $   & 3.3   &  0.28   \\  
140607A &  P  &  BAT        &  Concat     &            &      &               &    $0.38  \pm  0.12$   & 3.9   &  0.08   \\  \hline
140629A &  AC &  UVOT       &  2014-06-30 &  56838.87  & 4.0  & 1.27          &    $0.09  \pm  0.04$   & 3.2   &  0.05   \\  
140629A &  AC &  UVOT       &  2014-07-01 &  56839.87  & 4.0  & 2.28          &    $0.15  \pm  0.04$   & 4.3   &  0.05   \\  
140629A &  AC &  UVOT       &  2014-07-02 &  56840.88  & 3.7  & 3.28          &    $0.11  \pm  0.04$   & 4.3   &  0.04   \\  
140629A &  AC &  UVOT       &  2014-07-03 &  56841.90  & 4.0  & 4.31          &    $0.24  \pm  1.13$   & 2.3   &  0.06   \\  
140629A &  AC &  UVOT       &  2014-07-04 &  56842.87  & 6.0  & 5.27          &    $0.13  \pm  0.05$   & 3.4   &  0.06   \\  
140629A &  AC &  UVOT       &  2014-07-06 &  56844.74  & 4.0  & 7.14          &    $0.03  \pm  0.02$   & 2.3   &  0.04   \\  
140629A &  AC &  UVOT       &  2014-07-10 &  56848.76  & 6.0  & 11.16         &    $0.02  \pm  0.02$   & 2.2   &  0.04   \\  
140629A &  AC &  UVOT       &  2014-07-16 &  56854.72  & 4.0  & 17.12         &    $-0.04 \pm  0.03$   & 2.5   &  0.06   \\  
140629A &  AC &  UVOT       &  2014-07-22 &  56860.75  & 4.0  & 23.16         &    $-0.02 \pm  0.03$   & 1.8   &  0.05   \\  
140629A &  AC &  UVOT       &  Concat     &            &      &               &  * $0.10  \pm  0.02$   & 5.0   &  0.02   \\  \hline
\end{tabular}
\end{table}
\begin{table}
\contcaption{}
\begin{tabular}{lccccccccc} 
\\
\hline
GRB$^{\mathrm{~a}}$ & Radio$^{\mathrm{~b}}$ & Telescope$^{\mathrm{~c}}$ & Date$^{\mathrm{~d}}$ & Start$^{\mathrm{~e}}$ & Integration$^{\mathrm{~f}}$ & Days$^{\mathrm{~g}}$ & Peak Flux$^{\mathrm{~h}}$ & Sig$^{\mathrm{~i}}$ & RMS$^{\mathrm{~j}}$ \\
 & Flag & & (yyyy-mm-dd) & (MJD) & (hours) & (post-burst) & (mJy/beam) & & (mJy/beam) \\
\hline
140703A &  R  &  XRT        &  2014-07-03 &  56841.03  & 2.0  & 0.01 (11.52)  &    $0.02  \pm  0.02$   & 3.1   &  0.07   \\  
140703A &  R  &  XRT        &  2014-07-04 &  56842.22  & 4.0  & 1.19          &  * $0.32  \pm  0.06$   & 5.4   &  0.05   \\  
140703A &  R  &  XRT        &  2014-07-06 &  56844.08  & 4.0  & 3.05          &  * $0.49  \pm  0.06$   & 7.8   &  0.06   \\  
140703A &  R  &  XRT        &  2014-07-07 &  56845.13  & 4.0  & 4.11          &  * $0.24  \pm  0.04$   & 5.7   &  0.04   \\  
140703A &  R  &  XRT        &  2014-07-08 &  56846.09  & 6.0  & 5.07          &  * $0.26  \pm  0.04$   & 6.0   &  0.04   \\  
140703A &  R  &  XRT        &  2014-07-12 &  56850.14  & 4.0  & 9.11          &    $0.07  \pm  0.04$   & 2.7   &  0.07   \\  
140703A &  R  &  XRT        &  2014-07-14 &  56852.13  & 4.0  & 11.10         &  * $0.20  \pm  0.04$   & 4.7   &  0.04   \\  
140703A &  R  &  XRT        &  2014-07-15 &  56853.18  & 4.0  & 12.16         &    $0.17  \pm  0.04$   & 3.6   &  0.05   \\  
140703A &  R  &  XRT        &  2014-07-17 &  56855.06  & 6.0  & 14.03         &    $-0.03 \pm  0.03$   & 2.0   &  0.06   \\  
140703A &  R  &  XRT        &  2014-07-20 &  56858.15  & 4.0  & 17.12         &    $0.07  \pm  0.04$   & 2.6   &  0.05   \\  
140703A &  R  &  XRT        &  2014-07-28 &  56866.16  & 4.0  & 25.13         &    $0.04  \pm  0.03$   & 1.8   &  0.05   \\  
140703A &  R  &  XRT        &  Concat     &            &      &               &  * $0.20  \pm  0.02$   & 9.7   &  0.02   \\  \hline
140709A &  A  &  XRT        &  2014-07-09 &  56847.05  & 2.0  & 0.00 (4.32)   &    $0.20  \pm  0.63$   & 2.7   &  0.07   \\  
140709A &  A  &  XRT        &  2014-07-10 &  56848.04  & 4.0  & 0.99          &    $-0.05 \pm  0.07$   & -1.1  &  0.04   \\  
140709A &  A  &  XRT        &  2014-07-11 &  56849.93  & 4.0  & 2.88          &  * $0.46  \pm  0.05$   & 10.9  &  0.04   \\  
140709A &  A  &  XRT        &  2014-07-14 &  56852.96  & 4.0  & 5.91          &  * $0.29  \pm  0.05$   & 5.9   &  0.05   \\  
140709A &  A  &  XRT        &  2014-07-15 &  56853.93  & 4.0  & 6.88          &  * $0.28  \pm  0.05$   & 5.9   &  0.04   \\  
140709A &  A  &  XRT        &  2014-07-18 &  56856.03  & 4.0  & 8.97          &    $-0.12 \pm  0.06$   & -1.4  &  0.08   \\  
140709A &  A  &  XRT        &  2014-07-18 &  56856.97  & 4.0  & 9.92          &    $0.13  \pm  0.06$   & 2.0   &  0.06   \\  
140709A &  A  &  XRT        &  2014-07-21 &  56859.95  & 4.0  & 12.90         &    $0.04  \pm  0.03$   & 0.9   &  0.04   \\  
140709A &  A  &  XRT        &  2014-07-23 &  56861.86  & 4.0  & 14.81         &    $-0.01 \pm  0.02$   & -0.3  &  0.04   \\  
140709A &  A  &  XRT        &  2014-07-27 &  56865.98  & 4.0  & 18.93         &    $0.07  \pm  0.13$   & 1.5   &  0.05   \\  
140709A &  A  &  XRT        &  2014-07-31 &  56869.86  & 4.0  & 22.81         &    $0.08  \pm  0.04$   & 1.4   &  0.05   \\  
140709A &  A  &  XRT        &  Concat     &            &      &               &  * $0.17  \pm  0.02$   & 10.0  &  0.01   \\  \hline
140713A &  R  &  XRT        &  2014-07-13 &  56851.78  & 2.0  & 0.00 (5.76)   &    $0.11  \pm  0.06$   & 2.8   &  0.09   \\  
140713A &  R  &  XRT        &  2014-07-14 &  56852.79  & 4.0  & 1.01          &    $0.17  \pm  0.06$   & 3.5   &  0.06   \\  
140713A &  R  &  XRT        &  2014-07-16 &  56854.88  & 4.0  & 3.10          &  * $0.60  \pm  0.09$   & 7.1   &  0.08   \\  
140713A &  R  &  XRT        &  2014-07-17 &  56855.86  & 4.0  & 4.08          &    $0.16  \pm  0.08$   & 3.2   &  0.09   \\  
140713A &  R  &  XRT        &  2014-07-18 &  56856.79  & 4.0  & 5.01          &  * $0.78  \pm  0.09$   & 9.4   &  0.07   \\  
140713A &  R  &  XRT        &  2014-07-19 &  56857.94  & 4.0  & 6.16          &  * $0.84  \pm  0.07$   & 14.8  &  0.05   \\  
140713A &  R  &  XRT        &  2014-07-20 &  56858.94  & 4.0  & 7.16          &  * $0.82  \pm  0.09$   & 11.0  &  0.07   \\  
140713A &  R  &  XRT        &  2014-07-22 &  56860.92  & 4.0  & 9.14          &  * $1.37  \pm  0.08$   & 28.2  &  0.04   \\  
140713A &  R  &  XRT        &  2014-07-24 &  56862.86  & 4.0  & 11.08         &  * $1.31  \pm  0.10$   & 18.3  &  0.07   \\  
140713A &  R  &  XRT        &  2014-07-26 &  56864.89  & 4.0  & 13.11         &  * $1.65  \pm  0.10$   & 28.7  &  0.06   \\  
140713A &  R  &  XRT        &  2014-07-28 &  56866.78  & 4.0  & 15.00         &  * $0.87  \pm  0.07$   & 16.1  &  0.05   \\  
140713A &  R  &  XRT        &  2014-07-30 &  56868.81  & 4.0  & 17.03         &  * $0.69  \pm  0.07$   & 11.3  &  0.06   \\  
140713A &  R  &  XRT        &  2014-08-01 &  56870.86  & 4.0  & 19.08         &  * $0.89  \pm  0.07$   & 16.3  &  0.05   \\  
140713A &  R  &  XRT        &  2014-08-03 &  56872.86  & 4.0  & 21.08         &  * $1.05  \pm  0.07$   & 21.1  &  0.04   \\  
140713A &  R  &  XRT        &  2014-08-05 &  56874.82  & 4.0  & 23.04         &  * $0.70  \pm  0.07$   & 11.0  &  0.06   \\  
140713A &  R  &  XRT        &  2014-08-06 &  56875.87  & 4.0  & 24.09         &  * $0.79  \pm  0.06$   & 15.0  &  0.05   \\  
140713A &  R  &  XRT        &  2014-08-12 &  56881.79  & 3.0  & 30.01         &  * $0.71  \pm  0.07$   & 11.9  &  0.06   \\  
140713A &  R  &  XRT        &  2014-08-14 &  56883.87  & 3.0  & 32.09         &  * $0.53  \pm  0.07$   & 8.3   &  0.06   \\  
140713A &  R  &  XRT        &  2014-08-16 &  56885.87  & 2.0  & 34.09         &  * $0.40  \pm  0.06$   & 7.0   &  0.06   \\  
140713A &  R  &  XRT        &  2014-08-18 &  56887.78  & 4.0  & 36.00         &  * $0.49  \pm  0.07$   & 7.2   &  0.06   \\  
140713A &  R  &  XRT        &  2014-08-20 &  56889.79  & 2.0  & 38.01         &    $0.20  \pm  0.06$   & 4.0   &  0.06   \\  
140713A &  R  &  XRT        &  2014-08-23 &  56892.73  & 4.0  & 40.95         &  * $0.35  \pm  0.05$   & 8.0   &  0.04   \\  
140713A &  R  &  XRT        &  2014-08-27 &  56896.85  & 4.0  & 45.07         &  * $0.29  \pm  0.04$   & 7.2   &  0.04   \\  
140713A &  R  &  XRT        &  2014-08-29 &  56898.82  & 4.0  & 47.04         &  * $0.27  \pm  0.05$   & 5.9   &  0.04   \\  
140713A &  R  &  XRT        &  2014-08-31 &  56900.76  & 1.8  & 48.98         &    $-0.07 \pm  0.10$   & 1.8   &  0.07   \\  
140713A &  R  &  XRT        &  2014-09-01 &  56901.79  & 4.0  & 50.01         &  * $0.32  \pm  0.08$   & 4.2   &  0.07   \\  
140713A &  R  &  XRT        &  2014-09-02 &  56902.68  & 5.9  & 50.90         &  * $0.18  \pm  0.04$   & 4.5   &  0.04   \\  
140713A &  R  &  XRT        &  2014-09-05 &  56905.75  & 4.0  & 53.97         &    $0.09  \pm  0.04$   & 2.4   &  0.04   \\  
140713A &  R  &  XRT        &  2014-09-07 &  56907.78  & 3.9  & 56.00         &    $0.23  \pm  0.08$   & 3.6   &  0.07   \\  
140713A &  R  &  XRT        &  2014-09-10 &  56910.80  & 4.0  & 59.02         &    $0.21  \pm  0.05$   & 4.1   &  0.06   \\  
140713A &  R  &  XRT        &  2014-09-14 &  56914.72  & 4.0  & 62.94         &    $0.09  \pm  0.04$   & 2.8   &  0.05   \\  
140713A &  R  &  XRT        &  2014-09-17 &  56917.66  & 5.8  & 65.88         &    $0.10  \pm  0.03$   & 3.7   &  0.03   \\  
140713A &  R  &  XRT        &  2014-09-23 &  56923.77  & 4.0  & 71.99         &    $0.11  \pm  0.04$   & 3.8   &  0.04   \\  
140713A &  R  &  XRT        &  2014-10-02 &  56932.59  & 5.8  & 80.81         &    $-0.22 \pm  0.87$   & 1.8   &  0.05   \\  
140713A &  R  &  XRT        &  Concat     &            &      &               &  * $0.50  \pm  0.03$   & 28.9  &  0.01   \\  \hline
\end{tabular}
\end{table}
\begin{table}
\contcaption{}
\begin{tabular}{lccccccccc} 
\\
\hline
GRB$^{\mathrm{~a}}$ & Radio$^{\mathrm{~b}}$ & Telescope$^{\mathrm{~c}}$ & Date$^{\mathrm{~d}}$ & Start$^{\mathrm{~e}}$ & Integration$^{\mathrm{~f}}$ & Days$^{\mathrm{~g}}$ & Peak Flux$^{\mathrm{~h}}$ & Sig$^{\mathrm{~i}}$ & RMS$^{\mathrm{~j}}$ \\
 & Flag & & (yyyy-mm-dd) & (MJD) & (hours) & (post-burst) & (mJy/beam) & & (mJy/beam) \\
\hline
141121A &  A  &  UVOT       &  2014-11-21 &  56982.16  & 2.0  & 0.00 (5.76)   &    $-0.41 \pm  1.03$   & 1.5   &  0.18   \\  
141121A &  A  &  UVOT       &  2014-11-22 &  56983.05  & 4.0  & 0.89          &    $0.11  \pm  0.05$   & 3.1   &  0.06   \\  
141121A &  A  &  UVOT       &  2014-11-24 &  56985.17  & 4.0  & 3.01          &  * $0.37  \pm  0.07$   & 5.5   &  0.05   \\  
141121A &  A  &  UVOT       &  2014-11-25 &  56986.97  & 4.0  & 4.81          &    $-0.09 \pm  0.13$   & 1.3   &  0.12   \\  
141121A &  A  &  UVOT       &  2014-11-28 &  56989.04  & 4.0  & 6.88          &  * $0.32  \pm  0.06$   & 5.4   &  0.04   \\  
141121A &  A  &  UVOT       &  2014-11-29 &  56990.15  & 4.0  & 7.99          &  * $0.36  \pm  0.06$   & 6.7   &  0.04   \\  
141121A &  A  &  UVOT       &  2014-11-30 &  56991.13  & 4.0  & 8.97          &  * $0.31  \pm  0.07$   & 4.6   &  0.05   \\  
141121A &  A  &  UVOT       &  2014-12-01 &  56992.98  & 7.0  & 10.82         &    $0.13  \pm  0.04$   & 3.7   &  0.03   \\  
141121A &  A  &  UVOT       &  2014-12-02 &  56994.00  & 4.0  & 11.84         &  * $0.37  \pm  0.06$   & 6.8   &  0.04   \\  
141121A &  A  &  UVOT       &  2014-12-04 &  56995.13  & 4.0  & 12.97         &    $0.15  \pm  0.05$   & 3.7   &  0.04   \\  
141121A &  A  &  UVOT       &  2014-12-06 &  56997.01  & 4.0  & 14.85         &    $0.10  \pm  0.04$   & 3.9   &  0.03   \\  
141121A &  A  &  UVOT       &  2014-12-06 &  56997.98  & 4.0  & 15.82         &  * $0.23  \pm  0.04$   & 5.1   &  0.04   \\  
141121A &  A  &  UVOT       &  2014-12-08 &  56999.12  & 4.0  & 16.96         &    $0.20  \pm  0.11$   & 2.4   &  0.1    \\  
141121A &  A  &  UVOT       &  2014-12-08 &  57000.00  & 4.0  & 17.84         &    $0.13  \pm  0.04$   & 3.3   &  0.04   \\  
141121A &  A  &  UVOT       &  2014-12-10 &  57001.99  & 4.0  & 19.82         &    $0.14  \pm  0.05$   & 3.0   &  0.04   \\  
141121A &  A  &  UVOT       &  2014-12-13 &  57004.13  & 3.3  & 21.97         &    $0.11  \pm  0.05$   & 2.5   &  0.05   \\  
141121A &  A  &  UVOT       &  2014-12-15 &  57006.02  & 4.0  & 23.86         &    $0.17  \pm  0.05$   & 3.5   &  0.05   \\  
141121A &  A  &  UVOT       &  2014-12-19 &  57010.98  & 4.0  & 28.82         &    $0.05  \pm  0.03$   & 2.0   &  0.04   \\  
141121A &  A  &  UVOT       &  2014-12-23 &  57014.99  & 4.0  & 32.83         &    $0.10  \pm  0.05$   & 2.4   &  0.05   \\  
141121A &  A  &  UVOT       &  2014-12-29 &  57020.94  & 4.0  & 38.78         &  * $0.23  \pm  0.05$   & 4.6   &  0.04   \\  
141121A &  A  &  UVOT       &  2015-01-05 &  57027.95  & 4.0  & 45.79         &    $0.08  \pm  0.04$   & 2.7   &  0.04   \\  
141121A &  A  &  UVOT       &  2015-01-16 &  57038.87  & 4.0  & 56.71         &    $0.18  \pm  0.9 $   & 1.5   &  0.04   \\  
141121A &  A  &  UVOT       &  2015-01-22 &  57044.96  & 4.0  & 62.80         &    $-0.03 \pm  0.03$   & 1.9   &  0.05   \\  
141121A &  A  &  UVOT       &  Concat     &            &      &               &  * $0.20  \pm  0.02$   & 8.6   &  0.01   \\  \hline
150110B &  A  &  XRT        &  2015-01-11 &  57033.43  & 3.5  & 0.51          &    $0.02  \pm  0.02$   & 3.4   &  0.05   \\  
150110B &  A  &  XRT        &  2015-01-14 &  57036.47  & 4.0  & 3.54          &  * $0.41  \pm  0.07$   & 5.9   &  0.06   \\  
150110B &  A  &  XRT        &  2015-01-15 &  57037.46  & 2.4  & 4.54          &  * $0.31  \pm  0.08$   & 4.0   &  0.06   \\  
150110B &  A  &  XRT        &  2015-01-16 &  57038.42  & 4.0  & 5.49          &  * $0.53  \pm  0.06$   & 10.8  &  0.04   \\  
150110B &  A  &  XRT        &  2015-01-17 &  57039.44  & 4.0  & 6.52          &  * $0.16  \pm  0.03$   & 5.1   &  0.03   \\  
150110B &  A  &  XRT        &  2015-01-18 &  57040.45  & 4.0  & 7.52          &  * $0.25  \pm  0.04$   & 6.4   &  0.04   \\  
150110B &  A  &  XRT        &  2015-01-19 &  57041.42  & 4.0  & 8.50          &  * $0.26  \pm  0.05$   & 6.2   &  0.04   \\  
150110B &  A  &  XRT        &  2015-01-20 &  57042.40  & 1.9  & 9.47          &  * $0.29  \pm  0.06$   & 5.1   &  0.05   \\  
150110B &  A  &  XRT        &  2015-01-21 &  57043.47  & 4.0  & 10.54         &    $0.11  \pm  0.05$   & 3.4   &  0.05   \\  
150110B &  A  &  XRT        &  2015-01-22 &  57044.41  & 4.0  & 11.49         &  * $0.21  \pm  0.04$   & 4.7   &  0.04   \\  
150110B &  A  &  XRT        &  2015-01-25 &  57047.44  & 4.0  & 14.52         &  * $0.46  \pm  0.06$   & 7.8   &  0.05   \\  
150110B &  A  &  XRT        &  2015-01-29 &  57051.44  & 4.0  & 18.51         &  * $0.36  \pm  0.05$   & 6.8   &  0.05   \\  
150110B &  A  &  XRT        &  2015-02-06 &  57059.30  & 4.0  & 26.37         &  * $0.29  \pm  0.04$   & 7.9   &  0.04   \\  
150110B &  A  &  XRT        &  2015-02-08 &  57061.41  & 4.0  & 28.48         &    $0.13  \pm  0.04$   & 3.7   &  0.04   \\  
150110B &  A  &  XRT        &  2015-04-08 &  57120.11  & 4.0  & 87.19         &    $-0.16 \pm  0.43$   & 3.0   &  0.06   \\  
150110B &  A  &  XRT        &  2015-04-15 &  57127.23  & 4.0  & 94.30         &    $-0.01 \pm  0.01$   & 2.1   &  0.05   \\  
150110B &  A  &  XRT        &  Concat     &            &      &               &  * $0.24  \pm  0.02$   & 20.1  &  0.01   \\  \hline
150413A &  A  &  MASTER II  &  2015-04-13 &  57125.75  & 4.0  & 0.17          &    $0.03  \pm  0.04$   & 1.9   &  0.04   \\  
150413A &  A  &  MASTER II  &  2015-04-14 &  57126.75  & 5.5  & 1.17          &  * $0.21  \pm  0.05$   & 4.2   &  0.04   \\  
150413A &  A  &  MASTER II  &  2015-04-15 &  57127.80  & 4.0  & 2.22          &  * $0.23  \pm  0.05$   & 4.6   &  0.04   \\  
150413A &  A  &  MASTER II  &  2015-04-16 &  57128.95  & 4.0  & 3.37          &  * $0.20  \pm  0.04$   & 4.8   &  0.03   \\  
150413A &  A  &  MASTER II  &  2015-04-17 &  57129.81  & 3.8  & 4.23          &  * $0.19  \pm  0.05$   & 4.2   &  0.03   \\  
150413A &  A  &  MASTER II  &  2015-04-18 &  57130.95  & 4.0  & 5.37          &    $0.10  \pm  0.03$   & 3.0   &  0.03   \\  
150413A &  A  &  MASTER II  &  2015-04-19 &  57131.89  & 4.0  & 6.31          &    $0.19  \pm  0.05$   & 3.5   &  0.04   \\  
150413A &  A  &  MASTER II  &  2015-04-20 &  57132.96  & 4.0  & 7.38          &    $0.01  \pm  0.01$   & 5.3   &  0.04   \\  
150413A &  A  &  MASTER II  &  2015-04-21 &  57133.78  & 4.0  & 8.20          &    $0.06  \pm  0.04$   & 4.9   &  0.05   \\  
150413A &  A  &  MASTER II  &  2015-04-22 &  57134.85  & 4.0  & 9.27          &    $0.04  \pm  0.03$   & 2.4   &  0.04   \\  
150413A &  A  &  MASTER II  &  2015-04-23 &  57135.95  & 4.0  & 10.37         &    $0.05  \pm  0.03$   & 2.5   &  0.03   \\  
150413A &  A  &  MASTER II  &  2015-04-25 &  57137.89  & 4.0  & 12.31         &    $0.02  \pm  0.02$   & 3.0   &  0.11   \\  
150413A &  A  &  MASTER II  &  2015-04-26 &  57138.76  & 4.0  & 13.18         &    $0.10  \pm  0.04$   & 2.3   &  0.04   \\  
150413A &  A  &  MASTER II  &  2015-04-29 &  57141.91  & 4.0  & 16.33         &    $-0.19 \pm  0.81$   & 2.5   &  0.03   \\  
150413A &  A  &  MASTER II  &  2015-05-02 &  57144.75  & 4.0  & 19.17         &    $0.01  \pm  0.01$   & 2.9   &  0.04   \\  
150413A &  A  &  MASTER II  &  2015-05-03 &  57145.74  & 4.0  & 20.16         &    $0.15  \pm  0.07$   & 2.5   &  0.06   \\  
150413A &  A  &  MASTER II  &  2015-05-09 &  57151.91  & 4.0  & 26.33         &    $-0.17 \pm  0.55$   & 1.6   &  0.04   \\  
150413A &  A  &  MASTER II  &  Concat     &            &      &               &  * $0.11  \pm  0.02$   & 6.1   &  0.01   \\  
\hline
\end{tabular}
\\
\end{table}
\begin{table}
\contcaption{}
Note: All GRBs in this table are classed as long GRBs. All GRB observations from GRB 130907A onwards followed the updated ALARRM strategy (see Section 2.1).\\
$^{\mathrm{~a}}$ GRB discovery flag: $\dagger$: \textit{Fermi} discovered GRB; $\ddagger$: \textit{Integral} discovered GRB\\
$^{\mathrm{~b}}$ Radio detection flag: A - new radio GRB discovered with AMI; AC - new radio GRB discovered with AMI that was only detected in the concatenated image; R - radio afterglow first detected with another radio telescope; P - possible new candidate radio GRB discovered with AMI (see Section 4 for further details on these categories).\\
$^{\mathrm{~c}}$ Best \swift\ telescope position used to search for a radio counterpart: BAT - \swift\ Burst Alert Telescope; XRT - \swift\ X-ray Telescope; UVOT - \swift\ Ultraviolet/Optical Telescope. For GRB 150413A, the best optical position provided by the MASTER II robotic telescope (Ivanov \etal\ 2015, GCN, 17689) was used instead. \\
$^{\mathrm{~d}}$ Date of the AMI observation in yyyy-mm-dd. Concat - concatenation of all epochs with the same pointing (in the case of GRB 120320A, two observations were taken with different pointings so the concatenated image is the same as the 2012-04-05 epoch).\\
$^{\mathrm{~e}}$ Start date of the AMI observation in modified julian date (MJD) format. \\
$^{\mathrm{~f}}$ AMI observation length in hours. \\
$^{\mathrm{~g}}$ Number of days post-burst since the start of the AMI observation. The response time, in minutes, has also been included in brackets for those AMI observations that began $\leq0.01$ days post-burst. \\
$^{\mathrm{~h}}$ Peak flux density as reported by {\sc PySE}. The `*' symbol indicates those GRBs for which the listed AMI peak flux is the measured flux of a radio source detected above a $4\sigma_s$ significance, coincident with the best known \swift\ position of the GRB (i.e. within $3\sigma_p$). All other listed fluxes are derived from a forced Gaussian fit at the best known \swift\ position. The $1\sigma_s$ error bar is the flux error output by {\sc PySE}, added in quadrature to the AMI 5\% calibration error. \\
$^{\mathrm{~i}}$ The significance of the AMI flux reported by {\sc PySE}, which are measured in units of $\sigma_s$ above a local RMS. \\
$^{\mathrm{~j}}$ The global RMS of the middle quarter of the AMI image.
\end{table}
\end{center}


\begin{center}
\onecolumn
\begin{table}
\caption{The AMI 15.7 GHz GRB catalogue: GRBs that were not detected with AMI, have a possible steady source association or a concatenated detection}
\begin{tabular}{lcccccccccc} \label{tab:2}
\\
\hline
GRB$^{\mathrm{~a}}$ & Type$^{\mathrm{~b}}$ & Radio$^{\mathrm{~c}}$ & Telescope$^{\mathrm{~d}}$ & Date$^{\mathrm{~e}}$ & Start$^{\mathrm{~f}}$ & Integration$^{\mathrm{~g}}$ & Days$^{\mathrm{~h}}$ & Peak Flux$^{\mathrm{~i}}$ & Sig$^{\mathrm{~j}}$ & RMS$^{\mathrm{~k}}$ \\
 & & Flag & & (yyyy-mm-dd) & (MJD) & (hours) & (post-burst) & (mJy/beam) & & (mJy/beam) \\
\hline
120305A  &  S &  N   &  XRT     &  2012-03-05 &  55991.82 &  1.0  & 0.00  (4.32)  &      $ 0.16    \pm  0.25     $  &  2.6   &  0.17  \\    
120305A  &  S &  N   &  XRT     &  2012-03-06 &  55992.60 &  1.0  & 0.78          &      $ 0.07    \pm  0.05     $  &  1.9   &  0.10  \\     
120305A  &  S &  N   &  XRT     &  2012-03-07 &  55993.64 &  1.0  & 1.83          &      $ 0.63    \pm  2.77     $  &  1.6   &  0.17  \\    
120305A  &  S &  N   &  XRT     &  2012-03-08 &  55994.66 &  1.0  & 2.84          &      $ -0.0    \pm  0.0      $  &  2.3   &  0.11  \\    
120305A  &  S &  N   &  XRT     &  2012-03-09 &  55995.76 &  1.0  & 3.94          &      $ -0.04   \pm  0.05     $  &  2.9   &  0.12  \\
\hline  
\end{tabular}
\end{table}
\begin{table}
\contcaption{}
\begin{tabular}{lcccccccccc}
\\
\hline
GRB$^{\mathrm{~a}}$ & Type$^{\mathrm{~b}}$ & Radio$^{\mathrm{~c}}$ & Telescope$^{\mathrm{~d}}$ & Date$^{\mathrm{~e}}$ & Start$^{\mathrm{~f}}$ & Integration$^{\mathrm{~g}}$ & Days$^{\mathrm{~h}}$ & Peak Flux$^{\mathrm{~i}}$ & Sig$^{\mathrm{~j}}$ & RMS$^{\mathrm{~k}}$ \\
 & & Flag & & (yyyy-mm-dd) & (MJD) & (hours) & (post-burst) & (mJy/beam) & & (mJy/beam) \\
\hline
120514A  &  L &  N   &  XRT     &  2012-05-16 &  56063.17 &  1.0  & 2.12          &      $ 0.46    \pm  0.15     $  &  3.7   &  0.13  \\    
120514A  &  L &  N   &  XRT     &  2012-05-19 &  56066.13 &  1.0  & 5.07          &      $ -0.48   \pm  1.12     $  &  2.2   &  0.24  \\    
120514A  &  L &  N   &  XRT     &  2012-05-22 &  56069.08 &  1.0  & 8.03          &      $ -0.02   \pm  0.02     $  &  2.5   &  0.15  \\    
120514A  &  L &  N   &  XRT     &  2012-06-03 &  56082.00 &  1.0  & 20.95         &      $ -2.14   \pm  16.13    $  &  1.5   &  0.33  \\    
120514A  &  L &  N   &  XRT     &  Concat     &           &       &               &      $ 0.36    \pm  0.10     $  &  4.2   &  0.09  \\
\hline   
\end{tabular}
\end{table}
\begin{table}
\contcaption{}
\begin{tabular}{lcccccccccc}
\\
\hline
GRB$^{\mathrm{~a}}$ & Type$^{\mathrm{~b}}$ & Radio$^{\mathrm{~c}}$ & Telescope$^{\mathrm{~d}}$ & Date$^{\mathrm{~e}}$ & Start$^{\mathrm{~f}}$ & Integration$^{\mathrm{~g}}$ & Days$^{\mathrm{~h}}$ & Peak Flux$^{\mathrm{~i}}$ & Sig$^{\mathrm{~j}}$ & RMS$^{\mathrm{~k}}$ \\
 & & Flag & & (yyyy-mm-dd) & (MJD) & (hours) & (post-burst) & (mJy/beam) & & (mJy/beam) \\
\hline
120819A  &  L &  N   &  XRT     &  2012-08-23 &  56162.76 &  1.0  & 4.22          &      $ -0.09   \pm  0.07     $  &  3.0   &  0.19  \\    
120819A  &  L &  N   &  XRT     &  2012-08-26 &  56165.69 &  1.0  & 7.14          &      $ -53.18  \pm  5726.41  $  &  2.5   &  0.49  \\    
120819A  &  L &  N   &  XRT     &  2012-09-02 &  56172.79 &  1.0  & 14.24         &      $ 3.66    \pm  37.02    $  &  2.0   &  0.31  \\    
120819A  &  L &  N   &  XRT     &  2012-09-18 &  56188.71 &  1.0  & 30.17         &      $ 0.12    \pm  0.14     $  &  2.7   &  0.21  \\    
120819A  &  L &  N   &  XRT     &  Concat     &           &       &               &      $ -0.1    \pm  0.08     $  &  2.6   &  0.13  \\
\hline  
\end{tabular}
\end{table}
\begin{table}
\contcaption{}
\begin{tabular}{lcccccccccc}
\\
\hline
GRB$^{\mathrm{~a}}$ & Type$^{\mathrm{~b}}$ & Radio$^{\mathrm{~c}}$ & Telescope$^{\mathrm{~d}}$ & Date$^{\mathrm{~e}}$ & Start$^{\mathrm{~f}}$ & Integration$^{\mathrm{~g}}$ & Days$^{\mathrm{~h}}$ & Peak Flux$^{\mathrm{~i}}$ & Sig$^{\mathrm{~j}}$ & RMS$^{\mathrm{~k}}$ \\
 & & Flag & & (yyyy-mm-dd) & (MJD) & (hours) & (post-burst) & (mJy/beam) & & (mJy/beam) \\
\hline
121017A  &  L &  N   &  XRT     &  2012-10-18 &  56218.74 &  1.0  & 0.93          &      $ -0.29   \pm  0.18     $  &  1.7   &  0.28  \\    
121017A  &  L &  N   &  XRT     &  2012-10-20 &  56220.78 &  1.0  & 2.97          &      $ -0.40   \pm  0.87     $  &  1.8   &  0.19  \\    
121017A  &  L &  N   &  XRT     &  2012-10-25 &  56225.80 &  1.0  & 7.99          &      $ 0.31    \pm  0.87     $  &  2.0   &  0.15  \\    
121017A  &  L &  N   &  XRT     &  2012-11-09 &  56240.51 &  1.0  & 22.71         &      $ 0.07    \pm  0.07     $  &  1.7   &  0.28  \\    
121017A  &  L &  N   &  XRT     &  2012-11-15 &  56246.70 &  1.0  & 28.89         &      $ -0.49   \pm  2.22     $  &  2.2   &  0.11  \\
\hline 
\end{tabular}
\end{table}
\begin{table}
\contcaption{}
\begin{tabular}{lcccccccccc}
\\
\hline
GRB$^{\mathrm{~a}}$ & Type$^{\mathrm{~b}}$ & Radio$^{\mathrm{~c}}$ & Telescope$^{\mathrm{~d}}$ & Date$^{\mathrm{~e}}$ & Start$^{\mathrm{~f}}$ & Integration$^{\mathrm{~g}}$ & Days$^{\mathrm{~h}}$ & Peak Flux$^{\mathrm{~i}}$ & Sig$^{\mathrm{~j}}$ & RMS$^{\mathrm{~k}}$ \\
 & & Flag & & (yyyy-mm-dd) & (MJD) & (hours) & (post-burst) & (mJy/beam) & & (mJy/beam) \\
\hline
130102A  &  L &  N   &  XRT     &  2013-01-04 &  56296.57 &  1.0  & 1.82          &      $ -0.05   \pm  0.06     $  &  2.5   &  0.07  \\    
130102A  &  L &  N   &  XRT     &  2013-01-06 &  56298.43 &  1.0  & 3.68          &      $ -0.35   \pm  1.23     $  &  1.7   &  0.10  \\     
130102A  &  L &  N   &  XRT     &  2013-01-11 &  56303.44 &  1.0  & 8.69          &      $ -4.86   \pm  246.8    $  &  3.7   &  0.09  \\    
130102A  &  L &  N   &  XRT     &  2013-01-17 &  56309.57 &  1.0  & 14.82         &      $ 1.01    \pm  4.16     $  &  2.4   &  0.23  \\    
130102A  &  L &  N   &  XRT     &  2013-01-22 &  56314.35 &  1.0  & 19.60         &      $ -1.44   \pm  2.95     $  &  2.3   &  0.87  \\
\hline 
\end{tabular}
\end{table}
\begin{table}
\contcaption{}
\begin{tabular}{lcccccccccc}
\\
\hline
GRB$^{\mathrm{~a}}$ & Type$^{\mathrm{~b}}$ & Radio$^{\mathrm{~c}}$ & Telescope$^{\mathrm{~d}}$ & Date$^{\mathrm{~e}}$ & Start$^{\mathrm{~f}}$ & Integration$^{\mathrm{~g}}$ & Days$^{\mathrm{~h}}$ & Peak Flux$^{\mathrm{~i}}$ & Sig$^{\mathrm{~j}}$ & RMS$^{\mathrm{~k}}$ \\
 & & Flag & & (yyyy-mm-dd) & (MJD) & (hours) & (post-burst) & (mJy/beam) & & (mJy/beam) \\
\hline
130418A  &  L &  R   &  UVOT    &  2013-04-19 &  56401.82 &  2.0  & 1.03          &      $ 0.06    \pm  0.04     $  &  2.9   &  0.07  \\    
130418A  &  L &  R   &  UVOT    &  2013-04-20 &  56402.71 &  0.7  & 1.92          &      $ -0.55   \pm  2.19     $  &  2.2   &  0.15  \\    
130418A  &  L &  R   &  UVOT    &  2013-04-21 &  56403.74 &  3.3  & 2.95          &      $ 0.09    \pm  0.05     $  &  2.9   &  0.06  \\    
130418A  &  L &  R   &  UVOT    &  2013-04-23 &  56405.75 &  1.0  & 4.96          &      $ 0.18    \pm  0.10     $  &  2.3   &  0.13  \\    
130418A  &  L &  R   &  UVOT    &  2013-04-26 &  56408.79 &  1.0  & 8.00          &      $ 0.09    \pm  0.05     $  &  2.8   &  0.07  \\
\hline  
\end{tabular}
\end{table}
\begin{table}
\contcaption{}
\begin{tabular}{lcccccccccc}
\\
\hline
GRB$^{\mathrm{~a}}$ & Type$^{\mathrm{~b}}$ & Radio$^{\mathrm{~c}}$ & Telescope$^{\mathrm{~d}}$ & Date$^{\mathrm{~e}}$ & Start$^{\mathrm{~f}}$ & Integration$^{\mathrm{~g}}$ & Days$^{\mathrm{~h}}$ & Peak Flux$^{\mathrm{~i}}$ & Sig$^{\mathrm{~j}}$ & RMS$^{\mathrm{~k}}$ \\
 & & Flag & & (yyyy-mm-dd) & (MJD) & (hours) & (post-burst) & (mJy/beam) & & (mJy/beam) \\
\hline
130508A  &  L &  N   &  XRT     &  2013-05-08 &  56421.00 &  1.0  & 0.28          &      $ 0.55    \pm  0.21     $  &  3.0   &  0.14  \\    
130508A  &  L &  N   &  XRT     &  2013-05-10 &  56422.31 &  2.0  & 1.59          &      $ 0.13    \pm  0.08     $  &  2.4   &  0.07  \\    
130508A  &  L &  N   &  XRT     &  2013-05-11 &  56423.13 &  1.5  & 2.42          &      $ -0.15   \pm  0.22     $  &  1.9   &  0.17  \\    
130508A  &  L &  N   &  XRT     &  2013-05-12 &  56424.30 &  1.0  & 3.59          &      $ -0.0    \pm  0.0      $  &  2.5   &  0.08  \\    
130508A  &  L &  N   &  XRT     &  2013-05-13 &  56425.20 &  1.0  & 4.48          &      $ -0.02   \pm  0.02     $  &  1.6   &  0.07  \\
\hline 
\end{tabular}
\end{table}
\begin{table}
\contcaption{}
\begin{tabular}{lcccccccccc}
\\
\hline
GRB$^{\mathrm{~a}}$ & Type$^{\mathrm{~b}}$ & Radio$^{\mathrm{~c}}$ & Telescope$^{\mathrm{~d}}$ & Date$^{\mathrm{~e}}$ & Start$^{\mathrm{~f}}$ & Integration$^{\mathrm{~g}}$ & Days$^{\mathrm{~h}}$ & Peak Flux$^{\mathrm{~i}}$ & Sig$^{\mathrm{~j}}$ & RMS$^{\mathrm{~k}}$ \\
 & & Flag & & (yyyy-mm-dd) & (MJD) & (hours) & (post-burst) & (mJy/beam) & & (mJy/beam) \\
\hline
130606A  &  L &  C   &  XRT     &  2013-06-06 &  56449.88 &  1.0  & 0.00  (4.32)  &      $ -0.07   \pm  0.05     $  &  1.5   &  0.08  \\    
130606A  &  L &  C   &  XRT     &  2013-06-08 &  56452.00 &  1.0  & 2.12          &      $ 0.12    \pm  0.05     $  &  3.2   &  0.07  \\    
130606A  &  L &  C   &  XRT     &  2013-06-10 &  56453.88 &  1.0  & 4.00          &      $ 0.26    \pm  0.08     $  &  3.4   &  0.07  \\    
130606A  &  L &  C   &  XRT     &  2013-06-16 &  56459.93 &  1.0  & 10.06         &      $ 0.08    \pm  0.05     $  &  1.8   &  0.09  \\    
130606A  &  L &  C   &  XRT     &  2013-06-21 &  56464.00 &  1.0  & 14.13         &      $ 0.14    \pm  0.09     $  &  2.2   &  0.15  \\
\hline  
\end{tabular}
\end{table}
\begin{table}
\contcaption{}
\begin{tabular}{lcccccccccc}
\\
\hline
GRB$^{\mathrm{~a}}$ & Type$^{\mathrm{~b}}$ & Radio$^{\mathrm{~c}}$ & Telescope$^{\mathrm{~d}}$ & Date$^{\mathrm{~e}}$ & Start$^{\mathrm{~f}}$ & Integration$^{\mathrm{~g}}$ & Days$^{\mathrm{~h}}$ & Peak Flux$^{\mathrm{~i}}$ & Sig$^{\mathrm{~j}}$ & RMS$^{\mathrm{~k}}$ \\
 & & Flag & & (yyyy-mm-dd) & (MJD) & (hours) & (post-burst) & (mJy/beam) & & (mJy/beam) \\
\hline
130831A  &  L &  N   &  UVOT    &  2013-08-31 &  56535.93 &  3.0  & 0.38          &      $ -0.08   \pm  0.06     $  &  3.3   &  0.06  \\    
130831A  &  L &  N   &  UVOT    &  2013-09-01 &  56536.94 &  4.0  & 1.40          &      $ -0.02   \pm  0.02     $  &  3.0   &  0.05  \\    
130831A  &  L &  N   &  UVOT    &  2013-09-03 &  56538.92 &  4.0  & 3.38          &      $ -0.11   \pm  0.17     $  &  2.2   &  0.06  \\    
130831A  &  L &  N   &  UVOT    &  2013-09-07 &  56542.98 &  4.0  & 7.44          &      $ 0.12    \pm  0.07     $  &  2.6   &  0.07  \\    
130831A  &  L &  N   &  UVOT    &  2013-09-10 &  56545.97 &  4.0  & 10.43         &      $ 0.06    \pm  0.04     $  &  2.7   &  0.06  \\
\hline 
\end{tabular}
\end{table}
\begin{table}
\contcaption{}
\begin{tabular}{lcccccccccc}
\\
\hline
GRB$^{\mathrm{~a}}$ & Type$^{\mathrm{~b}}$ & Radio$^{\mathrm{~c}}$ & Telescope$^{\mathrm{~d}}$ & Date$^{\mathrm{~e}}$ & Start$^{\mathrm{~f}}$ & Integration$^{\mathrm{~g}}$ & Days$^{\mathrm{~h}}$ & Peak Flux$^{\mathrm{~i}}$ & Sig$^{\mathrm{~j}}$ & RMS$^{\mathrm{~k}}$ \\
 & & Flag & & (yyyy-mm-dd) & (MJD) & (hours) & (post-burst) & (mJy/beam) & & (mJy/beam) \\
\hline
140211A  &  L &  N   &  XRT     &  2014-02-11 &  56699.73 &  2.0  & 0.21          &      $ 0.01    \pm  0.01     $  &  1.5   &  0.10  \\     
140211A  &  L &  N   &  XRT     &  2014-02-13 &  56701.88 &  4.0  & 2.36          &      $ 0.08    \pm  0.03     $  &  3.0   &  0.03  \\    
140211A  &  L &  N   &  XRT     &  2014-02-16 &  56704.83 &  4.0  & 5.32          &      $ 0.10    \pm  0.03     $  &  2.5   &  0.04  \\    
140211A  &  L &  N   &  XRT     &  2014-02-18 &  56706.91 &  4.0  & 7.39          &      $ -0.02   \pm  0.03     $  &  2.3   &  0.04  \\    
140211A  &  L &  N   &  XRT     &  2014-02-21 &  56709.86 &  4.0  & 10.34         &      $ -0.67   \pm  10.87    $  &  2.2   &  0.03  \\
\hline 
\end{tabular}
\end{table}
\begin{table}
\contcaption{}
\begin{tabular}{lcccccccccc}
\\
\hline
GRB$^{\mathrm{~a}}$ & Type$^{\mathrm{~b}}$ & Radio$^{\mathrm{~c}}$ & Telescope$^{\mathrm{~d}}$ & Date$^{\mathrm{~e}}$ & Start$^{\mathrm{~f}}$ & Integration$^{\mathrm{~g}}$ & Days$^{\mathrm{~h}}$ & Peak Flux$^{\mathrm{~i}}$ & Sig$^{\mathrm{~j}}$ & RMS$^{\mathrm{~k}}$ \\
 & & Flag & & (yyyy-mm-dd) & (MJD) & (hours) & (post-burst) & (mJy/beam) & & (mJy/beam) \\
\hline
140508A  &  L &  C   &  UVOT    &  2014-05-17 &  56794.06 &  4.0  & 8.93          &      $ 0.08    \pm  0.04     $  &  3.1   &  0.05  \\    
140508A  &  L &  C   &  UVOT    &  2014-05-20 &  56797.07 &  4.0  & 11.95         &      $ 0.05    \pm  0.03     $  &  2.0   &  0.04  \\    
140508A  &  L &  C   &  UVOT    &  2014-05-21 &  56798.04 &  4.0  & 12.91         &      $ 0.07    \pm  0.03     $  &  2.9   &  0.04  \\    
140508A  &  L &  C   &  UVOT    &  2014-05-24 &  56801.92 &  4.0  & 16.79         &      $ 0.11    \pm  0.06     $  &  2.2   &  0.08  \\    
140508A  &  L &  C   &  UVOT    &  Concat     &           &       &               &    * $ 0.12    \pm  0.02     $  &  4.9   &  0.02  \\
\hline 
\end{tabular}
\end{table}
\begin{table}
\contcaption{}
\begin{tabular}{lcccccccccc}
\\
\hline
GRB$^{\mathrm{~a}}$ & Type$^{\mathrm{~b}}$ & Radio$^{\mathrm{~c}}$ & Telescope$^{\mathrm{~d}}$ & Date$^{\mathrm{~e}}$ & Start$^{\mathrm{~f}}$ & Integration$^{\mathrm{~g}}$ & Days$^{\mathrm{~h}}$ & Peak Flux$^{\mathrm{~i}}$ & Sig$^{\mathrm{~j}}$ & RMS$^{\mathrm{~k}}$ \\
 & & Flag & & (yyyy-mm-dd) & (MJD) & (hours) & (post-burst) & (mJy/beam) & & (mJy/beam) \\
\hline
140801A  &  L &  CS  &  XRT     &  2014-08-05 &  56874.03 &  4.0  & 3.24          &      $ 0.14    \pm  0.05     $  &  3.8   &  0.05  \\    
140801A  &  L &  CS  &  XRT     &  2014-08-06 &  56875.16 &  4.0  & 4.36          &      $ 0.05    \pm  0.04     $  &  3.0   &  0.07  \\    
140801A  &  L &  CS  &  XRT     &  2014-08-08 &  56877.11 &  4.0  & 6.32          &      $ 0.01    \pm  0.01     $  &  2.6   &  0.05  \\    
140801A  &  L &  CS  &  XRT     &  2014-08-09 &  56878.24 &  4.0  & 7.45          &      $ 0.10    \pm  0.05     $  &  3.5   &  0.05  \\    
140801A  &  L &  CS  &  XRT     &  2014-08-13 &  56882.08 &  4.0  & 11.28         &      $ 0.04    \pm  0.02     $  &  2.9   &  0.04  \\
\hline 
\end{tabular}
\end{table}
\begin{table}
\contcaption{}
\begin{tabular}{lcccccccccc}
\\
\hline
GRB$^{\mathrm{~a}}$ & Type$^{\mathrm{~b}}$ & Radio$^{\mathrm{~c}}$ & Telescope$^{\mathrm{~d}}$ & Date$^{\mathrm{~e}}$ & Start$^{\mathrm{~f}}$ & Integration$^{\mathrm{~g}}$ & Days$^{\mathrm{~h}}$ & Peak Flux$^{\mathrm{~i}}$ & Sig$^{\mathrm{~j}}$ & RMS$^{\mathrm{~k}}$ \\
 & & Flag & & (yyyy-mm-dd) & (MJD) & (hours) & (post-burst) & (mJy/beam) & & (mJy/beam) \\
\hline
141020A  &  L &  S   &  UVOT    &  2014-10-20 &  56950.33 &  2.0  & 0.00  (1.44)  &      $ 0.05    \pm  0.08     $  &  2.8   &  0.07  \\    
141020A  &  L &  S   &  UVOT    &  2014-10-22 &  56952.44 &  4.0  & 2.11          &      $ 3.40    \pm  184.72   $  &  2.2   &  0.05  \\    
141020A  &  L &  S   &  UVOT    &  2014-10-23 &  56953.54 &  4.0  & 3.21          &      $ -0.18   \pm  0.56     $  &  2.9   &  0.05  \\    
141020A  &  L &  S   &  UVOT    &  2014-10-24 &  56954.49 &  4.0  & 4.17          &    * $ 0.30    \pm  0.07     $  &  4.5   &  0.06  \\    
141020A  &  L &  S   &  UVOT    &  2014-10-27 &  56957.53 &  4.0  & 7.21          &      $ 0.13    \pm  0.05     $  &  3.0   &  0.05  \\
\hline 
\end{tabular}
\end{table}
\begin{table}
\contcaption{}
\begin{tabular}{lcccccccccc}
\\
\hline
GRB$^{\mathrm{~a}}$ & Type$^{\mathrm{~b}}$ & Radio$^{\mathrm{~c}}$ & Telescope$^{\mathrm{~d}}$ & Date$^{\mathrm{~e}}$ & Start$^{\mathrm{~f}}$ & Integration$^{\mathrm{~g}}$ & Days$^{\mathrm{~h}}$ & Peak Flux$^{\mathrm{~i}}$ & Sig$^{\mathrm{~j}}$ & RMS$^{\mathrm{~k}}$ \\
 & & Flag & & (yyyy-mm-dd) & (MJD) & (hours) & (post-burst) & (mJy/beam) & & (mJy/beam) \\
\hline
141225A  &  L &  N   &  UVOT    &  2014-12-25 &  57016.96 &  2.0  & 0.00  (2.88)  &      $ 0.03    \pm  0.03     $  &  1.8   &  0.15  \\    
141225A  &  L &  N   &  UVOT    &  2014-12-27 &  57018.96 &  4.0  & 2.00          &      $ -0.14   \pm  0.32     $  &  2.0   &  0.05  \\    
141225A  &  L &  N   &  UVOT    &  2014-12-28 &  57019.96 &  4.0  & 3.00          &      $ -0.01   \pm  0.02     $  &  2.8   &  0.04  \\    
141225A  &  L &  N   &  UVOT    &  2015-01-03 &  57025.03 &  4.0  & 8.07          &      $ -0.02   \pm  0.02     $  &  2.9   &  0.05  \\    
141225A  &  L &  N   &  UVOT    &  2015-01-04 &  57026.98 &  4.0  & 10.02         &      $ -0.0    \pm  0.0      $  &  2.7   &  0.04  \\
\hline 
\end{tabular}
\end{table}
\begin{table}
\contcaption{}
\begin{tabular}{lcccccccccc}
\\
\hline
GRB$^{\mathrm{~a}}$ & Type$^{\mathrm{~b}}$ & Radio$^{\mathrm{~c}}$ & Telescope$^{\mathrm{~d}}$ & Date$^{\mathrm{~e}}$ & Start$^{\mathrm{~f}}$ & Integration$^{\mathrm{~g}}$ & Days$^{\mathrm{~h}}$ & Peak Flux$^{\mathrm{~i}}$ & Sig$^{\mathrm{~j}}$ & RMS$^{\mathrm{~k}}$ \\
 & & Flag & & (yyyy-mm-dd) & (MJD) & (hours) & (post-burst) & (mJy/beam) & & (mJy/beam) \\
\hline
150317A  &  L &  N   &  XRT     &  2015-03-17 &  57098.68 &  2.0  & 0.50          &      $ -0.98   \pm  10.81    $  &  3.6   &  0.06  \\    
150317A  &  L &  N   &  XRT     &  2015-03-18 &  57099.88 &  4.0  & 1.69          &      $ 0.08    \pm  0.04     $  &  6.7   &  0.04  \\    
150317A  &  L &  N   &  XRT     &  2015-03-20 &  57101.88 &  4.0  & 3.69          &      $ 0.03    \pm  0.03     $  &  2.8   &  0.09  \\    
150317A  &  L &  N   &  XRT     &  2015-03-26 &  57107.85 &  4.0  & 9.67          &      $ 0.02    \pm  0.02     $  &  4.4   &  0.04  \\    
150317A  &  L &  N   &  XRT     &  Concat     &           &       &               &      $ 0.08    \pm  0.03     $  &  7.6   &  0.02  \\
\hline
\end{tabular}
\\
Note: All GRB observations from GRB 130813A onwards followed the updated ALARRM strategy (see Section 2.1).\\
$^{\mathrm{~a}}$ GRB discovery flag: $\dagger$: \textit{Fermi} discovered GRB; $\ddagger$: \textit{Integral} discovered GRB\\
$^{\mathrm{~b}}$ GRB Type: S - short, L - long \\
$^{\mathrm{~c}}$ Radio detection flag: R - radio afterglow first detected with another radio telescope; S - AMI detected coincident radio source is likely steady; C - coincident radio source detected in AMI concatenated image; CS - coincident radio source detected in AMI concatenated image that is confirmed to be a steady source; N - no detection. \\
$^{\mathrm{~d}}$ Best \swift\ telescope position used to search for a radio counterpart: BAT - \swift\ Burst Alert Telescope; XRT - \swift\ X-ray Telescope; UVOT - \swift\ Ultraviolet/Optical Telescope. For GRB 130521A, the best optical position provided by Skynet/PROMPT (James et al., 2013, GCN, 14713) was used instead. \\
$^{\mathrm{~e}}$ Date of the AMI observation in yyyy-mm-dd. Concat - concatenation of all epochs with the same pointing. \\
$^{\mathrm{~f}}$ Start date of the AMI observation in modified julian date (MJD) format. \\
$^{\mathrm{~g}}$ AMI observation length in hours. \\
$^{\mathrm{~h}}$ Number of days post-burst since the start of the AMI observation. The response time, in minutes, has also been included in brackets for those AMI observations that began $\leq0.01$ days post-burst. \\
$^{\mathrm{~i}}$ Peak flux density as reported by {\sc PySE}. The `*' symbol indicates those GRBs for which the listed AMI peak flux is the measured flux of a radio source detected above a $4\sigma_s$ significance, coincident with the best known \swift\ position of the GRB (i.e. within $3\sigma_p$). All other listed fluxes are derived from a forced Gaussian fit at the best known \swift\ position. The $1\sigma_s$ error bar is the flux error output by {\sc PySE}, added in quadrature to the AMI 5\% calibration error. \\
$^{\mathrm{~j}}$ The significance of the AMI flux reported by {\sc PySE}, which are measured in units of $\sigma_s$ above a local RMS. (Note that the {\sc PySE} forced fitting algorithm failed on the 2012-04-05 and 2012-04-08 observations of GRB 120311A and all of the GRB 121128A. The forced fits were conducted using MIRIAD and the reported significance is just the forced fitted peak flux divided by the global RMS reported in the last column.) \\
$^{\mathrm{~k}}$ The global RMS of the middle quarter of the AMI image. 
\end{table}
\end{center}

\twocolumn

\section{Discussions of AMI observed GRBs}

In this section we individually discuss those GRBs with \textit{confirmed} radio afterglow detections with AMI that obey the criteria presented in Section~\ref{criteria}. We then briefly comment on those GRBs that may have been detected in just one AMI epoch, which we classify as \textit{possible} radio afterglow detections, and those GRBs that are coincident with a radio source detected in their deep concatenated image only. We also list those GRBs that appear to be coincident with a steady radio source and discuss the ramifications of such associations. 

It is important to note that some of the radio detections discussed in this section were made at early times and may therefore suffer from scintillation effects \citep[for example see][]{frail97}. However, 15.7 GHz is usually above the scattering frequency in a given direction and therefore in the weak scattering regime. This implies that scintillation effects are likely to be minimal \citep[see Table 1 of][]{granot14}. Of course, this assumption is predicated upon the simple framework of \citet{walker98}, which assumes a single scattering screen at a fixed distance. It is therefore possible there are unaccounted scintillation effects that result in flux variations more significant than the quoted $1\sigma_s$ errors.

\subsection{New \textit{confirmed} GRB radio afterglows discovered with AMI}\label{newgrb}

This section individually discusses the six GRBs whose \textit{confirmed} radio afterglows were discovered with AMI. In five of the cases, the proposed counterparts fulfil the two criteria for a radio afterglow identification, while the other was identified via concatenating the early and late epochs separately (see Section~\ref{criteria}). 

\begin{figure*}
\begin{center}
\includegraphics[width=0.47\textwidth]{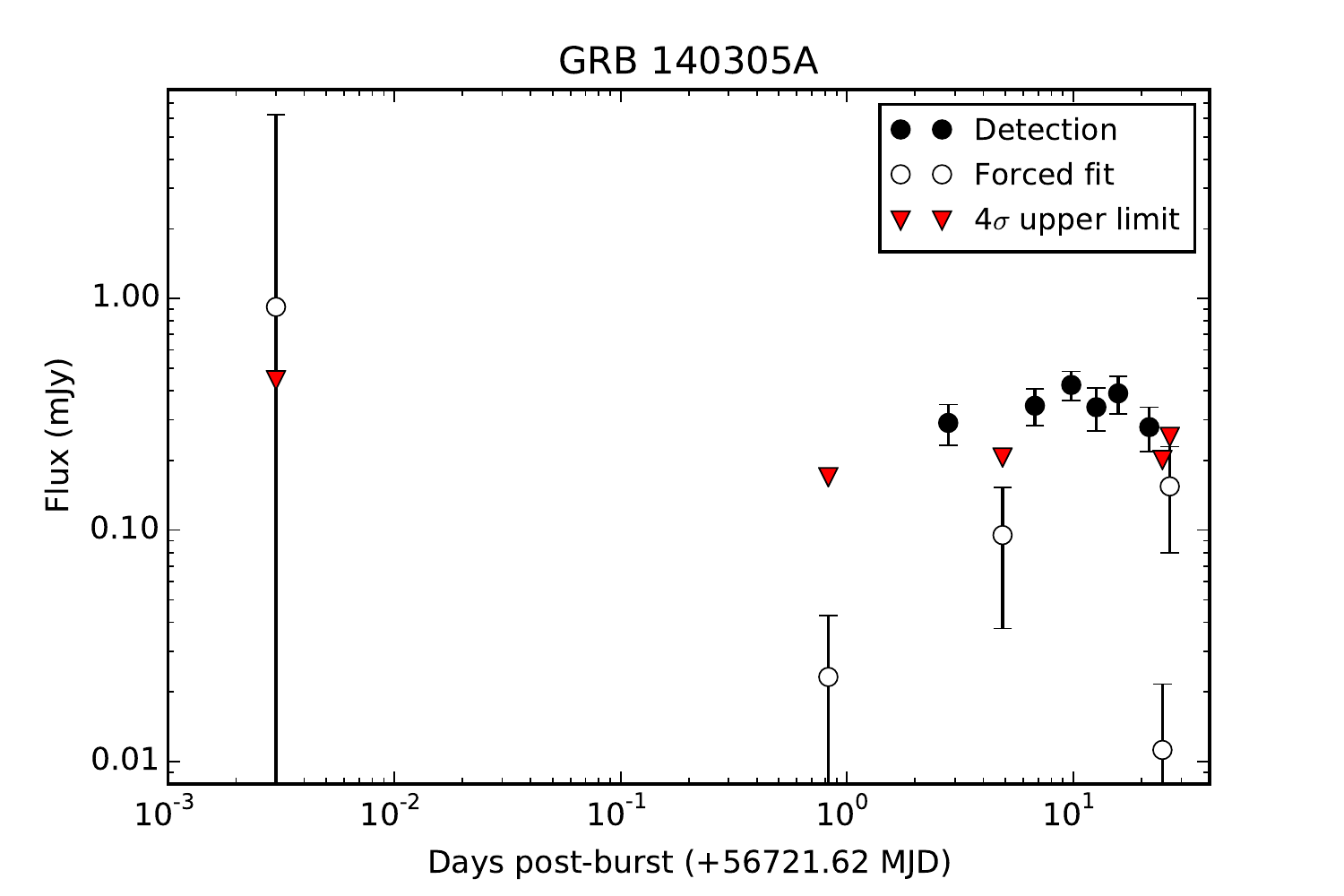}
\includegraphics[width=0.47\textwidth]{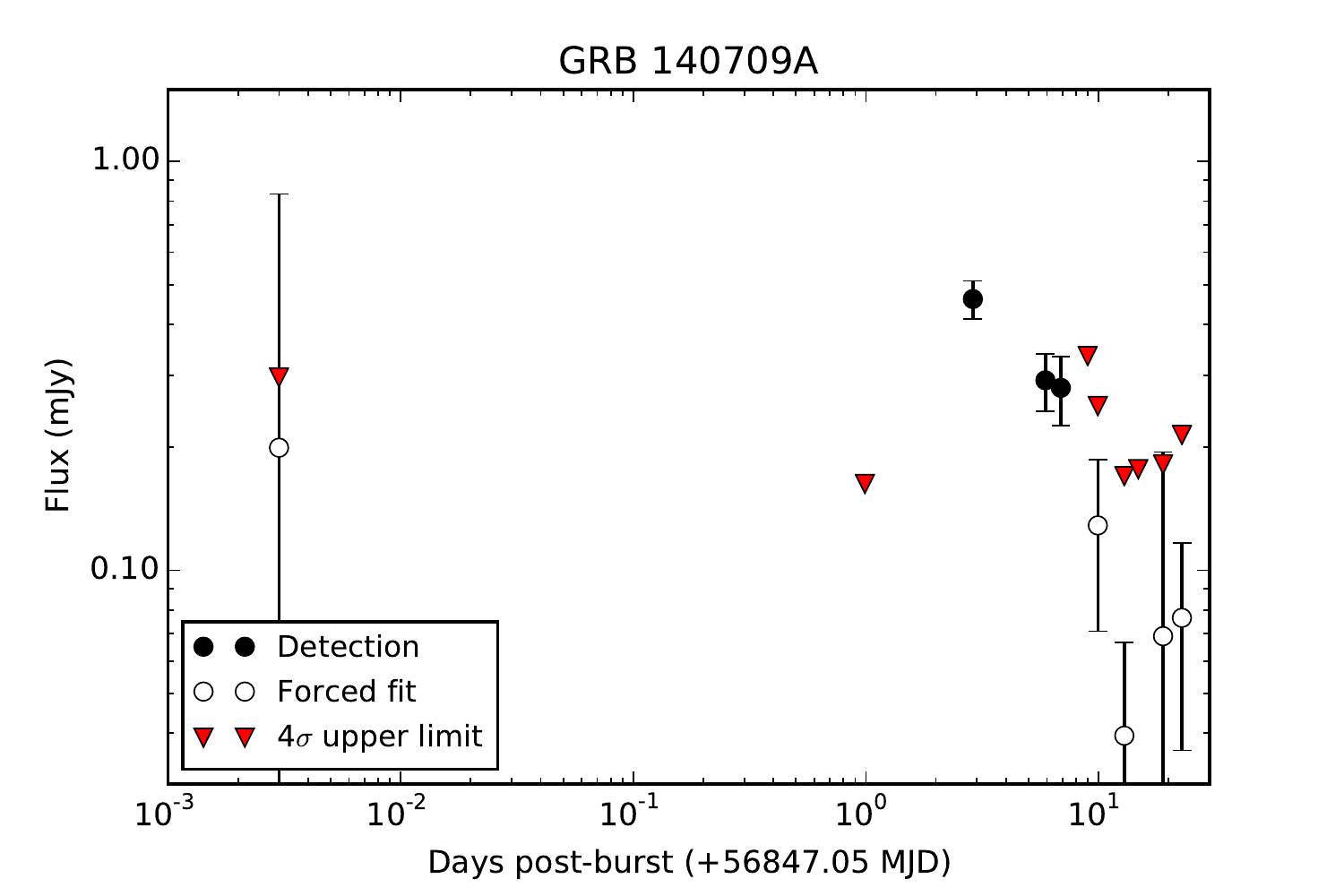}
\includegraphics[width=0.47\textwidth]{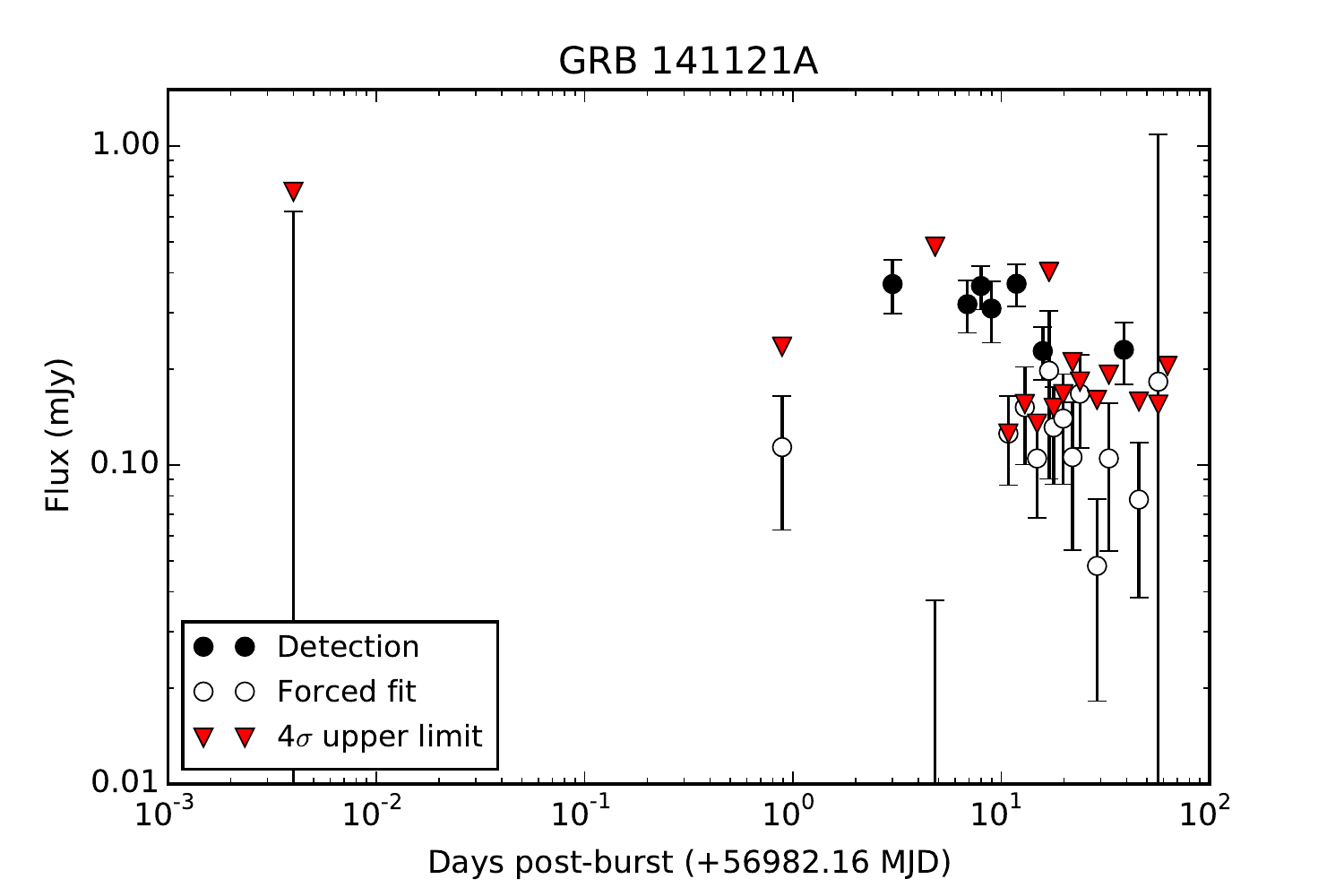}
\includegraphics[width=0.47\textwidth]{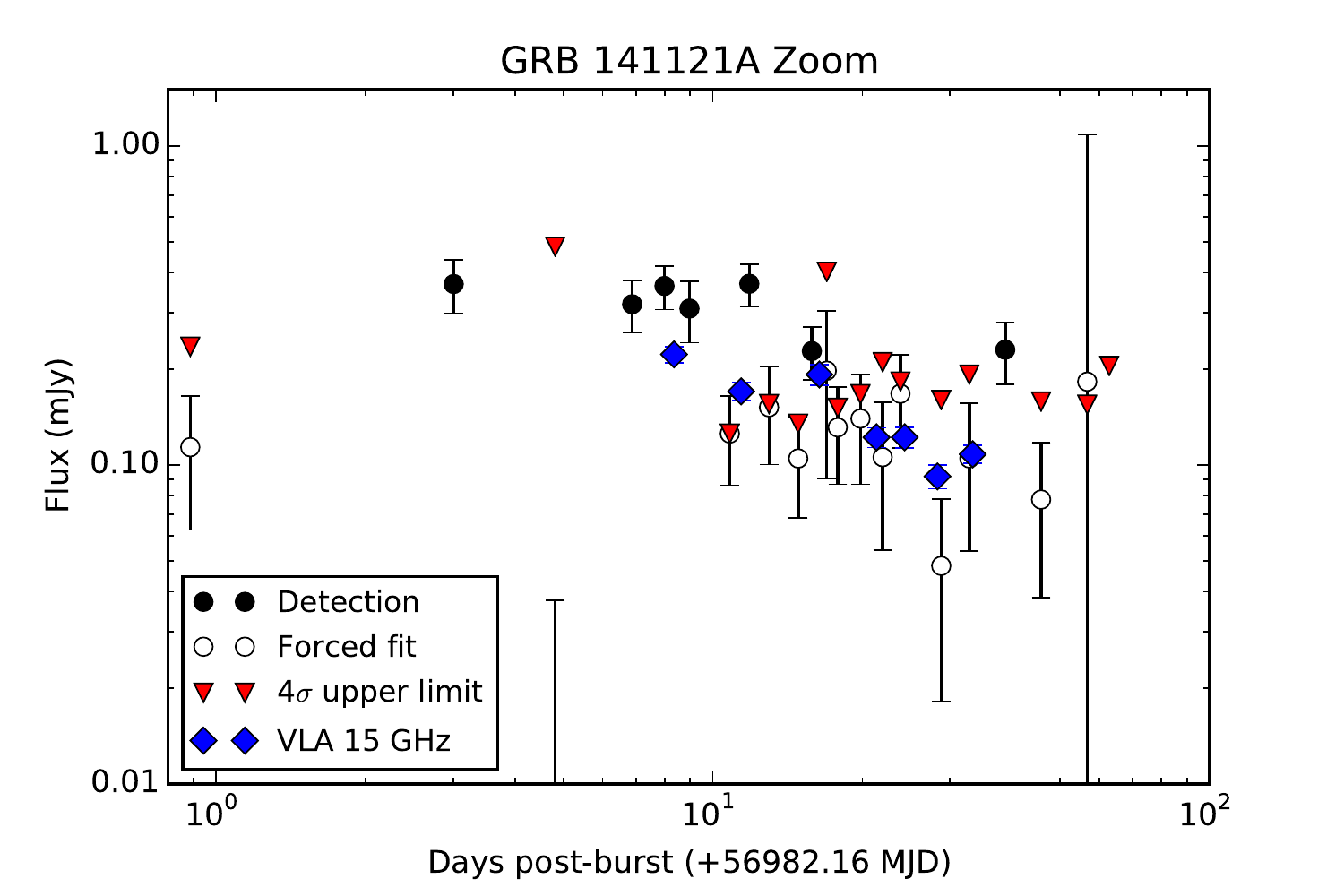}
\includegraphics[width=0.47\textwidth]{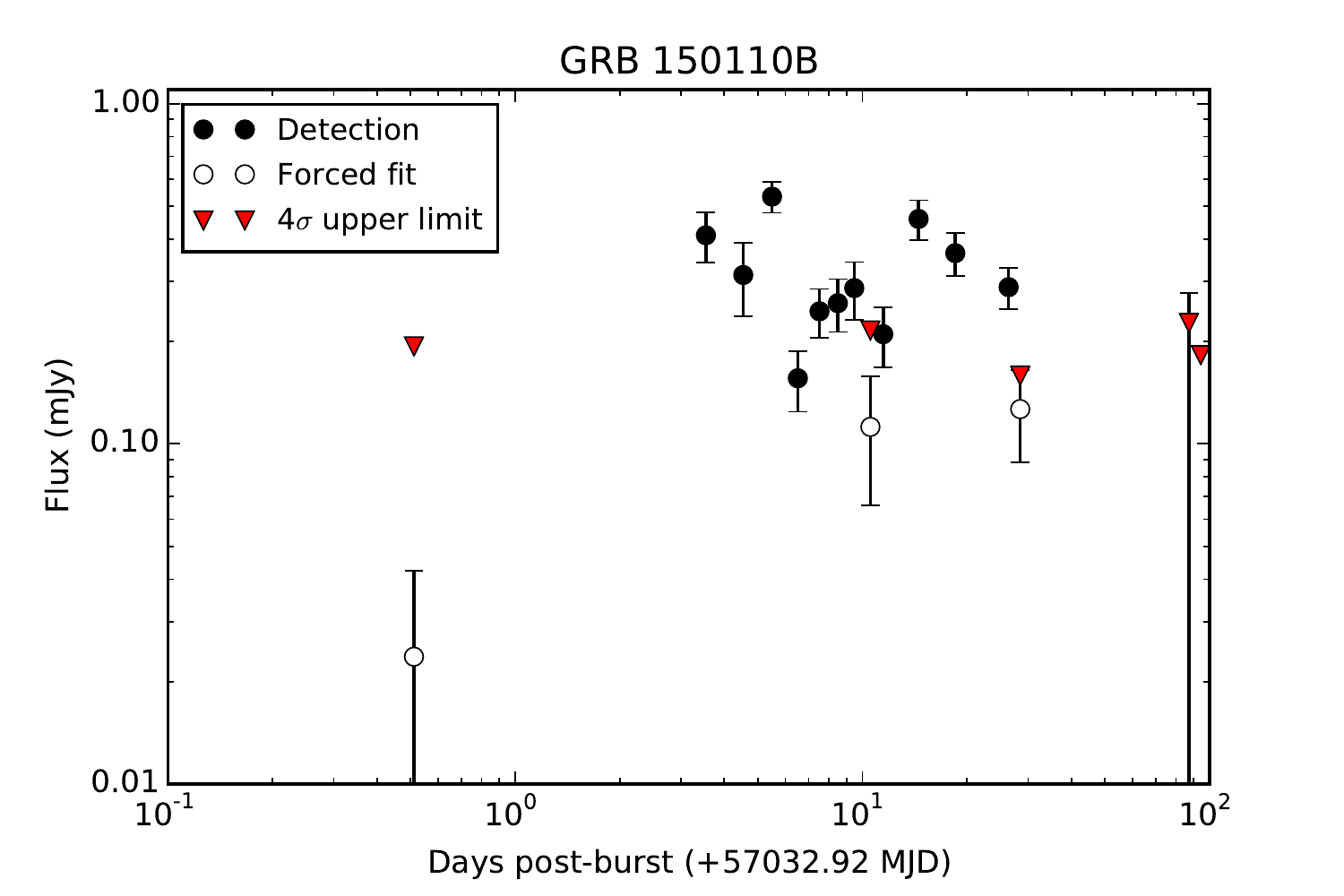}
\includegraphics[width=0.47\textwidth]{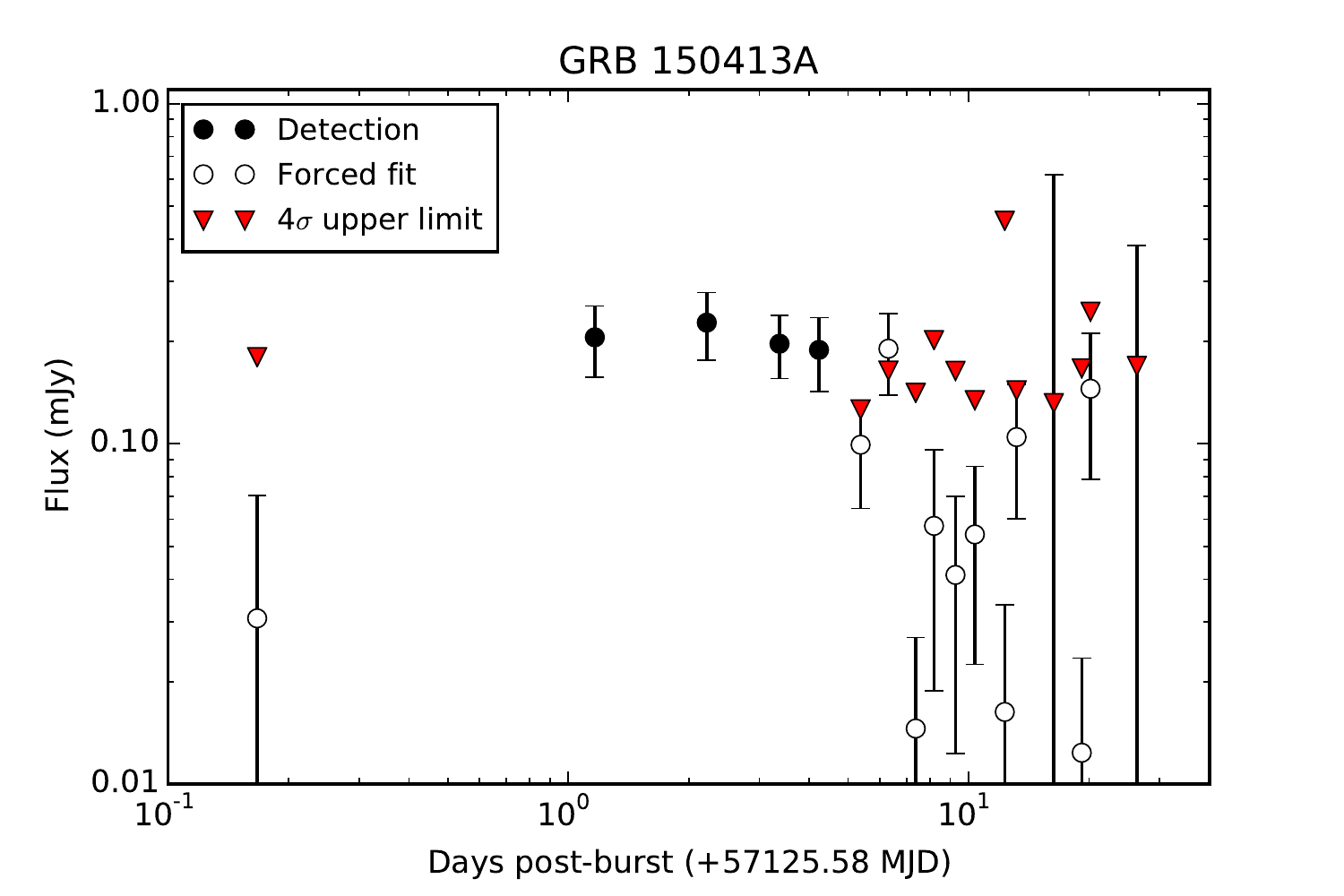}
\end{center}
\caption{The AMI 15.7 GHz light curves of new \textit{confirmed} GRBs discovered with AMI. The black data points are the $>4\sigma_s$ AMI detections of the GRB radio afterglows. The open circles are the fluxes measured from a point source force-fitted at the position of the GRB in those AMI datasets for which there was no detection. In some cases the flux measured from the forced fit, and/or the negative error-bar, are consistent with zero and were therefore either not depicted or extend lower than the y-axis). The $4 \sigma_s$ upper limits on the non-detections are also illustrated (red triangles). The light-curve of GRB 141121A is plotted twice, the second plot zooming in at late times ($\geq1$ day post-burst) so as to better discern the light-curve structure, while also including VLA 15\,GHz detections \citep{cucchinara15}. All error-bars are $1\sigma_s$.}
\label{fig:1}
\end{figure*}

\subsubsection{GRB 140305A}

GRB 140305A was detected by \swift\ BAT at 15:00:20 UT but, due to a Solar observing constraint, was not followed-up at X-ray or optical wavelengths \citep{cummings14c}. As a result only a BAT position with a $1.7'$ 90\% error circle was obtained for this event. AMI was on-target and observing this GRB for 2 hours within $5$ minutes post-burst, followed by 10 subsequent observations ranging from 3 to 5 hrs in duration up until 2014 April 1. The deep image resulting from the concatenation of 11 observation of this source showed the detection of 2 sources within the BAT error circle. One of these two sources was NVSS 225809+152439, which lies $3.2'$ SE of the BAT position. The second source was uncatalogued and lies only $7.8"$ from the BAT position at RA (J2000.0) = 22:57:59.32 ($\pm 1.70"$) and Dec (J2000.0) = +15:27:00.60 ($\pm 2.84"$). In the individual epochs this uncatalogued radio source was not detected until the third AMI observation on 2014 March 8 (2.80 days post-burst), with a $5\sigma_s$ detection of $0.29 \pm 0.06$ mJy/beam. A further five AMI observations detected this radio source with a flux significance between $4.6 < \sigma_s < 6.9$ until it dropped below detectability on 2014 March 30 (25 days post-burst). The light curve of GRB 140305A can be found in Figure~\ref{fig:1}. Given that the most sensitive AMI non-detection on 2014 March 30 would have detected the brightest flux measurement of $0.42 \pm 0.06$ mJy/beam (observed on 2014 March 15) with a $\sim6\sigma_s$ significance, we identify this uncatalogued radio source as transient and likely the radio counterpart to GRB 140305A.

\subsubsection{GRB 140629A}\label{grb140629a}

AMI monitored GRB 140629A nine times from 1.27 until 23.16 days post-burst. A blind search of the deep concatenated image detected a $5.0\sigma_s$ source with flux $0.10 \pm 0.02$ mJy/beam within $1.4\sigma_p$ of the UVOT position. However, no sources were blindly detected in the individual epochs. We therefore concatenated the four earliest AMI observations of GRB 140629A ranging between 2 to 6 days post-burst, and the four latest observations between 7 to 23 days post-burst, to search for transient behaviour using the technique described in Section 2.3.1. The first concatenated epoch detected a radio source coincident with that detected in the deep concatenated image but with a brighter flux of $0.15 +/- 0.03$ mJy/beam at a $5.2\sigma_s$ significance. The second concatenated epoch had a comparable RMS noise to the first concatenated epoch but no coincident radio source was detected down to a $4\sigma_s$ upper limit of 0.09 mJy/beam. This analysis demonstrates that GRB 140629A obeys criterion (ii) in Section~\ref{criteria}, suggesting the coincident radio source was transient in nature, and therefore likely the radio afterglow of GRB 140629A, which faded below detectability at $\gtrsim6$ days post-burst. 

\subsubsection{GRB 140709A}

The XRT position of GRB 140709A lies $30.1''$ from the NVSS 201841+511349, which has an integrated flux of $1.13 \pm 0.06$ mJy/beam at 15.7 GHz based on the concatenated AMI image. During the 2014 July 11 AMI observation of GRB 140709A, which took place 2.9 days post burst, a radio source began to appear at the GRB XRT position. However, due to the AMI-LA resolution of $30''$, this potential radio afterglow was blended with NVSS 201841+511349. In order to investigate whether this new radio source was real, {\sc CASA} was used to subtract a model of the contribution of the NVSS source from the visibilities. The resulting image revealed an unresolved point source at the position of the GRB with a peak flux of $0.46 \pm 0.05$ mJy/beam \citep{anderson14e}. This same technique was performed on all the AMI observations, showing the detection of a fading point source present in two subsequent observations (see Figure~\ref{fig:1}), fading below detectability by 2014 July 18. For those AMI observations for where radio afterglows were detected, forced fit fluxes were obtained manually using the radio reduction software {\sc MIRIAD} \citep{sault95}. The reported significance on these forced fits reported in Table~\ref{tab:1} correspond to the forced fit flux divided by the RMS noise (which is why some of these values are negative). We therefore identify this transient radio source as the likely radio afterglow of GRB 140709A. While no other radio detections were reported, the optical counterpart detected by \citet{castro-tirado14c} was faint, and further studies indicate that this GRB may be a member of the ``dark" burst population \citep{littlejohns15}. Several proposed suggestions for their dark nature may be due to an intrinsically faint optical counterpart, that they reside in high redshift galaxies, or that their optical emission is obscured by dust and gas extinction in their host galaxies \citep[e.g.][]{vanderhorst09}.

\subsubsection{GRB 141121A}

GRB 141121A is a member of the newly established class of ultra-long GRBs (UL-GRBs), which have prompt $\gamma$-ray emission lasting for $\gtrsim1000$s \citep[e.g.][]{virgili13,evans14,levan14}. Through the ALARRM trigger, AMI was on-target and observing GRB 141121A just 6 minutes post-burst with a follow-up observation occurring 0.9 days post-burst. It was not until 2014 Nov 24, 3 days post burst, that AMI detected the radio counterpart to GRB 141121A, resulting in a flux of $0.37 \pm 0.07$ mJy/beam. This is the first and earliest reported detection of the radio counterpart to GRB 141121A \citep{anderson14c}, which was then observed and detected with the WSRT at 4.9 GHz \citep{vanderhorst14b} and the VLA at 6.2 and 14 GHz \citep{corsi14c,corsi14d}. Figure~\ref{fig:1} shows the AMI light curve of GRB 141121A. A multi-wavelength analysis of GRB 141121A has also been conducted by \citet{cucchinara15}.

A zoomed in light curve beginning $\sim1$ days post-burst that includes the 15.7\,GHz VLA detections \citep{cucchinara15} can be found in Figure~\ref{fig:1}. The AMI detections, forced fits and upper limits are consistent with the VLA detections but the AMI observations also show evidence for radio re-brightening at $\sim12$ and $\sim39$ days post-burst (note that the AMI detection at 39 days is within $3\sigma_s$ of the VLA detection at 33 days but both detections show the 15.7\,GHz flux is increasing at late times). \citet{cucchinara15} demonstrated that around 3 days, when the 15.7\,GHz flux is near its peak, at least 50\% of the radio flux is being contributed by the reverse-shock. The flatness of the AMI light curve up to 12 days post-burst, followed by evidence for late time rebrightening, may support the detection of multiple peaks from the forward- and reverse-shock at 15.7\,GHz. Energy injections, which is supported by the detection of flares and plateaus in the XRB light curve of GRB 141121A \citep{cucchinara15}, could also cause radio light-curve flattening \citep{frail04}. However, this flatness may also be an artefact of the radio counterpart being close to the AMI detection limit. The VLA light curves presented in \citet{cucchinara15} also show late-time radio modulations, particularly at lower-frequencies, which they attribute to scintillation. While the AMI observing frequency is in the weak scattering regime, it may still be contributing towards the fluctuations. However, such scintillation is unlikely to be significant enough to explain the variations seen in the GRB 141121A light curve.

\subsubsection{GRB 150110B}

The ALARRM trigger on GRB 150110B resulted in the first AMI observation occurring at 0.5 days post-burst, when the source had risen above the horizon. AMI then detected the radio counterpart to GRB 150110B on 2015 Jan 14, corresponding to 3.54 days post-burst, with a flux of $0.41 \pm 0.07$ mJy/beam \citep{anderson15b}. No optical counterpart was detected for this GRB but the limits are not particularly constraining \citep{kuin15,mazaeva15}. This radio counterpart remained detectable up until 27 days post-burst during which it showed periods of rapid variability (see Figure~\ref{fig:1}). The brightest detection with AMI was on 2015 Jan 16, 5.5 days post-burst, displaying a flux of $0.53 \pm 0.06$ mJy/beam, which then rapidly dropped by a factor of 3 when observed only one day later, implying a steep temporal index of $\alpha=-9 \pm 1$ (for flux $F(t) \propto t^{\alpha}$). The radio flux then took another 8 days to peak again, 14.5 days post-burst at $0.46 \pm 0.06$ mJy/beam, before it faded below detectability $\sim28$ days post-burst. It is possible that this variability could have been caused by scintillation, which has been shown to be significant for many GRBs, in some cases with modulations $>100\%$ \citep[for example see][]{frail00c}. Scintillation is a reasonable assumption for this event as in the direction of GRB 150110B, the transition frequency at which the scattering strength is unity is $\nu_{0} \approx 20$ GHz, placing 15.7 GHz in the strong scattering regime \citep{walker98}. 

\subsubsection{GRB 150413A}

ALARRM had AMI on-target and observing GRB 150413A  when the source had risen above the horizon at 0.17 days post-burst. The follow-up observation that took place on 2015 April 14, just 1.2 days post-burst, resulted in the detection of a coincident radio source with a flux of $0.21 \pm 0.05$ mJy/beam \citep{anderson15a,anderson15}. The source remained detectable until 2015 April 17, after which it dropped below the AMI sensitivity (see Figure~\ref{fig:1}). In the GRB rest-frame these 4 detections occurred within $\sim1$ day post-burst so it is possible we detected the reverse-shock radio peak. The most sensitive late time AMI observation proceeding the detections on 2015 April 29 would have detected the brightest flux (from the 2015 April 15 observation) at a $6\sigma_s$ level, indicating transient behaviour. We therefore suggest that this coincident radio source is the radio afterglow of GRB 150413A. The only other radio observation reported for this event was using the Nanshan 25m radio dish in the pulsar search mode, the purpose of which was to search for an associated FRB(s), but no results were included \citep{xu15}. GRB 150413A also had a bright optical counterpart \citep[e.g.][]{tyurina15} and \citet{gorbovskoy16} searched for early time optical polarisation but none was detected. The flux forced fitting in the AMI non-detections was performed at the optical position provided by \citep{ivanov15}. 

\subsection{\textit{Confirmed} radio-detected GRBs detected with AMI}\label{knowngrb}

This section individually discusses the seven radio-detected GRBs that were detected by AMI. We define ``radio-detected GRBs" as those for which the radio afterglow was first reported from a detection made with a telescope other than AMI. The detection must have been reported in the GCN or in the literature, and not subsequently reclassified as a steady radio source. These GRBs also obey the \textit{confirmed} radio afterglow selection criteria presented in Section~\ref{criteria}.

\begin{figure*}
\centering
\includegraphics[width=0.47\textwidth]{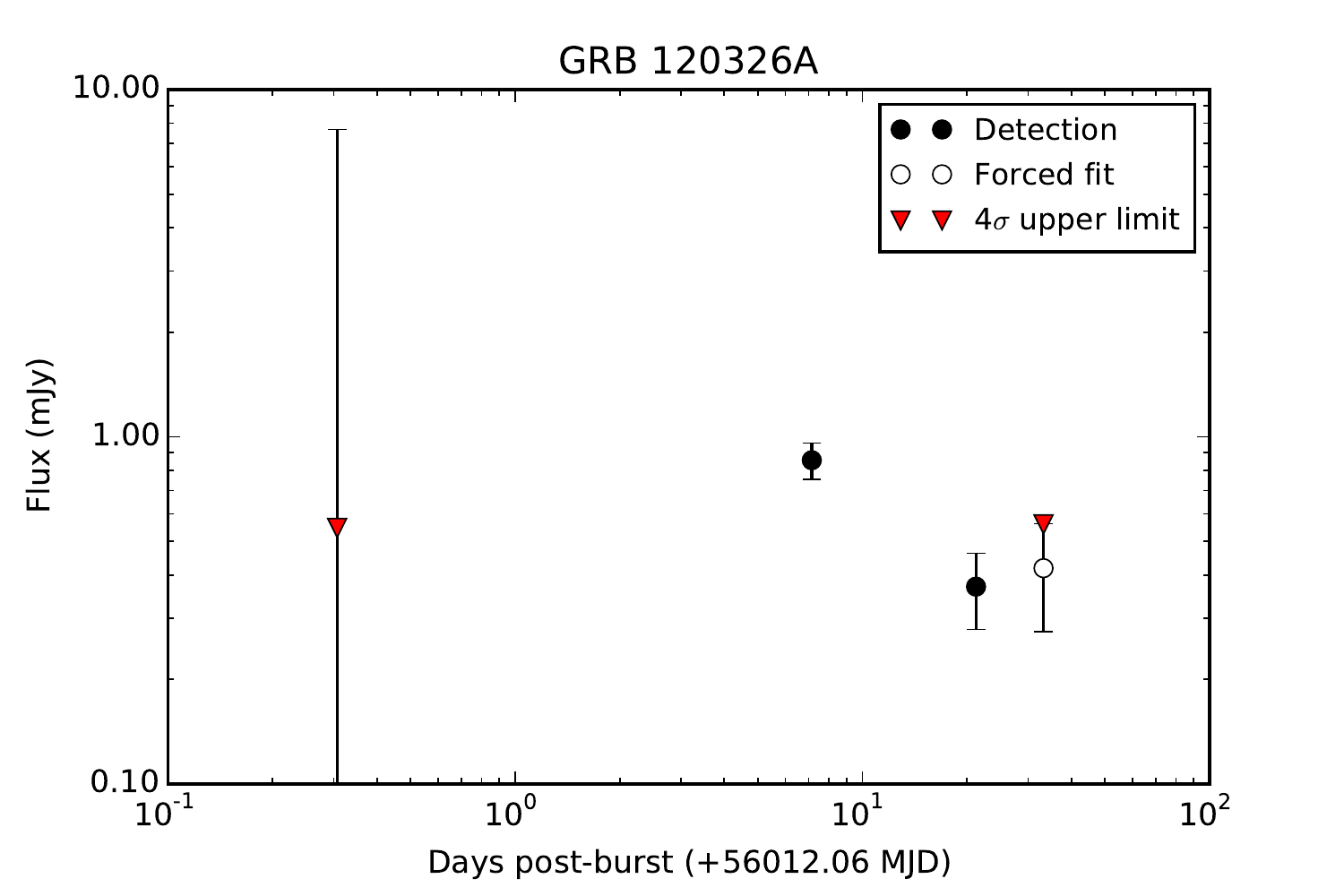}
\includegraphics[width=0.47\textwidth]{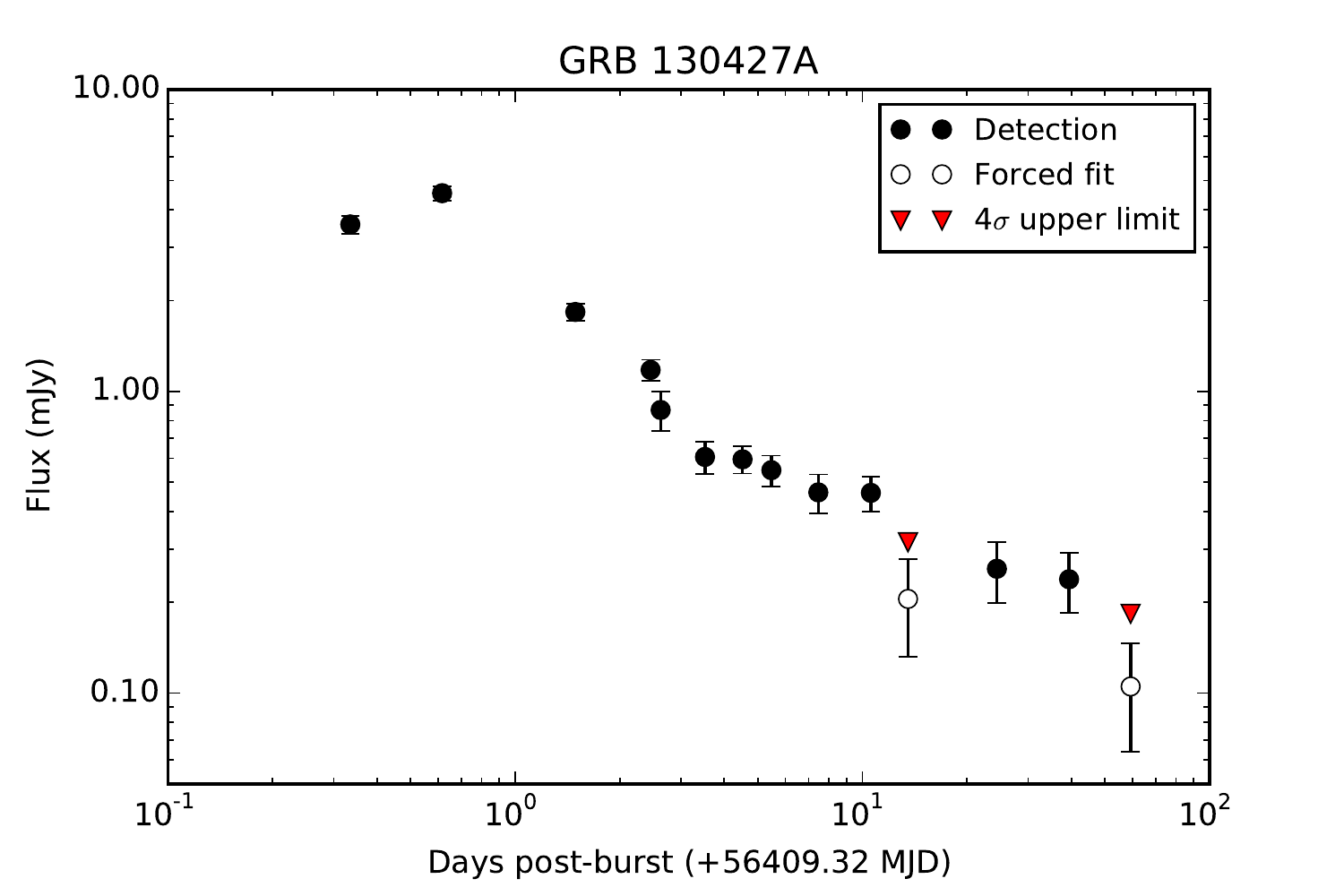}
\includegraphics[width=0.47\textwidth]{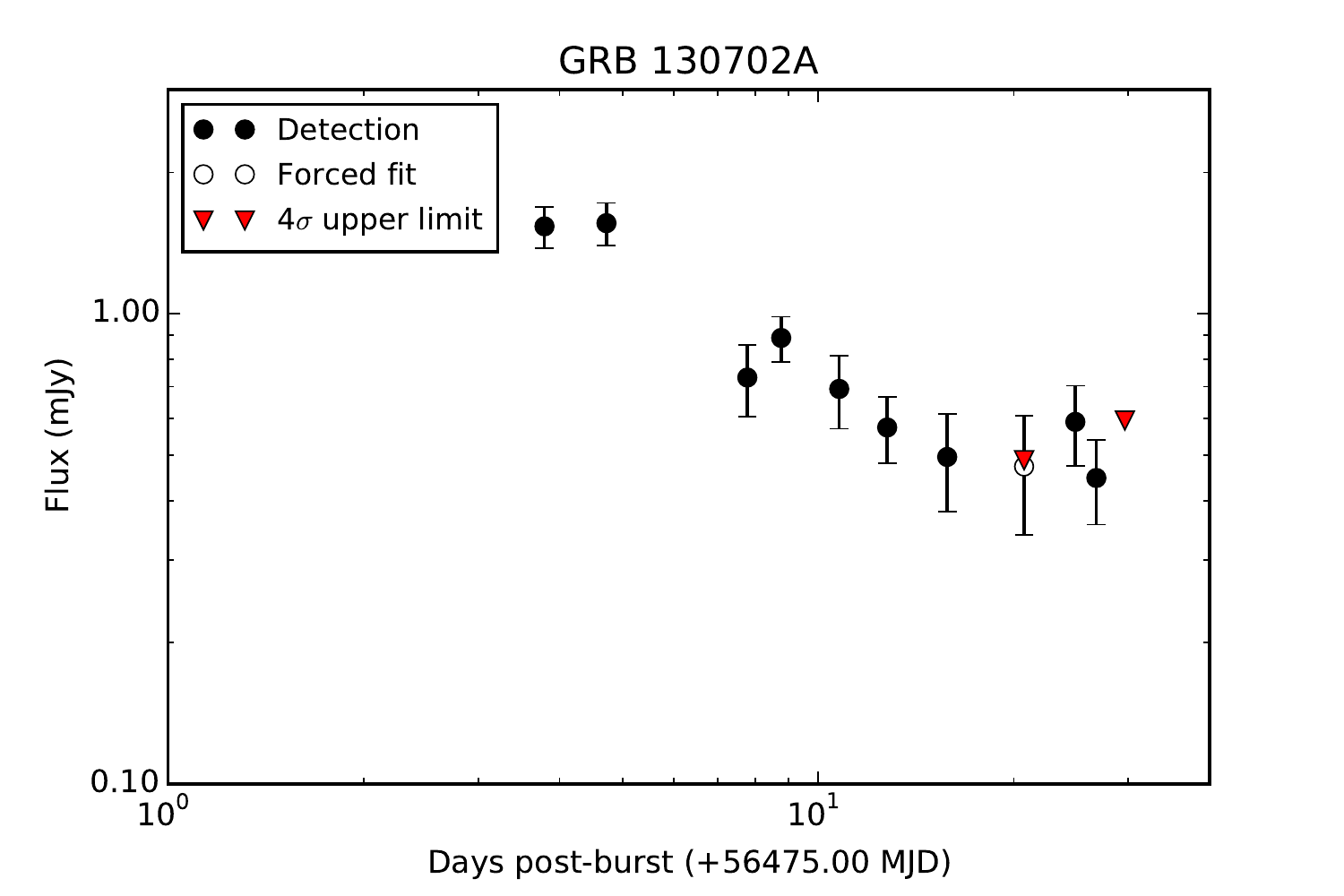}
\includegraphics[width=0.47\textwidth]{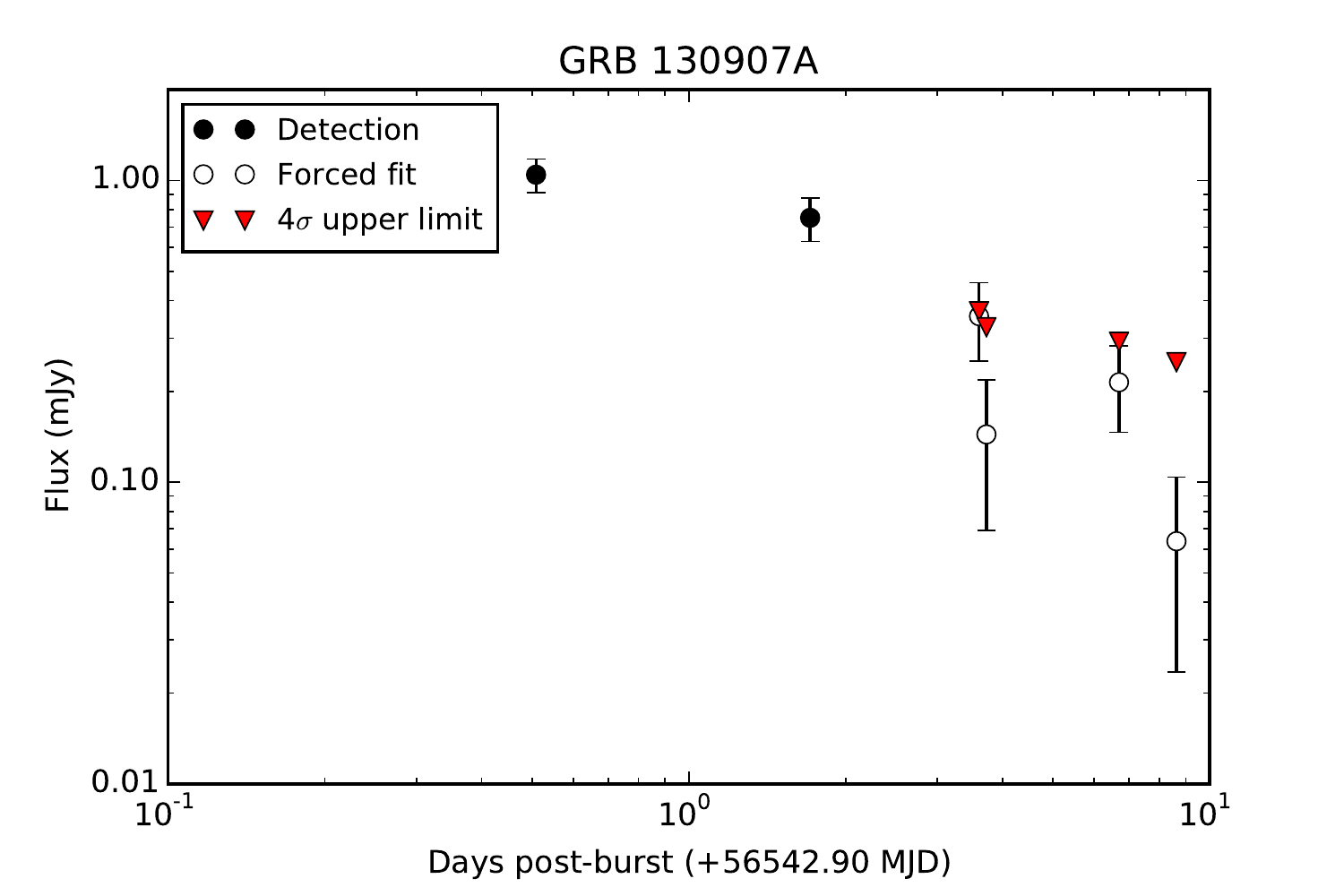}
\includegraphics[width=0.47\textwidth]{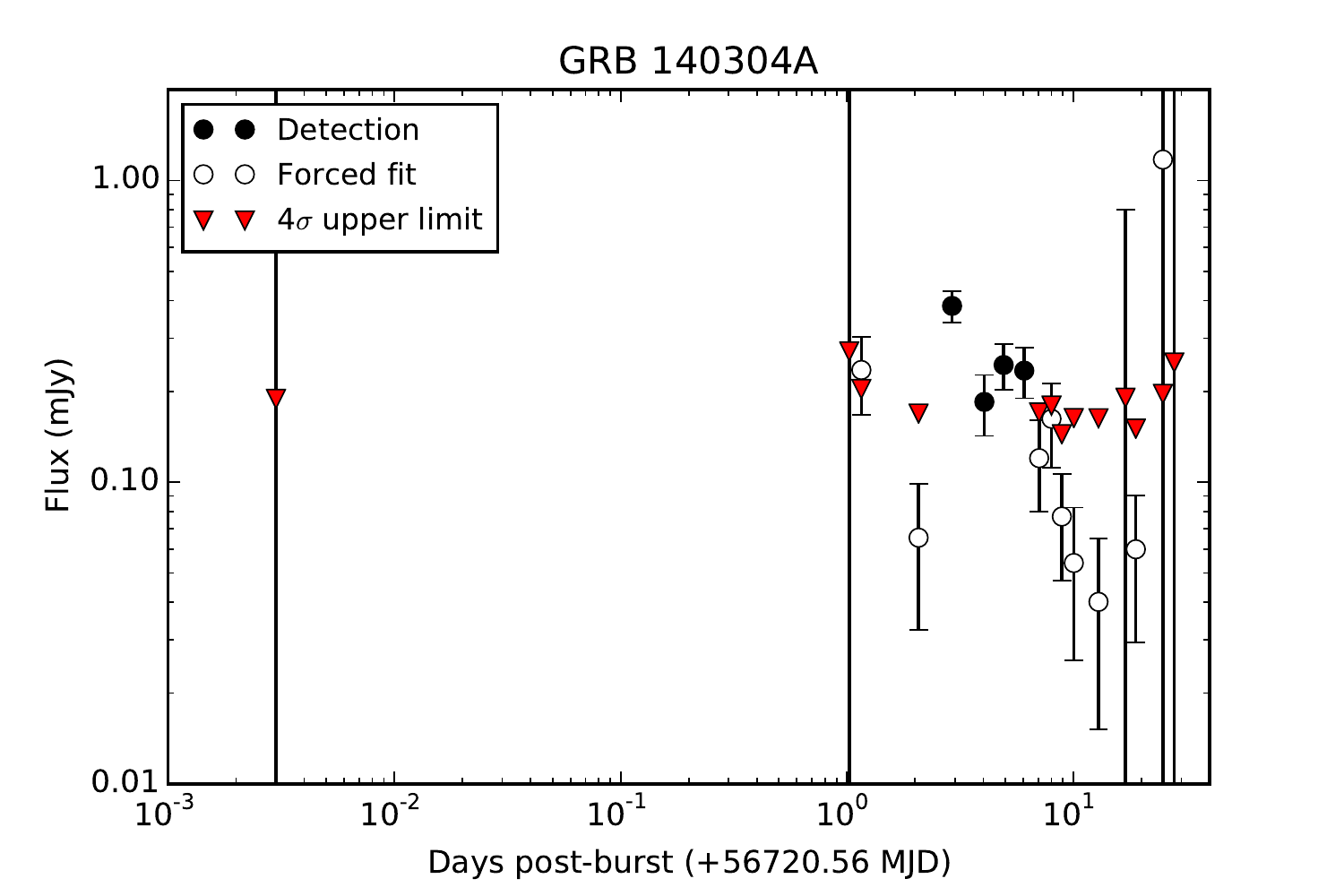}
\includegraphics[width=0.47\textwidth]{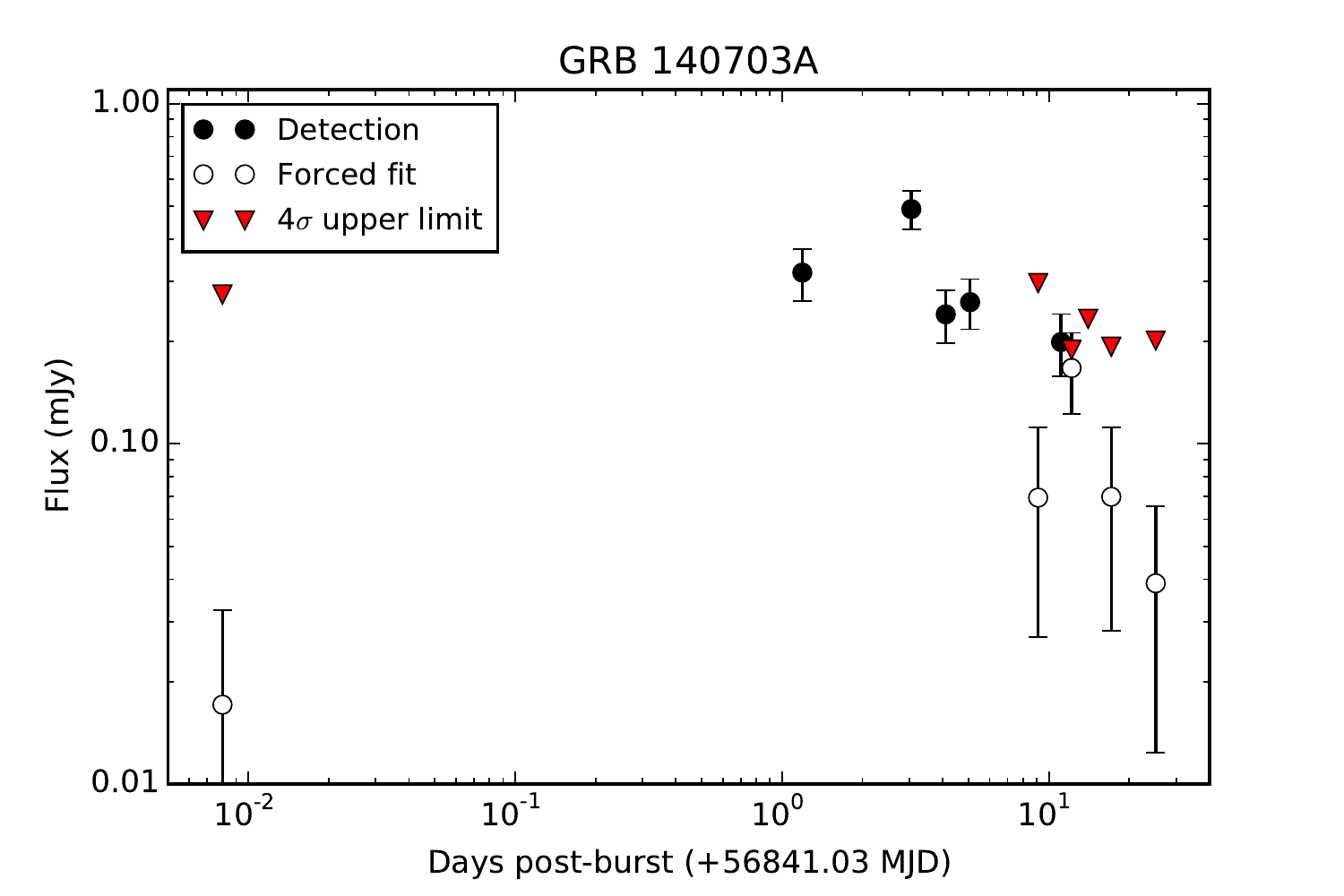}
\includegraphics[width=0.47\textwidth]{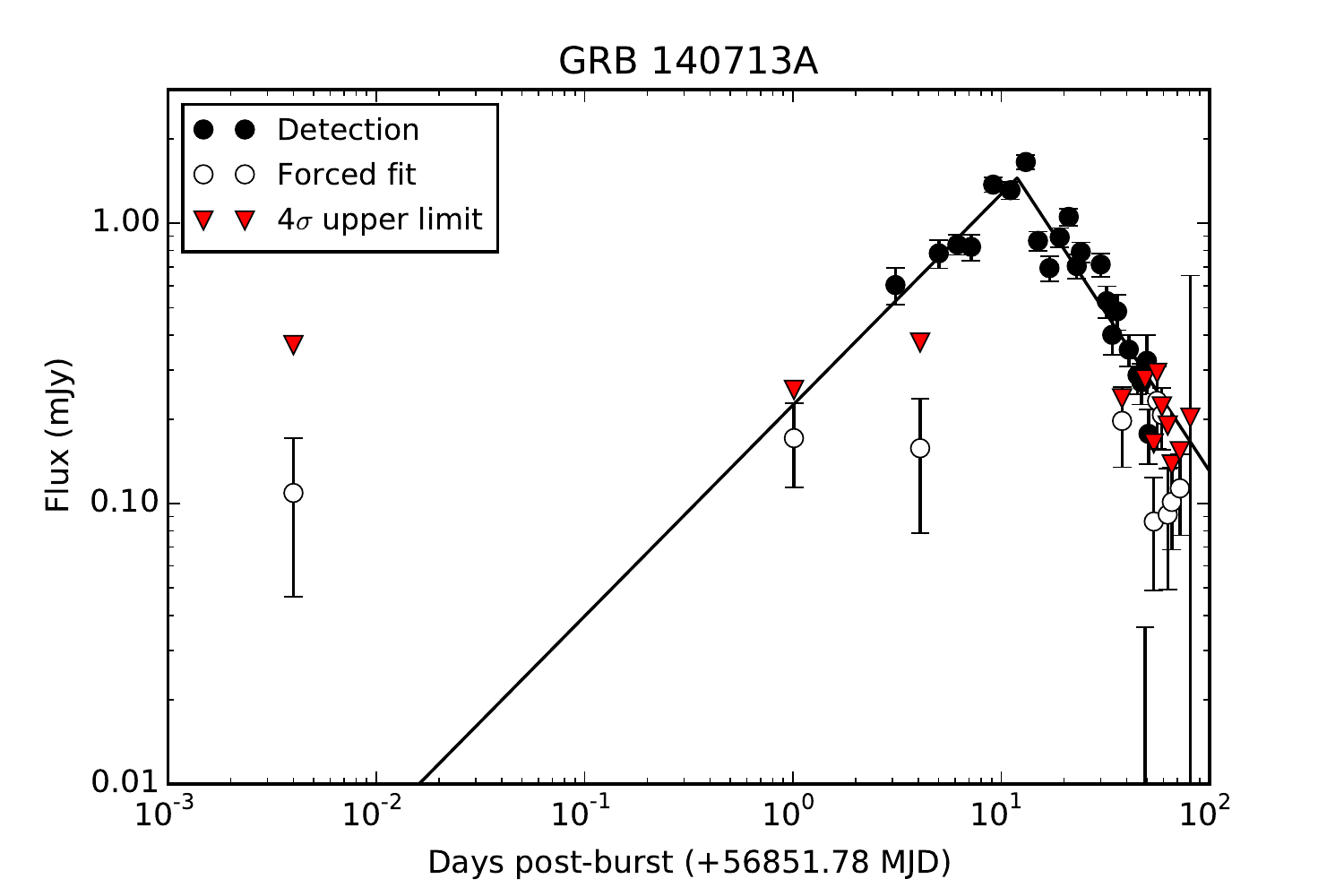}
\includegraphics[width=0.47\textwidth]{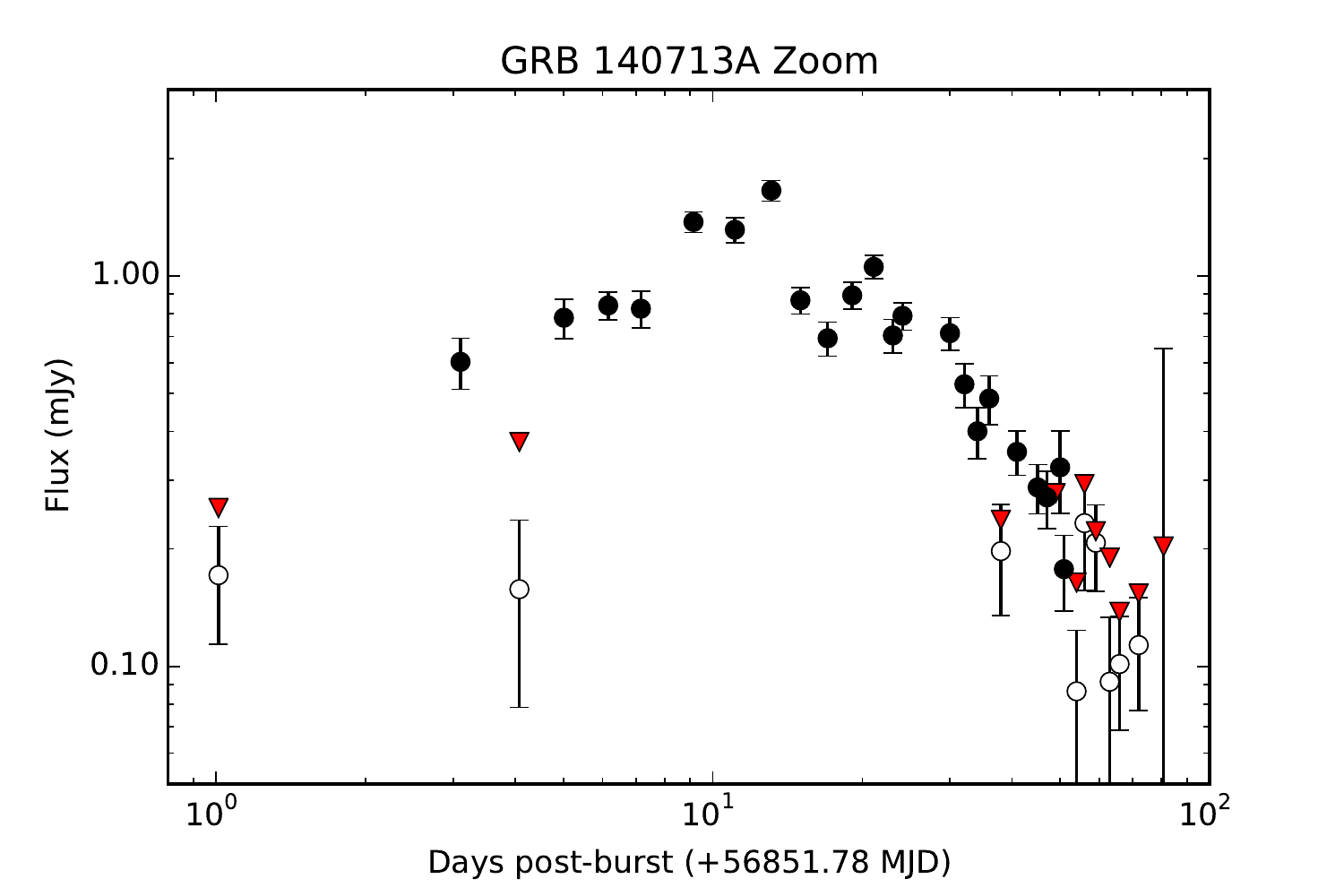}
\caption{As for Figure~\ref{fig:1} plotting AMI detections of the \textit{confirmed} radio-detected GRB afterglows. The light-curve of GRB 140713A includes a broken power-law fit to the radio detections. A zoomed in plot of GRB 140713A at late times ($\geq1$ day post-burst) is also included to better discern the light-curve structure.}
\label{fig:2}
\end{figure*}

\subsubsection{GRB 120326A}

GRB 120326A was the first GRB detected by AMI as part of the ALARRM programme. ALARRM triggered AMI on this object at 0.31 days post-burst, after this object had risen above the horizon, obtaining a $4\sigma_s$ upper limit of $0.56$ mJy/beam. It was then detected during the second AMI observation of this event at 7.15 days post-burst as seen in Figure~\ref{fig:2} (during the early stages of the ALARRM programme the follow-up AMI GRB monitoring observations were not very uniform in time). However, GRB 120326A was detected at very early times at 230 GHz with the Sub-Millimeter Array \citep[SMA;][]{urata14} just 0.53 days post-burst, with a flux of $2.84 \pm 0.86$ mJy/beam, yielding one of the earliest sub-mm detections of a GRB. Using this value and the first AMI limit (at 0.31 days) we can calculate a rough early time lower limit on the spectral index, which was $\beta >+0.6$ (for $F(\nu) \propto \nu^{\beta}$) at the $4\sigma_s$ level, indicating that self-absorption was playing a role above 15.7 GHz during this time. Radio detections were also made with CARMA at 92.5 GHz at 4.55 days post-burst \citep{perley12} and the VLA at 21.9 GHz at 5.45 days post-burst \citep{laskar12b}, yielding fluxes of $3.2 \pm\ 0.4$ mJy/beam and $1.36$ mJy/beam, respectively.  

The AMI flux measurements of GRB 120326A were first reported by \citet{staley13}. However, the flux values for GRB 120326A presented in this publication, calculated using our new and improved reduction and analysis pipeline, are slightly different due to updated scripts for the AMI data reduction process used in AMI-{\sc REDUCE} and the application of more stringent protocols for acceptable datasets. For example, the AMI observation of GRB 120326A that took place on 2012-04-04 and 2012-04-08, which are mentioned in \citet{staley13}, were rejected from further analysis due to being rain affected data-sets. 

\subsubsection{GRB 130427A}

The close proximity \citep[$z=0.34$;][]{levan13,xu13a,flores13} of GRB 130427A, and therefore its bright multi-wavelength counterpart, made it an excellent candidate for studying the forward-reverse shock scenario, particularly in the radio band \citep[i.e.][]{anderson14,vanderhorst14,perley14p,laskar13p}. Through the ALARRM programme, AMI triggered on this GRB and was on-target and observing 0.34 days post burst, as soon as GRB 130427A had risen above the horizon, resulting in one of the earliest published radio detections of a long GRB \citep{anderson14}. Further AMI observations beginning at 0.62 and 1.49 days post-burst then revealed a peak and rapid decay, demonstrating the early-time radio emission was likely dominated by the reverse-shock component at 15.7\,GHz (see the light curve in Figure~\ref{fig:2}). The AMI fluxes for GRB 130427A displayed in Table~\ref{tab:1} are slightly different to those quoted in \citet{anderson14} as they were calculated using the pipelined automated technique with a $4\sigma_s$ detection significance described in this paper rather than through a manual analysis using {\sc MIRIAD}. However, the flux values from both analyses agree within their $1\sigma_s$ flux errors. 

\subsubsection{GRB 130702A}

GRB 130702A was detected with both the \fermi-GBM and \fermi-LAT \citep{cheung13} and quickly localised by the Intermediate Palomar Transient Factory \citep[iPTF;][]{rau09,kulkarni13} through the discovery of its optical counterpart iPTF13bxl, which is the first afterglow identification based solely on a \fermi-GBM detection \citep{singer13p}. GRBs detected with \fermi-LAT are quite unusual as this instrument is sensitive to events that produce much higher energy $\gamma$-rays than those produced by GRBs detected by \swift-BAT and the \fermi-GBM.  As its counterpart localisation was not reported until 1.3 days post-burst, many of the usual GRB multi-wavelength follow-up programmes did not commence until 2 days post-burst. The first radio detection was with CARMA between 2.0 - 2.2 days post-burst, reporting a flux of $\sim2$ mJy/beam at 93 GHz \citep{perley13}, and was quickly followed by detections with the WSRT, VLA, and GMRT \citep{vanderhorst13,corsi13,chandra13}. See \citet{singer13p} for further details and the analysis of the VLA and CARMA observations. 

The first observation, and consequently detection, of GRB 130702A obtained with AMI was of two hours duration and took place $3.8$ days post-burst, with follow-up observations being regularly obtained every 2 or 3 days. The radio counterpart continued to be detectable with AMI up until 2013 July 28, with the final observation taking place on 2013 July 31. A light curve of the AMI detections of GRB 130702A can be found in Figure~\ref{fig:2} and a detailed analysis of the AMI and WSRT radio light-curves will be presented in van der Horst et al. in prep. 

\subsubsection{GRB 130907A}

GRB 130907A is another one of the small number of \swift\ detected GRBs that has also been detected with \fermi-LAT \citep{vianello13a,vianello13b}. It has also been classified as a dark burst, where the optical flux attenuation was likely caused by a high quantity of dust extinction \citep[$A_{V}>1$;][]{littlejohns15}. A VLA observation at 24.5 GHz was obtained just 4 hrs post-burst detecting the counterpart with a flux of $\sim1.2$ mJy/beam \citep{corsi13sept}. This represents one of the earliest radio detections of a GRB to date \citep[see Table 2 of][]{anderson14} and the earliest VLA detection of a GRB \citep{veres15}. The ALARRM triggered AMI observation began when the source had risen above the horizon 0.51 days post-burst. This observation was two hours in duration and detected the radio counterpart with a flux of $1.04 \pm 0.13$ mJy/beam, again providing one of the earliest recorded radio GRB detections \citep{anderson13}. Follow-up AMI observations were then manually scheduled every one to two days for a duration of 2 to 4 hours until 2013 Sept 16, yet only the observations on 2013 Sept 8 and 9 provided a firm detection (see Figure~\ref{fig:2}). 

The early time radio detections obtained with AMI \citep{anderson13} and the VLA \citep{corsi13sept} made GRB 130907A a prime candidate for investigating contributions from the reverse-shock. However, radio modelling of multi-frequency VLA observations by \citet[][see their Figure 1]{veres15} show that while the data can be modelled by the combination of a forward- and reverse-shock, it is not constraining enough to determine if the reverse-shock component is truly necessary. However, based on the 15.7\,GHz extrapolated light curve in Figure~6 of \citet{veres15} it appears that our first AMI detection at 0.51 days post burst occurred around the peak in that radio band. This peak time is even earlier than the 15.7\,GHz peak observed from GRB 130427A, which occurred around $0.6-0.9$ days post-burst and was well described by a reverse-shock model component \citep{anderson14,perley14}, so such a scenario should not be ruled out.
  
\subsubsection{GRB 140304A}

Following the initial detection of GRB 140304A with \swift, early time follow-up observations were conducted with several radio telescopes. ALARRM triggered on this GRB and had AMI on-target and observing GRB 140304A less than 5 minutes post-burst, resulting in a $4\sigma_s$ flux upper limit of 0.19 mJy/beam. Further AMI observations of GRB 140304A were scheduled daily until 2014 March 14, when a reduce observing cadence was implemented until 2014 April 1. This high redshift GRB \citep[$z=5.283$;][]{jeong14} was also simultaneously detected with the VLA and CARMA, just 0.45 days post-burst resulting in 5.8 GHz and 85 GHz fluxes of $\sim0.05$ mJy/beam and $\sim0.5$ mJy/beam, respectively \citep{laskar14,zauderer14}. 

AMI first detected GRB 140304A at 2.91 days post-burst with all following detections showing the flux to be decreasing. The non-detection less than one day prior to the first detection, with a $4\sigma_s$ upper limit of 0.17 mJy/beam, suggests a power law temporal index of $\alpha>+2.4$ for flux $F(t) \propto t^{\alpha}$. The full light-curve can be found in Figure~\ref{fig:2}. The high redshift of this GRB implies that the rest-frame time for the first AMI detection occurred at $\sim0.5$ days post-burst. Given this steep flux increase, it is possible that AMI may have detected the reverse-shock peak in the radio band. Further broadband modelling is required to investigate this suggestion. 

\subsubsection{GRB 140703A}

ALARRM rapidly triggered on this \swift\ event and pointed AMI at the position of GRB 140703A within 12 minutes post-burst, resulting in a $4\sigma_s$ upper limit of $0.27$ mJy/beam after 2 hours of integration \citep{anderson14d}. GRB 140703A was then rapidly detected in the radio band with the VLA and CARMA, obtaining radio afterglow detections at 19 GHz and 93 GHz just 0.35 and 0.67 days post-burst, resulting in fluxes of $\sim0.28$ mJy/beam and $\sim2$ mJy/beam, respectively \citep{corsi14,perley14b}. Follow-up AMI observations between 4 and 6 hours in duration were then manually scheduled to occur every one or two days up until 2014 July 20, with a final observation on 2014 July 28. The radio afterglow was clearly detected ($>4\sigma_s$) with AMI between 1.19 and 11.10 days post-burst, and appeared to peak $\sim3$ days post-burst. This peak corresponds to a rest-frame time of 0.7 days post-burst, once again suggesting a possible reverse-shock origin. The light curve then steadily declined until $\sim12$ days post-burst, at which point it dropped below detectability (see Figure~\ref{fig:2}). 

\subsubsection{GRB 140713A}

GRB 140713A has been detected in both the radio \citep{anderson14b,zauderer14b} and X-ray bands \citep{stamatikos14} but not at optical wavelengths. Deep photometric observations yielding non-detections suggest that this GRB may be a dark burst \citep{castro-tirado14}. AMI was on-target and observing GRB 140713A less than 6 minutes post-burst, obtaining a $4\sigma_s$ upper limit of 0.37 mJy/beam. The earliest radio detection was obtained with CARMA at 85 GHz, just 0.49 days post-burst, yielding a peak flux of $1.5 \pm 0.3$ mJy/beam \citep{zauderer14b}. The first AMI detection of GRB 140713A was 3.1 days post-burst and we continued to observe this event every 1 to 3 days for 2 to 6 hours until 2014 Sept 23, with the last observation taking place on 2014 October 2 (see the full light curve in  Figure~\ref{fig:2}). The brightest AMI detection occurred at 13 days post-burst, with a flux of $1.65 \pm 0.10$ mJy/beam, with a possible second peak at 21 days (see the zoomed version of the light curve in Figure~\ref{fig:2}). A broken power-law fit to the AMI light curve gives a rising slope of $\alpha =+0.75 \pm 0.11$ and a decay slope of $\alpha =-1.13 \pm 0.06$ for $F(t) \propto t^{\alpha}$, showing a peak flux of $1.45 \pm 0.36$~mJy/beam at $11.92 \pm 1.10$ days post-burst ($\chi^{2}_{red}=5.17$). This peak time is typical of the forward-shock emission often observed at 15.7\,GHz \citep[see modelling in][]{ghirlanda13}. A full modelling of the AMI light-curve, in conjunction with WSRT observations, will appear in van der Horst et al. in prep, which will give a thorough multi-wavelength analysis of this GRB.

\subsection{New \textit{possible} GRB radio afterglows discovered with AMI}\label{possgrb}

There were six GRBs that yielded a radio detection in only one AMI epoch. In each case the detection had a flux significance of $4 \leq \sigma_s < 5$, with late time constraining upper limits that would have detected this source with a $\geq5 \sigma_s$ significance. Even so, we only consider these events as \textit{possible} GRB radio afterglows. The GRBs that fall into this category include GRB 120320A, GRB 130625A, GRB 140209A, GRB 140318A, GRB 140320C, and GRB 140607A (the latter 4 were observed during the updated observing strategy that began in 2013 August). The most convincing \textit{possible} radio afterglow detections are from GRB 130625A, GRB 140318A, and GRB 140320C. GRB 140318A has also been classified as an optically dark burst \citep{littlejohns15}, and its AMI light curve can be found in Figure~\ref{fig:3}. A summary of each of these six GRBs can be found in Appendix~\ref{appendixa}.

\begin{figure}
\begin{center}
\includegraphics[width=0.5\textwidth]{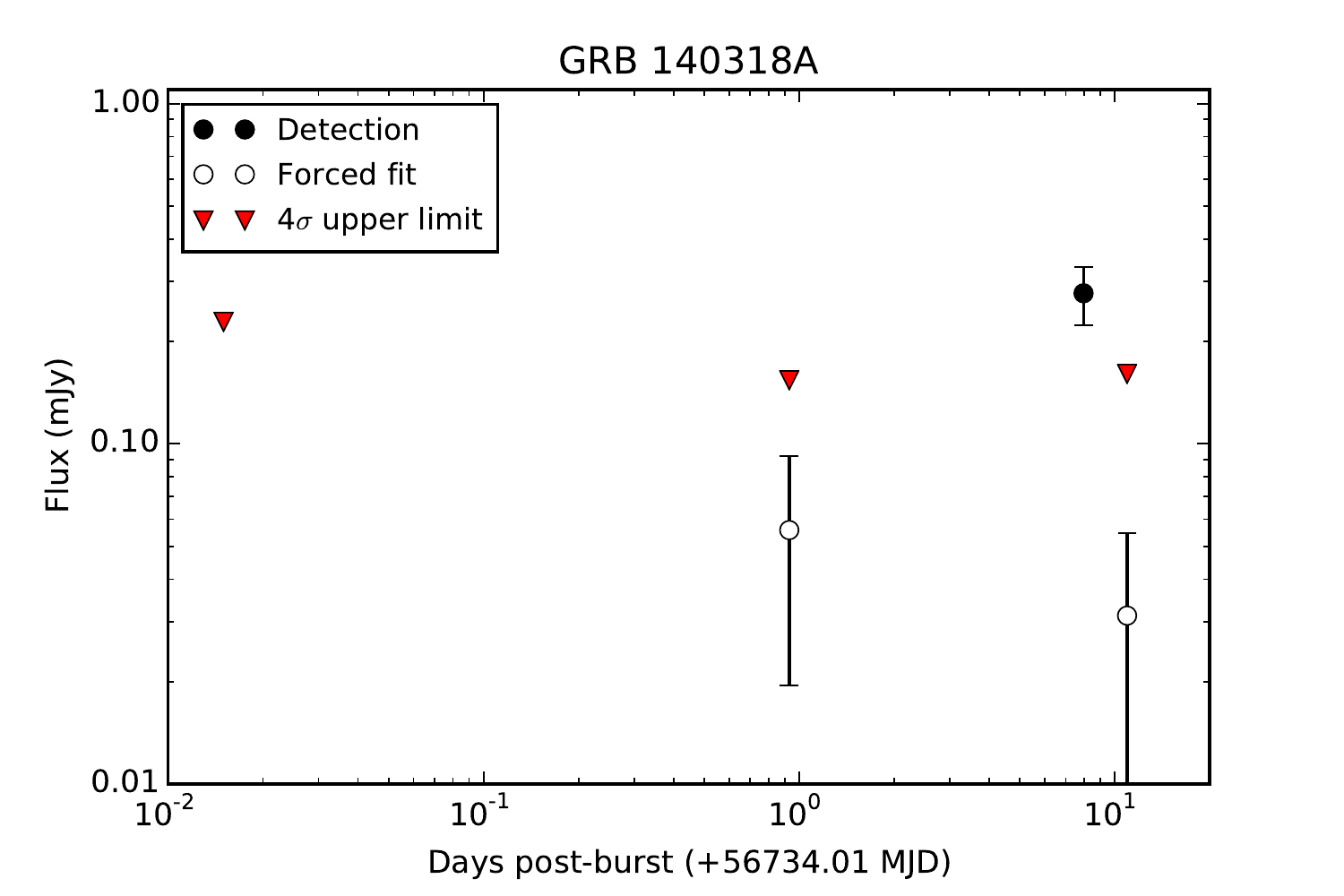}
\end{center}
\caption{The AMI 15.7 GHz light curve of GRB 140318A. The black data point is the $>4\sigma_s$ AMI detections of GRB 140318A. The red triangles show the $4 \sigma_s$ RMS noise levels of all four AMI epochs. All errors are $1\sigma_s$.}
\label{fig:3}
\end{figure}

\subsection{AMI concatenated detections}\label{concatgrb}

There are also several GRBs for which no coincident radio source was detected in a single epoch but one was detected in the deep concatenated image. In such cases it is possible that combining the individual epochs to create a deep concatenated image provided the sensitivity required to detect the radio afterglow. Combining the early- and late-time AMI epochs into two concatenated images allowed us to confirm the detection of the radio afterglow from GRB 140629A (see Sections~\ref{criteria} and \ref{grb140629a}). However, this technique did not work for GRB 130606A, GRB 140508A, GRB 140801A, and GRB 150309A as the resulting images were either not sensitive enough to detect the radio source identified in the full concatenated image, or the flux of the coincident radio source agreed between the two concatenated images indicating that it was likely steady. These GRBs are described in Appendix~\ref{appendixb}. 

If not radio afterglows, these coincident radio sources are likely serendipitous rather than radio emission from star formation in their host galaxies  \citep[for example see][]{berger03h}. Most claimed associations have been recently ruled out by sensitive late-time radio observations, which demonstrate that previous radio detections were likely of long-lived radio afterglows, or noise fluctuations or reduction artefacts, rather than from star formation \citep{perley17}.

\subsection{Radio-detected GRBs that were not detected with AMI}\label{nodetgrb}

There were eight \swift\ radio-detected GRBs identified during the old ALARRM observing strategy that ran between 2012 March and 2013 August. However, of these eight only GRB 120326A and GRB 130427A were detected with AMI. The radio-detected GRBs that were not detected with AMI during this period were GRB 120404A, GRB 120422A, GRB 120521C, GRB 130131A, GRB 130418A, and the short burst GRB 130603B \citep{zauderer12c,zauderer12,laskar14b,laskar13c,zauderer13,perley13c,fong14b}. It is highly likely that our lack of AMI detections of radio GRBs during this period is owed to the short 1~hr integrations that were specific to the old ALARRM observing strategy. The radio afterglows were likely below the $4\sigma_s$ sensitivity limit, which is usually between $0.3-0.7$ mJy/beam for a 1~hr observation. 

However, only four out of the 13 \swift\ radio-detected GRBs were not detected with AMI during the period of the new ALARRM strategy, between 2013 August and 2015 April. GRB 140419A, GRB 140515A, and GRB 141026A likely went undetected due to the AMI 4~hr sensitivity limit of $\sim0.2$ mJy/beam ($4\sigma_s$). The fourth event, GRB 140309A, had an unconfirmed detection during the first AMI epoch at 0.93 days post-burst, but its proximity to NVSS 155207+273501 means we cannot rule the possibility of it being an artefact. This flux density, along with the forced fitted flux densities derived for the rest of the observations of these GRBs, can be found in Table~\ref{tab:2}. Individual descriptions of these four GRBs, along with GRB 130603B (an interesting case), can be found in Appendix~\ref{appendixc}.

\subsection{Coincident steady sources}\label{ssgrb}

There are five GRBs that are coincident, or in close proximity of, an uncatalogued radio source detected with AMI, which are likely steady. These include GRB 130216A, GRB 140320B, GRB 140606A, GRB 141015A and GRB 141020A. The measured source flux densities of these GRBs are included in Table~\ref{tab:2} and a detailed description of these observations can be found in Appendix~\ref{appendixd}. Once again, these associations are likely to be serendipitous chance coincidences with radio cores or lobes from unrelated galaxies.

\section{Radio GRB Statistical Discussions}

\subsection{Radio GRB detection rates}\label{grbstat}

The current best estimate of the radio detection rate of \swift\ GRBs is 29\%, which is based on a comprehensive investigation of all GRBs observed in the radio band between 1997 and 2011 by \citet{chandra12}. However, as previously mentioned, it is possible that this detection rate may be biased due to radio follow-up often being based on some prior knowledge of the GRB's properties, with often the additional incentive of performing multi-wavelength studies with data at other wavelengths. In fact, only 28\% of \swift\ GRBs in their sample were observed at radio wavelengths, which means only 8\% of \swift\ GRBs before 2011 have confirmed radio afterglows.

The ALARRM project has now gathered a far more unbiased sample of radio observations of \swift\ GRBs where the only constraint is whether the explosion is within the observable AMI sky. However, the early strategy, though triggering on all \swift\ GRBs with a declination $>-10^{\circ}$, did not yield many detections. All triggered and monitoring AMI observations were 1 hour in duration, resulting in an RMS of $\sim0.1$ mJy/beam. We know from Figure 4 of \citet{chandra12} that the majority of radio-detected GRBs prior to 2011 had peak fluxes between $0.1-0.2$ mJy/beam so 1 hour AMI observations were not sensitive enough to detect most radio afterglows. As a result, we only detected 2 out of the 8 \swift\ radio-detected GRBs in the AMI sky, which corresponds to a detection rate of 3\% for the 67 monitored \swift\ GRBs between 2012 March and 2013 August. 

After the ALARRM strategy update in 2013 August, all \swift\ GRBs with a declination $>15^{\circ}$ were observed for 2~hrs with AMI following the $\gamma$-ray trigger, with 4~hr follow-up observations at 24 hours, 3, 7, and 10 days post-burst (with a higher cadence for those events where we identified the radio afterglow). These longer observations bought the RMS noise down to $\sim0.03-0.04$\,mJy/beam, making the programme sensitive to GRBs with radio counterparts $\geq0.12$ mJy/beam. Between 2013 August and 2015 April, AMI detected 10 ($\sim70$\%) out of the 14 radio-detected \swift\ GRBs in the AMI observable sky, 6 of which were discovered with AMI as part of this project (see Section~\ref{newgrb}), thus increasing the rate of GRB radio afterglow detections in this declination range by a factor of $\sim1.5$ over an 18 month period. As the only discriminant for AMI observations was the declination range, we can assume that the sample of the 65 events observed as part of this project between 2013 August and 2015 April are representative of the entire \swift\ GRB sample. Therefore, the radio detection rate is 15\% down to a conservative sensitivity limit of $\sim0.2$ mJy/beam (RMS noise for most 4~hr observations are $0.03-0.04$~mJy/beam), which is based on radio observations that were not informed by prior knowledge of any GRB multi-wavelength properties. By also including the 4 radio-detected \swift\ GRBs not detected with AMI since the start of the new strategy (see Section~\ref{nodetgrb}), the radio detection rate increases to 22\% but to an unknown completeness limit. If we also include the 4 \textit{possible} AMI-detected \swift\ GRBs since the start of the new strategy (see Section~\ref{possgrb}), then the total radio detection rate for \swift\ GRBs may be as high as 28\%.

\citet{chandra12} report several GRBs with radio peak fluxes $\geq0.2$\,mJy/beam, the majority of which were detected with the VLA prior to its upgrade (typical GRB sensitivity limits of $\sim0.1-0.15$\,mJy/beam, see their Table~4). If we assume that the broadband spectrum of all GRB afterglows in the \citet{chandra12} sample follow ``Spectrum 1" in the modelling performed by \citet{granot02}, which is the most relevant for describing GRB afterglows with detectable radio emission in the first days to weeks post-burst \citep{granot14}, then the spectral index at 15.7\,GHz is likely to be $\beta$ $\sim2$ or 1/3 \citep[for $F(\nu) \propto \nu^{\beta}$, note that this model assumes a spherical ultra-relativistic blast wave impacting a uniform or wind-like circumstellar medium,][]{blandford76}. As this model requires that $\beta>0$ in the radio band, we consider AMI capable of detecting all GRBs with reported peak fluxes $\geq0.2$\,mJy/beam for an observing frequency $\leq15$ GHz. AMI would therefore have only detected $\sim50$\% of the radio-detected \swift\ GRBs in the \citet{chandra12} sample. This suggests that the radio GRB detection rate at 15.7\,GHz may double if the sensitivity is improved by a factor of two, from $\sim0.2$\,mJy/beam to $\sim0.1$\,mJy/beam. Based on the AMI statistics, the radio-detection rate for \swift\ GRBs could be as high as $\sim44-56$\%, down to a sensitivity of $\sim0.1-0.15$\,mJy/beam. This result is supported by assuming the typical $\mathrm{log}N-\mathrm{log}S$ relationship for radio sources \citep[$N \propto S^{-3/2}$;][]{fomalont68}, where $N$ is the number of GRBs with fluxes greater than $S$. Using the ratio of the detection limits, $N \propto (0.125/0.2)^{-3/2} \approx 2$.

Note that there are several caveats associated with this implied AMI detection rate. \citet{chandra12} did not report GRB radio peaks in their Table 4 that occurred within 3 days post-burst to avoid contamination from the reverse shock in their statistics. This means that including early-time ($<3$ days) radio peaks from historical GRBs in this flux comparison may change the potential $\sim50$\% AMI detection rate of \citet{chandra12} GRBs. The assumed model ``Spectrum 1" may also only be relevant for radio emission generated by the forward shock so does not take into account the possibility of the early-time AMI detections arising from the reverse shock. This model is also only valid for a uniform or wind-like (density drops off with radius) medium and may therefore not be relevant for GRBs that occur in low density environments similar to that of the interstellar medium. GRBs are also now widely accepted to have jet-like outflows rather than spherical \citep{sari99jet}. This estimate also does not consider how incomplete sampling may affect the numbers \citep[i.e. GRB radio flux peaks missed due to the AMI monitoring cadence or the radio follow-up criteria applied to events in the][sample]{chandra12}. 

The implied AMI detection rate of $\sim44-56$\% is consistent with the GRB sample that \citet{ghirlanda13} and \citet{burlon15} base their synthesised population, which was used to demonstrate that the radio afterglows from all \swift-GRBs should be detectable with SKA1-MID (Band 5). Their sample consists of the radio observed GRBs is the BAT6 sample \citep[BAT6 being the brightest \swift-BAT-detected long GRBs that are considered complete with respect to the flux limit,][]{salvaterra12}, thus it can be argued that they represent an unbiased sample of GRB radio afterglow properties \citep[for details on \swift\ GRB selection effects and comparisons to GRBs detected with other instruments see][]{qin13,lien16}.

The advantage of the AMI derived GRB radio detection rate is that it is not biased by prior knowledge of the other electromagnetic properties. \citet{chandra12} demonstrated that the optical brightness of a GRB afterglow is potentially a positive indicator for radio detectability. Since radio follow-up of GRBs is historically biased to those with bright optical counterparts, with the goal of obtaining simultaneous observations across the spectrum, the \citet{chandra12} detection rate is likely to be positively biased. Two possible conclusions can be drawn from the higher implied radio detection rate obtained with AMI.
\begin{enumerate}
\item The \citet{chandra12} sample does represent the radio detection rate of GRBs at 8.5 GHz, and that the AMI rate is only higher as the GRB radio afterglow emission peaks more brightly at 15.7\,GHz \citep{granot02}. 
\item The AMI rapid-response system allows for the detection of more radio afterglows as it obtains at least one observation within 24~hours of the \swift\ GRB trigger.
\end{enumerate}

In order to investigate the first option, extensive multi-wavelength modelling over multiple timescales are required, which is beyond the scope of this paper. However, given the selective nature of the radio follow-up of historical GRBs, the \citet{chandra12} rate is likely affected by incomplete sampling and may therefore not be directly scalable to 15.7\,GHz.

The most likely explanation for the difference between the \citet{chandra12} and implied AMI detection rates may be related to the early-time ($<1$ day) radio observations made possible by the AMI rapid-response system (second option). Such early-time observations are particularly sensitive to the reverse-shock emission peak, which we know can be at least an order of magnitude brighter than the forward-shock peak \citep[as is the case for GRB 130427A;][]{anderson14}. Of the 18 GRBs we have classed as \textit{confirmed} or \textit{possible} AMI-detected GRBs (Sections~\ref{newgrb}, \ref{knowngrb}, and \ref{possgrb}) that were observed with AMI at early times, 7 ($37\%$) were detected in the radio band $<2$ days post-burst, 4 ($20\%$) of which were detected only within this time frame. \citep[The upper-limit of 2 days in the observer frame was a natural choice given modelling by][only explore the radio emission from the forward-shock between $2-10$ days post-burst.]{ghirlanda13,burlon15} Therefore, radio observations at early-times may increase the radio detection rate at least another $\sim20\%$, which would raise the \citep{chandra12} detection rate to a value consistent with the implied AMI detection rate (down to $\sim0.1-0.15$~mJy/beam). 

The AMI-ALARRM results therefore demonstrate that early-time radio observations of GRBs play an important role in constraining the radio afterglow detection rates as they are particularly sensitive to the reverse-shock emission. This strongly suggests that contributions from the reverse-shock component must be considered in future population synthesis and modelling. As the AMI and \citet{chandra12} GRB samples both display radio detections and upper-limits with comparable luminosities (see Section~5.3), it is highly unlikely we have reached the sensitivities necessary to discern the true radio GRB detection-rate \citep[as also concluded by][]{chandra12}. Assuming that the GRB sample in the \citet{ghirlanda13} and \citet{burlon15} simulations are representative of an unbiased sample of \swift-GRBs, then SKA1-MID (using Band 5), should be sensitive to the radio afterglow produced by the forward-shock from all \swift-GRBs, allowing us to determine if there is a dual population of radio-bright and radio-faint GRBs as suggested by \citet{hancock13}. 

\subsection{Early time radio properties of GRBs\label{earlytime}}

\begin{figure}
\begin{center}
\includegraphics[width=0.5\textwidth]{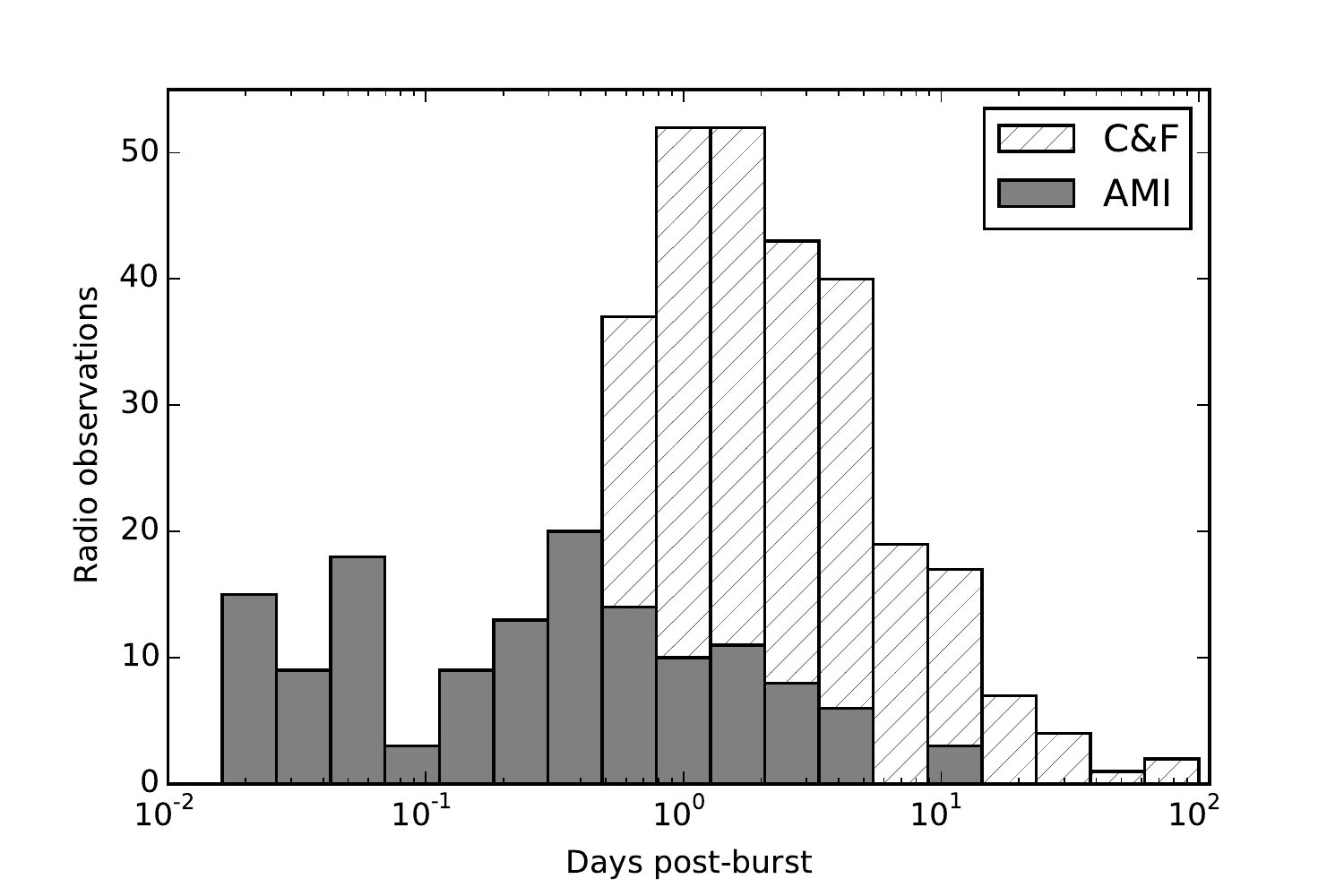}
\end{center}
\caption{Histogram showing the time post-burst (based on the mid-time of the observation in the observer frame) of the first radio observation of each GRB in the \citet{chandra12} sample (C \& F; cross hatched bars) and the AMI sample (grey bars). Neither of these samples include the radio triggering experiments mentioned in Section 1.}
\label{fig:4}
\end{figure}

\begin{figure}
\begin{center}
\includegraphics[width=0.5\textwidth]{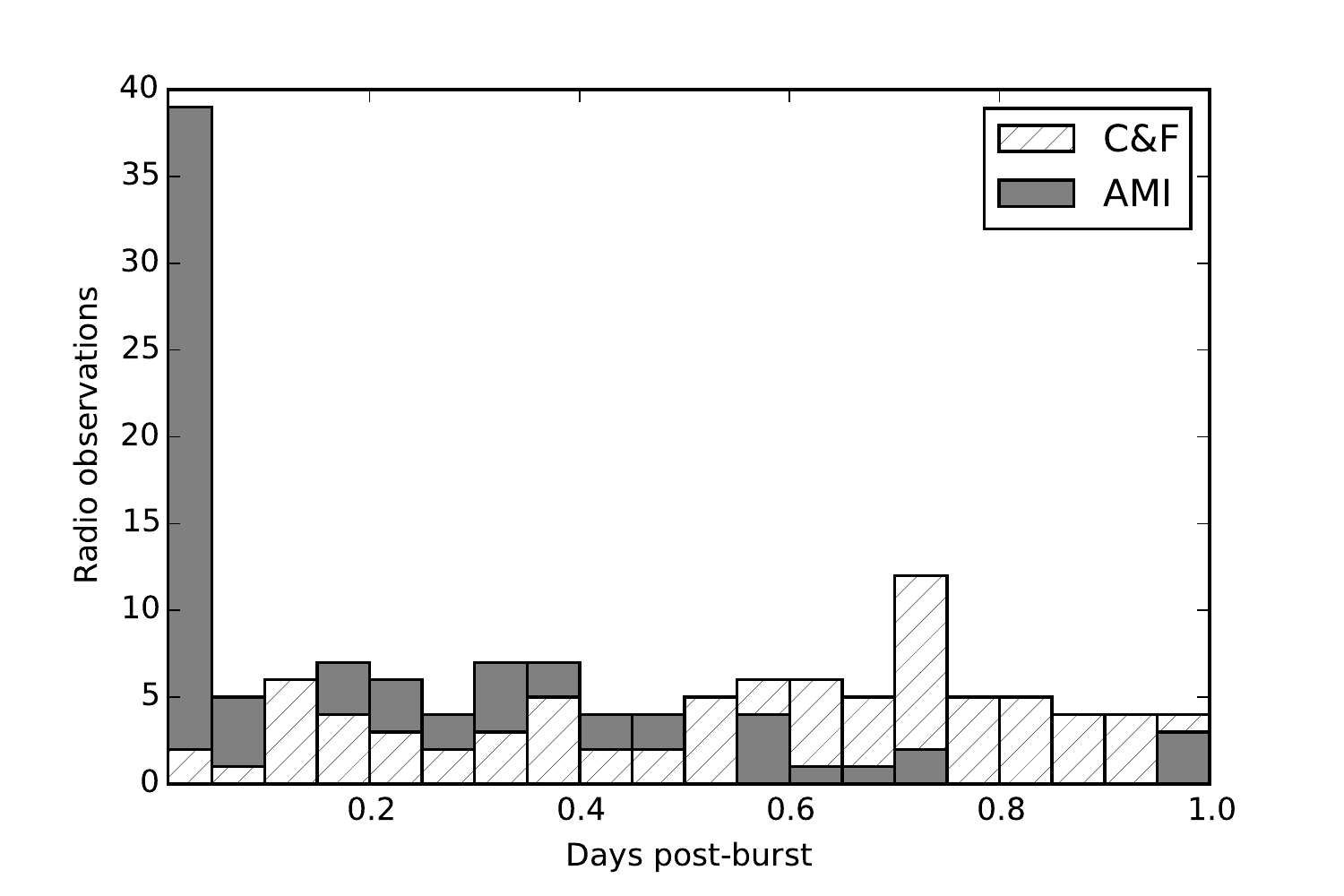}
\end{center}
\caption{As for Figure~\ref{fig:4} but zoomed in to see those GRBs that were first observed within a day post-burst. It can be clearly seen that the ALARRM programme has enabled some of the earliest responses to GRB alerts on record.}
\label{fig:5}
\end{figure}

The work of \citet{chandra12} clearly shows that for the last 20 years radio observations of GRBs have been obtained starting from a few minutes up until 1339 days post-burst. However, the majority of these observations were taken at times later than 10 days post-burst \citep[see Figure 1 of][]{chandra12}, with the earliest observation of each GRB usually occurring after 1 day post-burst \citep[see Figure~\ref{fig:4} and Figure 4 of][]{veres15}. However, since the inception of the ALARRM programme we have obtained 124 AMI observations of 106 GRBs within 1 day post-burst up until 2015 April as indicated in Figure~\ref{fig:4}. In fact, 39 of these observations occurred within the first hour following the \swift\ trigger (all based on the mid-time of the observation) using our rapid-response mode \citep[see Figure~\ref{fig:5}, which compares the first AMI observations of each GRB observed within 1 day post-burst to the sample from][note that Figures~\ref{fig:4} and \ref{fig:5} do not include the rapid-response triggers from programmes mentioned in Section 1 designed to search for prompt, coherent emission, and not the fireball, incoherent emission being probed with AMI]{chandra12}. This demonstrates the strength of rapid-response telescope modes compared to relying on manual scheduling.

None of the ALARRM triggered observations with response times within 1~hr of the \swift-BAT trigger resulted in a detection, with the lowest $4\sigma_s$ limit being 0.19 mJy/beam. The rapid-response observations are also usually less sensitive than the manually scheduled follow-up observations due to shorter exposure times at less optimum hour angles. However, it is highly likely that the reverse (and the forward) shock are too self-absorbed at 15.7 GHz at such early times to be detectable. The only rapid-response AMI triggers that did result in a detection were GRB 130427A at 0.34 days, GRB 130907A at 0.51 days, and possibly GRB 120320A at 0.6 days post-burst. These represent some of the earliest radio detections of long GRBs along with a few early VLA detections \citep[see Table 2 of][]{anderson14} including GRB 130907A just 0.193 days post-burst at 19.2 and 24.5 GHz \citep{veres15}. Based on these results it is likely that the radio emission from GRBs does not switch on at 15.7 GHz for several hours following the \swift-BAT trigger. Of course, the low redshifts for both GRB 130427A \citep[0.34;][]{levan13,xu13a,flores13} and GRB 130907A \citep[1.24][]{deugarte13} may also contribute to such early-time detections as the emission arrival time will be less delayed due to redshift. However, there are also several cases \citep[for example GRB 140703A and GRB 150413A, which were both at a slightly higher redshift of 3.14,][]{castro-tirado14d,deugarte15} where the manually scheduled follow-up AMI observation at around 24 hours post-burst did make a detection. 

While GRB 140703A and GRB 150413A were at high redshifts, we could have potentially missed the initial radio rise. In order to catch the possible radio turn-on, particularly for low redshift GRBs (for example $z < 2$), it may be necessary to further update the ALARRM observing strategy to conduct the rapid-response observation between 4 and 16~hrs post-bursts (consistent with the early-time AMI detections of GRB 130427A and GRB 130907A), when the potential radio afterglow (forward- or reverse-shock) becomes optically thin at 15.7~GHz. This could be achieved by programming the rapid-response observation to occur when the GRB is at its optimum hour angle, which would also negate some of the sensitivity complications mentioned above. Another options would be to perform a dense number of observations over a 24~hr period, which would better capture the GRB switching-on and the potential reverse-shock evolutions \citep[as was seen for GRB 130427A;][]{anderson14}.

\begin{figure}
\begin{center}
\includegraphics[width=0.5\textwidth]{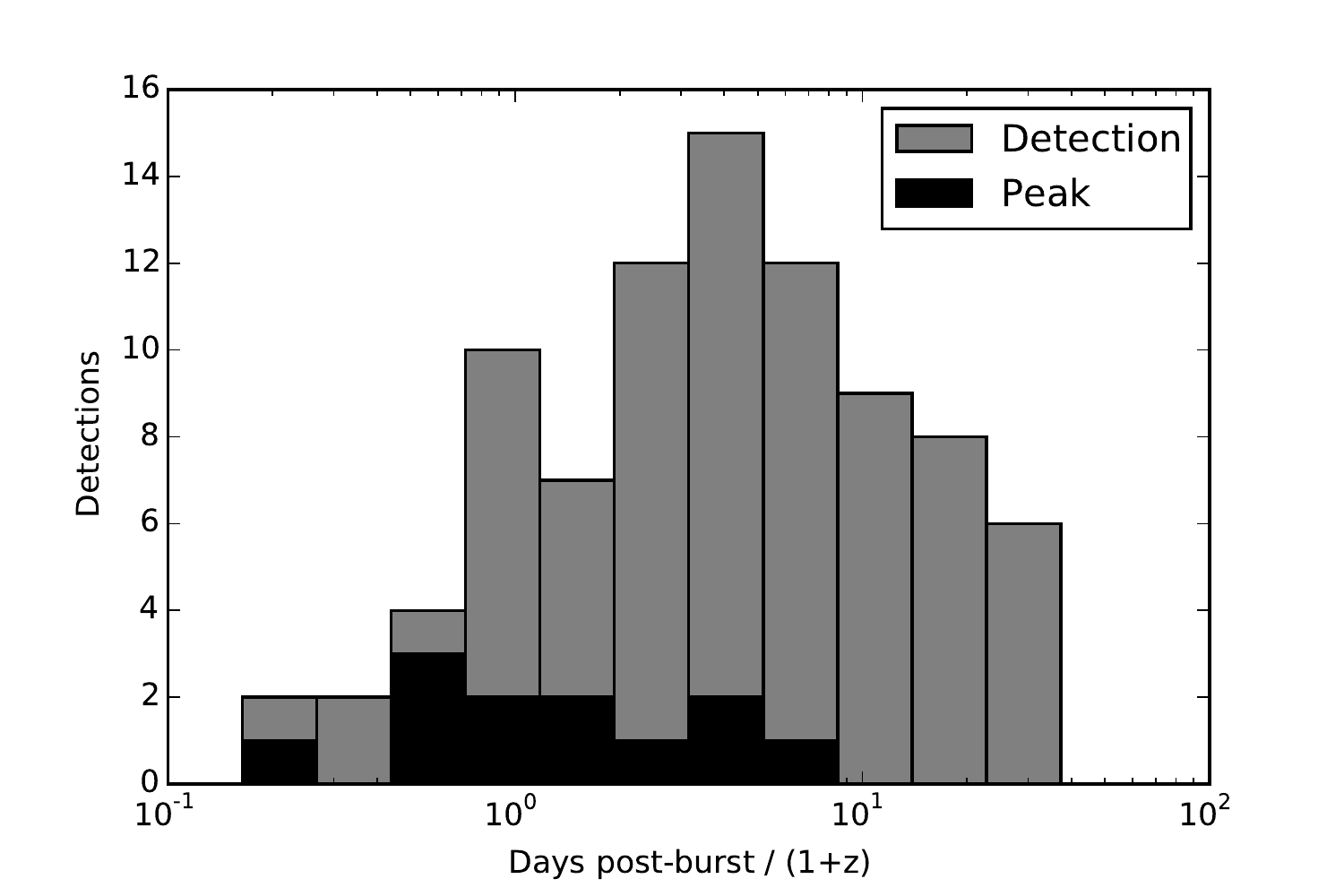}
\end{center}
\caption{Histogram of the number of GRB AMI detections for a given time post-burst in the rest-frame (days post-burst / (1+z)). The brightest or ``peak'' detection of each GRB is overlaid in black (which is a subset of the full sample of detections represented in grey). Only 9 out of 12 radio GRBs detected with AMI in two or more epochs have a known redshifts. For those 3 without redshift measurements we assume the average redshift of $z=2$ for radio detected \swift\ GRBs \citep{chandra12}.}
\label{fig:6}
\end{figure}

In order to estimate the rate of GRB reverse-shock detections made with AMI, we generated a histogram of all the 12 \textit{confirmed} AMI-detected GRBs with respect to days post-burst in the rest frame (days post-burst / (1+z)), where the peak (brightest AMI detection) for each GRB is indicated in black (see Figure~\ref{fig:6}). For those AMI detected GRBs without a redshift, we assumed the average \swift\ redshift of $z = 2.0$ \citep[see][]{chandra12}. This demonstrates that the majority of detections occurred within 10 days post-burst (in the rest frame) with all of the peak detections occurring within 7 days. To investigate whether any of the AMI radio detections were from the reverse-shock, we summarised the rest frame time of the peak, and the earliest and latest radio detections of the 12 AMI detected GRBs (Table~\ref{tab:3}). In the rest frame 6 out of these 12 GRBs peak before 1 day post-burst, including GRB 130427A, which is known to have a reverse-shock radio flare occurring around 0.6 days post-burst \citep{anderson14}. It is therefore possible that AMI has detected the reverse-shock flare of at least 6 GRBs. Another 3 of these 12 GRBs peaked between 1 and 3 days post-burst, with the other 3 peaking at times $>3$ days post-burst. GRB 130907A, GRB 140304A and GRB 150413A also faded below detectability in 1 day relative to the rest-frame. This may suggest that the reverse-shock dominated the radio afterglow for these events \citep[note that multi-frequency radio modelling of GRB 130907A was unable to confirm a preference for a forward-reverse shock scenario over a forward-shock-only model,][]{veres15}. However, this assumption is an over simplification as the forward-shock peak time ($t_{p}$) can occur anywhere within a day to several weeks following the explosions as described by: 

\begin{align}\label{eq:1}
\begin{split}
t_{p} = 1.93 ~F_p^{2/3} ~n_{0}^{-1/3}  ~\varepsilon_{e,-1}^{4/3}  ~E_{52}^{-1/3}  ~d_{L,28}^{4/3} \bigg(  \frac{(1+z)}{2} \bigg)  ^{-1/3}
\end{split}
\end{align}
for a homogeneous medium and:
\begin{align}\label{eq:2}
\begin{split}
t_{p} = 1.55 ~F_p^{2/3} ~A_{*}^{-2/3}  ~\varepsilon_{e,-1}^{4/3}  ~d_{L,28}^{4/3} \bigg(  \frac{(1+z)}{2} \bigg)  ^{-1/3}
\end{split}
\end{align}
for a stellar wind medium where $F_{p}$ is the peak flux in mJy, $n_{0}$ is the circumstellar density for a homogeneous medium, $A_{*}$ is the density parameter for a stellar wind medium, $\varepsilon_{e,-1}$ is the energy electron fraction divided by 0.1, $E_{52}$ is the isotropic energy in units of $10^{52}$ ergs, and $d_{L,28}$ is the luminosity distance in units of $10^{28}$~cm \citep{vanderhorst07}. These equations assume that  the self-absorption frequency is below 15.7 GHz \citep[for more details on GRB radio light curve modelling see][and references therein]{granot14}. Equations \ref{eq:1} and \ref{eq:2} demonstrate that the forward-shock peak is highly dependent on many parameters that can only be calculated through broadband modelling. 

\begin{table}
\begin{center}
 \caption{Rest frame times of peak, earliest and latest radio detection of the 12 AMI detected GRBs}
 \label{tab:3}
 \begin{tabular}{lccc}
 \hline
Detection$^{\mathrm{~a}}$ & $\leq1$ day$^{\mathrm{~b}}$ & $1<$ days $\leq3$$^{\mathrm{~b}}$  & $>3$ days$^{\mathrm{~b}}$  \\
\hline
Peak & 6 & 3 & 3 \\
Earliest & 7 & 4 & 1 \\
Latest & 3 & 2 & 7 \\
\hline
\end{tabular}
\end{center}
Note: We assume $z=2$ for those GRBs without known redshift measurements \citep{chandra12}.\\
$^{\mathrm{~a}}$ Refers to the peak (brightest), earliest and latest detection of the radio afterglow for each of the 12 \textit{confirmed} AMI-detected GRBs. \\
$^{\mathrm{~b}}$ The number of \textit{confirmed} AMI-detected GRBs for which the peak, earliest and latest detection occurred within 1 day, between 1 and 3 days, and later than 3 days post-burst in the rest frame.
\end{table}

\subsubsection{Brightness Temperature and Minimum Lorentz Factor} \label{btlf}

With the growing sample of radio detected GRBs, along with the early time ($<1$ day post-burst) detections and upper-limits we are obtaining with AMI, we are able to explore the distribution of GRB brightness temperatures ($T_{\rm{b}}$) and minimum bulk Lorentz factors ($\Gamma$) in the radio domain. These values can then be compared to the physical parameters of the blast-wave environment derived from broadband modelling of GRB afterglows obtained from multi-wavelength monitoring campaigns. If we assume that the radiation observed from a GRB is a non-relativistic flow emitted from a region of size $ct$ then its brightness temperature is:
\begin{equation}  T_{\rm{b}} = 1.153 \times 10^{-8}\,d^{2}\,F_{\nu}\,\nu^{-2}\,t^{-2}\,(1+z)^{-1}~\mathrm{K}, \label{eq:3}
\end{equation}
where $d$ is the luminosity distance to the GRB in cm, $F_{\nu}$ the flux density in Jy, $\nu$ is the observing frequency in Hz, $t$ is time in seconds since the $\gamma$-ray trigger, and $z$ the redshift \citep{rybicki79}. (We assume a $\Lambda$CDM cosmology using $H_{0}=68$~km~s$^{-1}$Mpc$^{-1}$ and $\Omega_{m}=0.30$ based on the findings by \citet{ade16}.) However, if the maximum source size $ct$ results in a brightness temperature that exceeds the inverse-Compton limit $T_{B} \approx 10^{12}$~K in the rest-frame, then this assumption is wrong and the GRB outflow is likely relativistic. Given that $I_{\nu}/\nu^{3}$ (where $I_{\nu}$ is the specific intensity) is a relativistic invariant \citep{mihalas84}, the observed brightness temperature is related to the minimum Lorentz factor such that $T_{b}/T_{B} = \Gamma^{3}$ \citep[see similar arguments made by][]{kulkarni98,galama99}.

\begin{figure}
\centering
\includegraphics[width=0.5\textwidth]{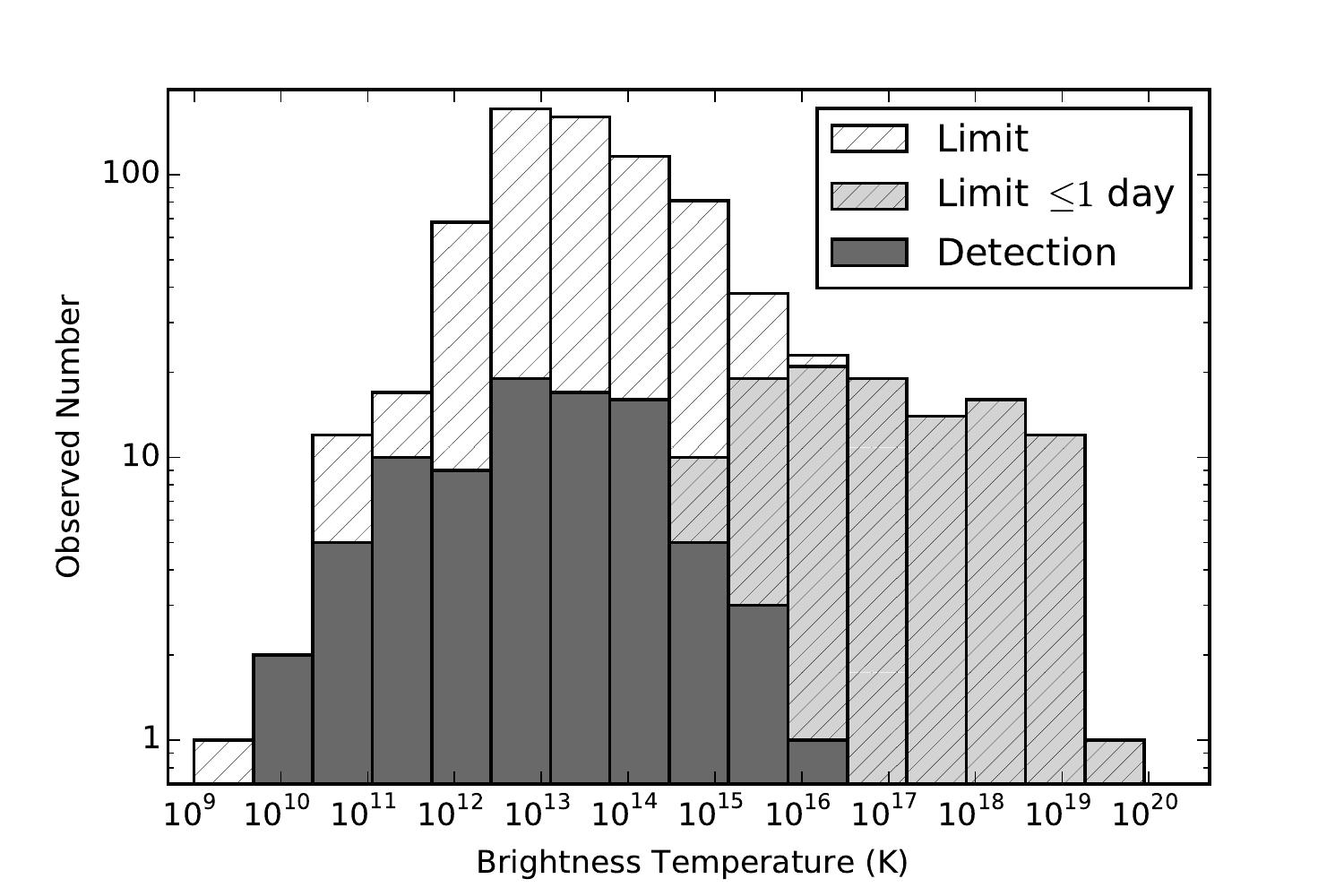}
\caption{Brightness temperature distribution of the AMI radio afterglow detections for the 12 \textit{confirmed} AMI-detected GRBs (grey filled). We assuming $z=2$ for those GRBs without redshift measurements. The upper limits on the brightness temperature derived from the $4\sigma_s$ flux limits of the AMI non-detections are also included (white hatches), with the subset of AMI upper-limits obtained within 1 day post-burst indicated (grey hatches).}
\label{fig:7}
\end{figure}

\begin{figure}
\centering
\includegraphics[width=0.5\textwidth]{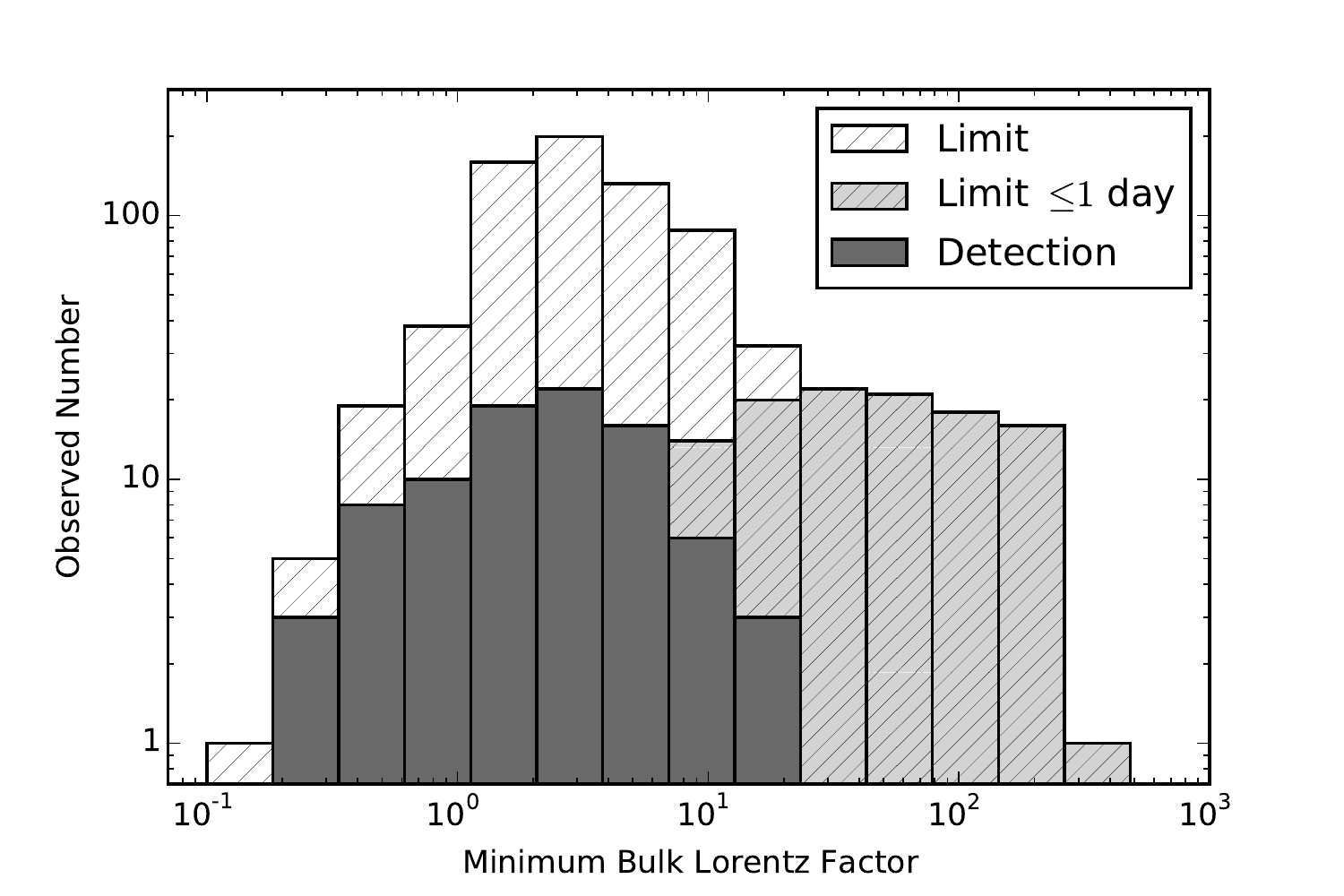}
\caption{As for Figure~\ref{fig:7} but for the minimum Lorentz factor.}
\label{fig:8}
\end{figure}

Figures~\ref{fig:7} and \ref{fig:8} are histograms of the brightness temperature and minimum Lorentz factor obtained from the AMI detections and upper-limits. The radio detections resulted in brightness temperatures ranging from $\sim2 \times10^{10}$ to $\sim8\times 10^{15}$ K \citep[which is consistent with the range of $10^{13}-10^{16}$ K proposed by][who used peak luminosity and variability timescales to constrain the range of $T_{\rm{b}}$ for a wide variety of radio flaring objects]{pietka15} and $0.3$ to 20.4 for the minimum Lorentz factors. The upper limits on the brightness temperatures and minimum Lorentz factors calculated using the $4\sigma_s$ flux limits from the AMI non-detections, probe a wider range of values, from $\sim4 \times 10^{9}$ to $\sim7 \times 10^{21}$ K and 0.2 to nearly 2000, respectively. Those upper-limits obtained from AMI non-detections within 1 day of the initial GRB trigger are represented by grey filled hatched histograms in Figures~\ref{fig:7} and \ref{fig:8}, dominating the higher end of the upper-limit distributions.

It can be seen that a few of the GRB detections resulted in minimum Lorentz factors ($\Gamma$) $>10$, which have been summarised (along with the corresponding brightness temperature) in Table~\ref{tab:4}. Each represents the first detection of the GRB (with the exception of GRB 130427A for which the first two detections resulted in $\Gamma>10$). In the case of GRB 130427A (the first entry in Table~\ref{tab:4}) and GRB 130907A, these radio detections came from rapid-response observations that were delayed several hours due to the source being below the AMI horizon at the time of the \swift\ detection. These represent some of the earliest detections of a radio afterglow from a long GRB. For comparison with radio GRB detections taken with other telescopes see Table~2 of \citet{anderson14}. While these earliest detections of the AMI GRBs in Table~\ref{tab:4} range from 0.3 to 3 days post-burst in the observer frame, they correspond to 0.2 and 0.5 days in the rest frame. It is therefore likely that AMI detected the radio counterpart of each of these GRBs just as they were becoming optically thin at 15.7 GHz, with the possibility of it being reverse-shock emission (see caveat in Section~\ref{earlytime}). Overall this is an excellent demonstration of how rapid-response systems on radio telescopes allow us to probe early time  brightness temperatures and minimum Lorentz factors in the radio band.

\begin{table}
\begin{center}
 \caption{Highest minimum Lorentz factors obtained from AMI detections}
 \label{tab:4}
 \begin{tabular}{lcccccc}
 \hline
GRB & Obs frame$^{\mathrm{~a}}$ & Rest frame$^{\mathrm{~b}}$ & $T_{b}$$^{\mathrm{~c}}$ & $\Gamma$$^{\mathrm{~d}}$ & $z^{\mathrm{~e}}$\\
& (days) & (days) &  ($\times 10^{15}$ K) & & \\
\hline
130427A & 0.36 & 0.27 & 4.29 & 16.24 & 0.34 \\
130427A & 0.64 & 0.48 & 1.70 & 11.94 & 0.34 \\
130907A & 0.55 & 0.25 & 7.21 & 19.32 & 1.24 \\
140304A & 2.99 & 0.48 & 1.07 & 10.22 & 5.28 \\ 
140703A & 1.27 & 0.31 & 2.16 & 12.93 & 3.14 \\
150413A & 1.28 & 0.31 & 1.38 & 11.14 & 3.14 \\
\hline
\end{tabular}
\end{center}
Note: All AMI detections for which the minimum Lorentz factors ($\Gamma) \geq 10$.\\
$^{\mathrm{~a}}$ Time of the AMI observation in days post-burst in the observer frame.\\
$^{\mathrm{~b}}$ Time of the AMI observation in days post-burst in the rest (GRB) frame.\\
$^{\mathrm{~c}}$ Brightness temperature\\
$^{\mathrm{~d}}$ Minimum Lorentz factor\\
$^{\mathrm{~e}}$ Redshift: Measurements taken from \citet{levan13,xu13a,flores13,deugarte13,deugarte14,jeong14,castro-tirado14d,deugarte15}.\\
\end{table}

\subsection{Radio Spectral Luminosity}

In order to investigate the radio GRB population dichotomy suggested by \citet{hancock13}, we chose to generate a histogram similar to Figure 4 of  \citet{chandra12} but instead looking at the distribution in luminosity (rather than the flux) of the radio detections and $4\sigma_s$ upper-limits. (Only 4 hour AMI observations with no GRB radio detection are used for the luminosity limits as these are the best illustration of the sensitivity we are achieving with AMI using our current strategy.) Following the same approach as \citet{chandra12}, the flux is converted into a spectral luminosity $L$ (erg s$^{-1}$ Hz$^{-1}$), using $L=4 \pi F d^{2}_{L}/(1+z)$, where $F$ is the measured radio flux density and $d_{L}$ is the luminosity distance calculated from the redshift ($z$) assuming the same $\Lambda$CDM cosmology as described in Section~\ref{btlf}. A $k$-correction factor of $(1+z)^{\alpha-\beta}$ is also applied, where $\alpha$ and $\beta$ are the temporal and spectral indices defined by $F \propto t^{\alpha}\nu^{\beta}$, resulting in a $k$-corrected radio spectral luminosity of $L=4 \pi F d^{2}_{L}/(1+z)^{\alpha - \beta - 1}$. Similar to \citet{chandra12}, we also assume $\alpha=0$ and $\beta=1/3$, which is appropriate for an optically thin, flat, post-jet-break light curve \citep[see their Section 3.2 and][]{frail06}.

\begin{figure}
\begin{center}
\includegraphics[width=0.5\textwidth]{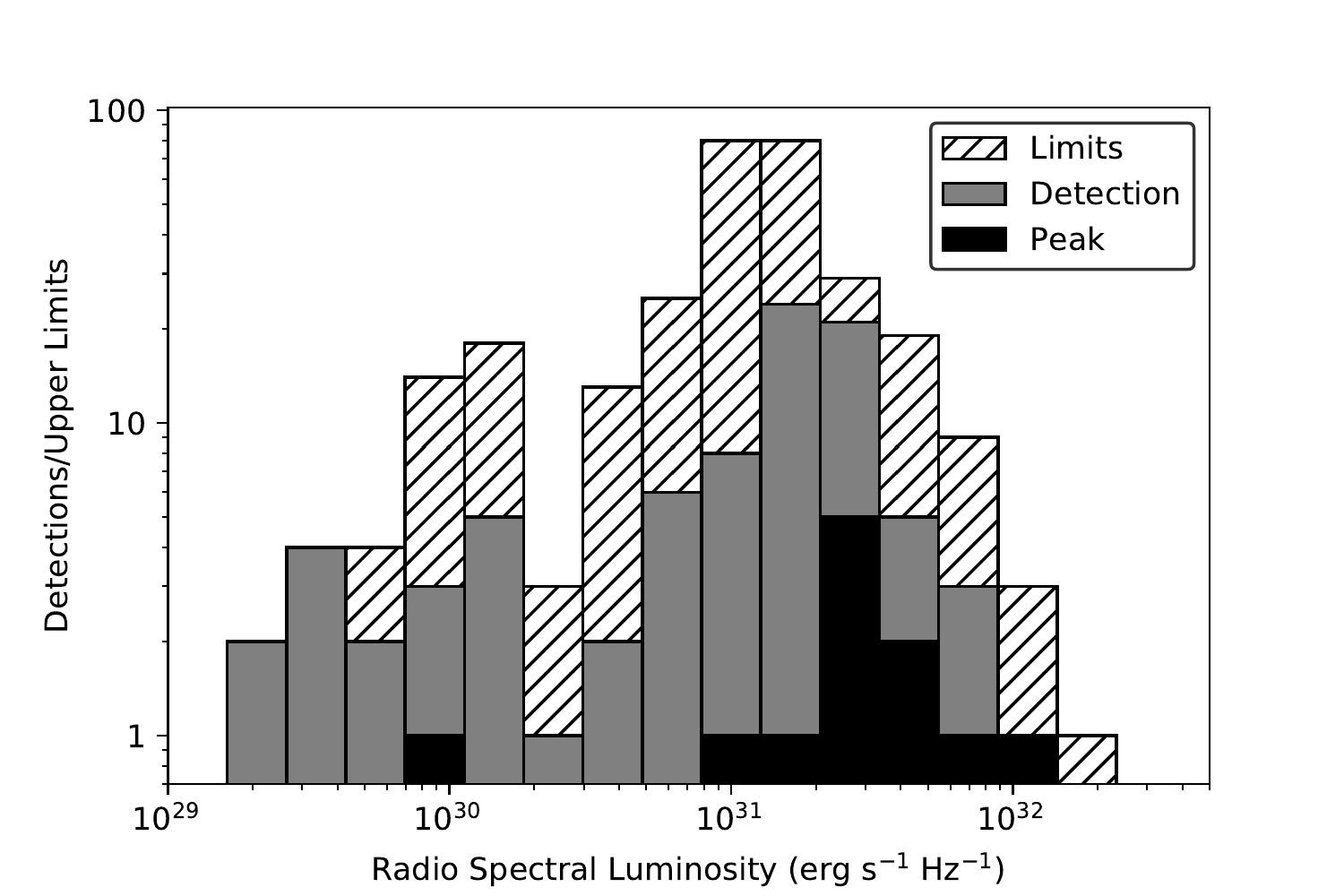}
\end{center}
\caption{Histogram of $k$-corrected spectral luminosities and $4\sigma_s$ upper-limits of the AMI GRB sample. We assume $z=2$ for those events without a known redshift. The 12 \textit{confirmed} AMI-detected GRBs are shown in grey. The peak (brightest) AMI detection for each of these 12 GRBs (black; which represent a subset of the detection distribution show in grey) are also included. The $4\sigma_s$ luminosity upper-limits (white hatched) are from the AMI observations that are at least 4 hours in duration with no detection.}
\label{fig:9}
\end{figure}

Figure~\ref{fig:9} shows the luminosity histogram of the radio detections and $4\sigma_s$ upper-limits. The luminosity of the brightest detection (peak) of each of the 12 \textit{confirmed} AMI-detected GRBs are also indicated. If such a dichotomy is possible to discern in AMI data then the peak from the luminosity distribution of the detections should be clearly separate to the peak of the luminosity upper limit distribution. However, it is clear from Figure~\ref{fig:9} that the peak in the detection, brightest (peak) detection and upper-limit luminosity distributions are consistent, and show no evidence for a fainter class of radio GRBs down to a sensitivity of 0.2 mJy/beam at 15.7 GHz. This does not preclude a population dichotomy as the limits on radio faint GRBs presented by \citet{hancock13} using visibility stacking are better than 0.1 mJy/beam (at 8.46 GHz) for all times with the exception of those within 0.3 days of the initial outburst.

\begin{figure*}
\begin{center}
\includegraphics[width=1.0\textwidth]{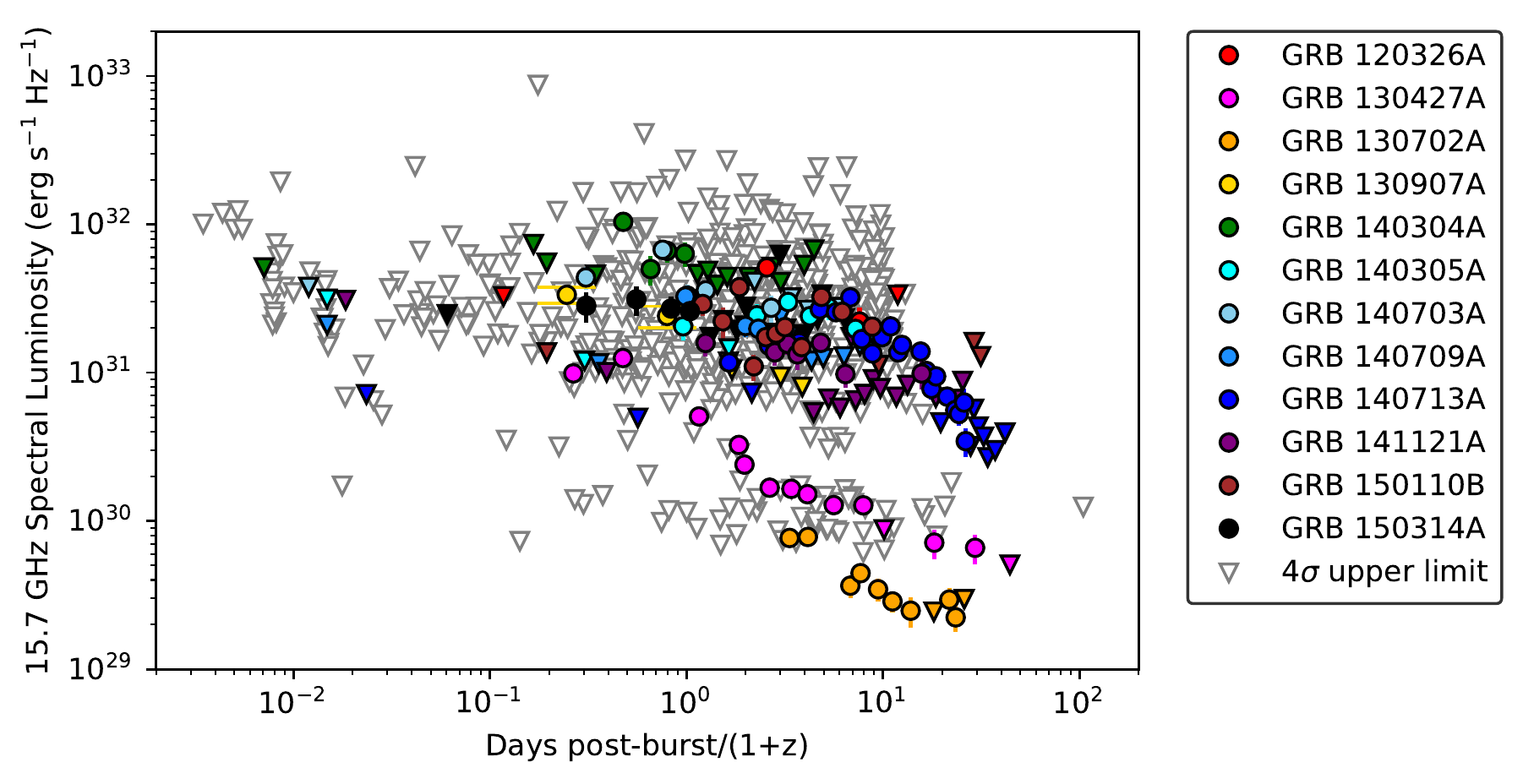}
\end{center}
\caption{The $k$-corrected spectral luminosity light curve of the 12 \textit{confirmed} AMI-detected GRBs in the rest frame. Each of these 12 GRBs are colour coded with their detections represented as dots and their $4\sigma_s$ upper limits on the non-detections represented by triangles (all errors are $1\sigma_s$).  The luminosity $4\sigma_s$ upper limits on all other non-detected GRBs are represented by white triangles. We again assume $z=2$ for those GRBs without a known redshift.
Redshift references: GRB 120326A: \citet{tello12}, GRB 120404A: \citet{cucchiara12}, GRB 120422A: \citet{schulze12}, GRB 120521C: \citet{tanvir12}, GRB 120722A: \citet{delia12}, GRB 120724A: \citet{cucchiara12b}, GRB 120729A: \citet{tanvir12b}, GRB 120802A: \citet{tanvir12c}, GRB 120811C:\citet{thoene12,fynbo12}, GRB 120907A: \citet{sanchez-ramirez12}, GRB 121128A: \citet{tanvir12d}, GRB 121211A: \citet{perley12b}, GRB 130131B: \citet{fynbo13}, GRB 130418A: \citet{deugarte13b,kruehler13}, GRB 130420A: \citet{deugarte13c}, GRB 130427A\citet{levan13,xu13a,flores13}, GRB 130505A: \citet{tanvir13b}, GRB 130511A: \citet{cucchiara13}, GRB 130514A: \citet{schmidl13}, GRB 130603B: \citet{foley13,sanchez-ramirez13,cucchiara13b,xu13c}, GRB130604A: \citet{cenko13}, GRB 130606A: \citet{castro-tirado13,xu13b}, GRB 130610A: \citet{smette13}, GRB 130612A: \citet{tanvir13c}, GRB 130701A: \citet{xu13d}, GRB 130702A: \citet{singer13p}, GRB 130831A: \citet{cucchiara13d}, GRB 130907A: \citet{deugarte13}, GRB 140206A: \citet{malesani14,delia14}, GRB 140304A: \citet{deugarte14,jeong14}, GRB 140318A: \citet{tanvir14}, GRB 140419A: \citet{tanvir14b}, GRB 140423A: \citet{tanvir14c}, GRB 140428A: \citet{perley14d}, GRB 140430A: \citet{kruehler14}, GRB 140508A: \citet{malesani14b,wiersema14}, GRB 140515A: \citet{chornock14}, GRB 140518A: \citet{chornock14b}, GRB 140606B: \citet{perley14e}, GRB 140629A: \citet{moskvitin14,davanzo14}, GRB 140703A: \citet{castro-tirado14d}, GRB 140710A: \citet{tanvir14d}, GRB 140713A: van der Horst et al. (in prep.), GRB 140801A: \citet{deugarte14b,moskvitin14b}, GRB 140903A: \citet{cucchiara14}, GRB 140907A: \citet{castro-tirado14e}, GRB 141026A: \citet{deugarte14c}, GRB 141121A: \citet{perley14f}, GRB 141212A: \citet{chornock14c}, GRB 141220A: \citet{deugarte14d}, GRB 141225A: \citet{gorosabel14}, GRB 150120A: \citet{chornock15}, GRB 150314A: \citet{deugarte15b}, GRB 150323A: \citet{perley15}, GRB 150413A: \citet{deugarte15}.}
\label{fig:10}
\end{figure*}

As a further investigation of the luminosity distribution of the AMI detected GRBs, we plotted the 15.7 GHz spectral luminosity against the time post-burst in  the rest frame (see Figure~\ref{fig:10}). The AMI luminosity of each of the 12 \textit{confirmed} AMI-detected GRBs, and the corresponding luminosity of the $4\sigma_s$ upper limits, are represented in different colours. The $4\sigma_s$ upper limit luminosities of all other GRB non-detections are included in this plot as white triangles. Once again there is no clear evidence for a dichotomy between the luminosity limits and the GRB detections. This plot also demonstrates that AMI has detected GRBs with wide ranges in spectral luminosities, between $\sim10^{29}$ to $\sim10^{32}$ erg s$^{-1}$ Hz$^{-1}$. However, compared to a similar luminosity plot constructed by \citet[][see their Figure~6]{chandra12}, we note that AMI has not detected the much fainter GRBs with radio spectral luminosities between $10^{27}-10^{29}$ erg s$^{-1}$ Hz$^{-1}$. Such events are known as sub-energetic or low luminosity GRBs \citep{soderberg04nat} and are extremely rare with only a handful known \citep[see][and references therein]{margutti13}, possibly representing a different GRB population \citep[for example see][]{liang07}.

The light curves in Figure~\ref{fig:10} clearly demonstrates that the AMI rapid-response programme is probing the early-time parameter space of GRBs, between $10^{-3}$ to $0.1$ days post-burst in the rest-frame, which has been very poorly sampled until now. The pile up of limits at $10^{-2}$ days post-burst in the rest frame is likely due to the most common rapid-response time (4.32 minutes) combined with assuming $z=2$ for GRBs without a known redshift. The deepest early-time limits ($<0.1$ days post-burst) are around $\sim10^{30}$erg s$^{-1}$ Hz$^{-1}$. This limit can be improved with longer observing times and the use of more sensitive instruments. Overall, this plot demonstrates a proof of concept for rapid-response telescope triggering on GRBs.

\subsection{\swift\ GRB subpopulations}

Within the AMI sample there is a sub-sample of \swift\ GRBs that were also detected by \fermi-LAT. These include GRB 120729A, GRB 121011A, GRB 130427A, GRB 130702A, GRB 130907A, and GRB 150314A \citep{ackermann13}. Of these 6 \fermi-LAT GRBs, three were detected with AMI, including GRB 130427A and GRB 130702A, which were the two least radio luminous AMI detected GRBs (see Figure~\ref{fig:10}), and GRB 130907A, which was only detected at early times and also radio faint (as previously mentioned both GRB 130427A and GRB 130907A have the earliest radio afterglow detections with AMI). 

Several dark bursts were also observed with AMI, including 11 of the 13 \swift\ GRBs classified as dark bursts by \citet[][including GRB 130420A, GRB 130502A, GRB 130514A, GRB 130606A, GRB 130609A, GRB 130907A, GRB 140114A, GRB 140318A, GRB 140518A, GRB 140709A, and GRB 140710A]{littlejohns15}, along with GRB 140713A \citep[][note that this list is not necessarily exhaustive]{castro-tirado14}. Of the 12 dark bursts observed with AMI, only three were detected, including GRB 130907A, GRB 140709A, and GRB 140713A, representing 25\% of this sub-sample. If we also include the \textit{possible} AMI detection of GRB 140318A and the concatenated detection of GRB 130609A, then the AMI radio detection rate of dark bursts may be as high as $\sim40-50\%$. While the detection rate for both the \fermi-LAT and dark burst populations are consistent with the implied AMI detection rate of $\sim44-56\%$ (see Section~\ref{earlytime}), with the \fermi-LAT GRBs appearing to favour low luminosity radio afterglows, these conclusions are based on a small number of events so further GRB monitoring is required to determine if their radio properties differ to the larger \swift\ GRB population.

\section{Summary and Conclusions}

Through the AMI GRB follow-up programme, we have produced the first catalogue of radio afterglows that is representative (i.e. not biased by target selection informed by prior knowledge of the event) of the radio properties of  \swift-detected GRBs down to 0.2\,mJy/beam at 15.7\,GHz. This catalogue includes 139 GRBs, 132 of which were detected with \swift, and is made up of AMI observations up to $>90$ days post-burst. This catalogue is also unique in including observations with response times on the order of minutes following the \swift\ GRB trigger, which were performed using the AMI-LA Rapid Response Mode (ALARRM). AMI is therefore the first radio telescope to target early-time incoherent (afterglow) emission from GRBs at high radio frequencies ($>2.3$\,GHz) via automatic triggering on \swift-BAT detection alerts. As a result, 39 GRBs were observed with AMI using the rapid-response system within $<1$\,hr post-burst (mid-time of observation), providing some of the most stringent early-time upper-limits of $\sim0.2$~mJy/beam at 15.7\,GHz. 

Using AMI rapid-response and monitoring observations, we have detected radio afterglows from 13 GRBs, 6 of which were discovered as part of this project and thus increasing the rate of GRBs with observed radio afterglows by 50\% within an 18 month period. This catalogue also includes a further 6 \textit{possible} AMI-detected GRBs, which cannot be confirmed with our datasets. Based on these results, AMI provides a lower-limit of 15\% on the radio detection rate of GRBs down to $\sim0.2$~mJy/beam. By including radio GRBs observed but not detected with AMI, as well as the \textit{possible} AMI-detected GRBs, we get a $22-28$\% detection rate (to an unknown completeness limit), which is more consistent with the $\sim30\%$ detection rate obtained by \citet{chandra12}. However, if we consider that AMI would have only detected $\sim50\%$ of the \swift\ GRBs in the \citet{chandra12} sample, the detection rate could be as high as $\sim44-56$\% down to $\sim0.1-0.15$\,mJy/beam. We suggest that the early-time ($<1$ day) observations provided by the AMI rapid-response mode are probing the fast evolving reverse-shock emission, which we know can have peak fluxes an order of magnitude brighter than the forward-shock \citep[e.g.][]{anderson14}, thus accounting for the $\sim20\%$ detection rate increase from the \citet{chandra12} value. Further support comes from 6 AMI-detected GRBs that peaked within 1 day post-burst in the rest-frame. This provides strong evidence that GRB radio afterglow simulations predicting detection rates with the SKA must also take into account contributions from the reverse-shock emission. 

The radio afterglows from both GRB 130427A and GRB 130907A were detected during their rapid-response observations, which took place 0.34 and 0.51\,days post-burst, respectively, when they had risen above the AMI horizon. Both detections have also resulted in some of the highest recorded minimum Lorentz factors obtained in the radio band \citep[see Table~\ref{tab:4} and][]{anderson14}. As they represent the earliest detections obtained with AMI, and also some of the earliest radio detections of long GRBs, it may be prudent to adjust the ALARRM strategy to perform automatic observations between $4-16$\,hours post-burst. This slight delay may allow the observations to better coincide with the afterglow emission becoming optically thin at 15.7\,GHz.

Luminosity investigations of the AMI data supports the conclusion by \citet{chandra12} that GRB radio detection rates are limited by the sensitivity of current radio telescope facilities. However, simulations by \citet{ghirlanda13} and \citet{burlon15} demonstrate that the forward-shock radio afterglow of \swift\ GRBs will be detectable with SKA1-MID (Band 5) provided that their sample (based on BAT6 GRBs) are representative of the population. Therefore, GRB radio follow-up conducted with the SKA will be able to determine if there is a radio-bright and radio-faint population of GRBs in the \swift-detected sample \citep{hancock13}.

Since the installation of the new AMI correlator at the end of 2015, the AMI-ALARRM GRB programme has continued, with all new radio afterglow detections and upper-limits being reported on the AMI-GRB database\footnote{https://4pisky.org/ami-grb/} and on the GCN. As the programme continues, AMI will build a larger sample to enable more thorough statistical studies of different \swift\ GRB subpopulations. We will be able to further investigate the radio properties of \fermi-LAT-detected GRBs and dark bursts, by establishing their radio detection rates in comparison with the larger \swift\ sample, and confirming whether \fermi-LAT GRBs tend to have lower-luminosity radio afterglows.

Rapid-response observing systems represent a new phase in radio transient astronomy. Installing such observing modes on radio telescopes is allowing astronomers to directly test strategies for the SKA facilities, which are baselined to have the same capabilities. Through utilising the VOEvent network \citep{staley16}, ALARRM is allowing us to probe the unusual parameter space of the early-time ($<1$~day) radio properties of high energy transients, and in-turn exploring the scientific payoff of interrupting SKA observing programmes to trigger on these events. This programme has also showed the value of rapid-response systems for radio transient science beyond GRBs, with the early-time detections of the flare star DG CVn \citep[on source within 6 minutes post-burst;][]{fender15} and the black hole X-ray binary V404 Cyg \citep[on source within 2 hours post-burst;][]{mooley15}. ALARRM also illustrates the benefits of simultaneous multi-wavelength observations \citep[e.g. triggered AMI observations of the flare star DG CVn, which were quasi-simultaneous with X-ray/$\gamma$-ray \swift\ observations;][]{fender15} and will hopefully encourage the implementation of other simultaneous multi-wavelength experiments (see Middleton et al. submitted). 

The ALARRM programme has also prompted the writing of the {\sc chimenea} and {\sc AMIsurvey} software packages \citep{staley15a}, which utilise Python and mature radio reduction software to automate the data reduction and analysis of multi-epoch radio observations. This is the first step to identifying transient sources in real time (using software like the {\sc TraP} and {\sc PySE}), which could be immediately reported to the astronomical community. Overall, the AMI-ALARRM programme is already demonstrating the exciting science that can be probed through early-time detections of high-energy transients, which in-turn shows the value in performing real-time transient triggering. We therefore encourage other radio telescopes to equip similar rapid-response capabilities in-order to demonstrate that there are no significant barriers to implementing such technologies on the SKA facilities.

\section*{Acknowledgements}

We thank the staff of the Mullard Radio Astronomy Observatory for their invaluable assistance in the operation of AMI. GEA, TDS, RPF, JWB acknowledge the support of the European Research Council Advanced Grant 267697 ``4 Pi Sky: Extreme Astrophysics with Revolutionary Radio Telescopes." AJvdH, AR, and RAMJW acknowledge support from the European Research Council via Advanced Investigator Grant no. 247295 ``AARTFAAC." KPM's research is supported by the Oxford Centre for Astrophysical Surveys which is funded through the Hintze Family Charitable Foundation. CR acknowledges the support of an STFC studentship.

\appendix

\section{Description of new \textit{possible} GRB radio afterglows discovered with AMI}
\label{appendixa}

\subsection{GRB 120320A}

The ALARRM triggered observation of GRB 120320A, which occurred 0.6 days post-burst, detected a $4\sigma_s$ coincident radio source with a flux of $0.38 \pm 0.09$ mJy/beam. However, it is also possible this source is an artefact produced by the nearby radio sources NVSS 141009+084149 and NVSS 140957+084108. This single epoch detection was not reported in \citet{staley13} likely due to differences in the reduction procedure and our use of an automatic source finder. No source was blindly detected in the follow-up observation that occurred 15.6 days later or in the concatenated image, with $4\sigma_s$ upper limits of 0.36 and 0.25 mJy/beam, respectively. However, the quality of both epochs are poor, likely resulting from terrestrial interference due to the low declination of $8.7$ deg for this GRB. It is therefore not possible to determine if the coincident radio source seen in the first epoch is real. No other radio observations of this event have been reported but there is a single report of a possible detection of a very faint optical counterpart \citep{levan12}. 

\subsection{GRB 130625A}

A possible radio source was detected within the XRT 90\% position error of GRB 130625A in the 2013 July 22 AMI observation, 27 days post-burst, with a flux of $0.59 \pm 0.14$ mJy/beam ($4.4\sigma_s$ flux significance). This source was not detected in any of the 5 earlier epochs, nor in the concatenated image. If such a source were steady it should have been detectable with a $>4\sigma_s$ significance in all the AMI observations of this event (with the exception of the 2013 July 3 observation) and at a $>12\sigma_s$ significance in the concatenated image. It is therefore possible that AMI detected the radio counterpart to GRB 130625A over a month post-burst. No optical counterpart was detected for this GRB but the limits are not particularly constraining \citep{xu13,linevsky13,yurkov13}. 

\subsection{GRB 140209A}

AMI possibly detected the radio counterpart to GRB 140209A with a flux of $0.43 \pm 0.10$mJy/beam, corresponding to a $4.2\sigma_s$ significance, during the first AMI observation on 2014 Feb 10 (1.36 days post-burst). The position of this coincident radio source agrees within $2\sigma_p$ of the optical counterpart position \citep{perley14c}.  All three follow-up AMI observations that occurred 2.5, 4.4 and 9.4 days post-burst did not detect this source even though their RMS noise level was improved by at least factor of $\sim2$ when compared to the first observation. All observations were four hours in duration. The deep concatenated image also did not detect this source but shows a complex region with some evidence for extended emission or uncleanable sidelobes due to the nearby NVSS sources that lie within $2.7' - 3.6'$ from the XRT position. It is therefore difficult to determine if the coincident radio source seen in the 2014 Feb 10 observation is real. No other radio observations have been reported for this event.

\subsection{GRB 140318A}

The radio counterpart to GRB 140318A was possibly detected with AMI on 2014 March 25 (8 days post-burst), corresponding to a flux of $0.28 \pm 0.05$ mJy/beam with a $5.0\sigma_s$ significance. In the concatenated AMI image of GRB 140318A, the possible radio counterpart was also detected with a flux of $0.15 \pm 0.03$ mJy/beam and a significance of $4.9\sigma_s$. The position of this radio source agrees with the position of the optical counterpart \citep{schulze14b}. The light curve in Figure~\ref{fig:3} shows the detection on 2014 March 25 and the $4\sigma_s$ RMS noise levels of the other three AMI epochs of GRB 140318A. The concatenated detection, with a similar significance to that of the detection on 2014 March 25, is fainter by a factor of $\sim2$, which is expected for a non-steady source detected in only one of four epochs. The final observation on 2014 March 28 should also have detected the coincident source with a significance of $4.8\sigma_s$ if it were still the same brightest as that measured during the 2014 March 25 epoch. It is therefore quite possible that AMI detected the radio counterpart to GRB 140318A. There have been no other reports of radio observations of this event. However, the faintness of its optical counterpart does suggest that GRB 140318A was a dark burst, where the optical attenuation was likely caused by moderate dust extinction \citep[$0.25<A_{V}<1$;][]{littlejohns15}.

\subsection{GRB 140320C}

GRB 140320C was initially detected by \textit{INTEGRAL} \citep{mereghetti14b} and later localised with the XRT \citep{pagani14b}. AMI detected a $4.1\sigma_s$ significant radio source with a flux of $0.14 \pm 0.03$mJy/beam within the 90\% XRT error circle of GRB 140320C on 2014 March 22 (2.1 days post-burst). If this source were steady then it should have been detected at a similar significance during the following AMI observation two days later on 2014 March 24, and at a significance of $7.8\sigma_s$ in the concatenated image. The position of this radio source agrees within $1.04\sigma_p$ of the optical afterglow position \citep{volnova14a,volnova14b}. No other radio observations were reported for this even. It is therefore possible that AMI detected the radio afterglow of GRB 140320C. 

\subsection{GRB 140607A}

AMI detected a $4.8\sigma_s$ significant radio source within $2.7\sigma_p$ of the best BAT position of GRB 140607A \citep{krimm14} on 2014 June 9 (1.75 days post-burst). The other two epochs were far less sensitive so would not have detected this source. However, this radio source was also not detected in the concatenated image, which has comparable sensitivity to the first AMI epoch. It is therefore possible that this radio source could be the afterglow of GRB 140607A.

\section{Description of AMI concatenated detections}
\label{appendixb}

\subsection{GRB 130606A}

GRB 130606A lies $21.3''$ west of NVSS 163736+294742 and has been classed as a dark burst, the optical flux attenuation likely caused by its high redshift of 5.913  \citep{castro-tirado13,xu13b,littlejohns15}. In the concatenated AMI observation of GRB 130606A, there appears to be two or even three blended sources at the position of this NVSS source and therefore the position of the GRB. Dividing the observations into two separate concatenated images does not show conclusive evidence for variability from any of the blended sources, likely due to the lack of sensitivity. Given the high redshift of GRB 130606A, it is also unlikely that the host galaxy would be resolved by AMI. However, the EVLA detected radio emission from GRB 130606A at 21.8GHz just 0.6 days post-burst measuring a flux of $\sim0.1$ mJy/beam \citep{laskar13}. Further follow-up at mm wavelengths using the Plateau de Bure Interferometer \citep[PdBI;][]{guilloteau92} detected a $\sim1.5$ mJy/beam source 3.30 days post-burst at 86.7 GHz but had faded below detectability at 7.50 days post-burst \citep{castro-tirado13pp}. Given the sensitivity and spatial resolution of the AMI data it is not possible to conclude anything about the nature of the coincident radio source. 

\subsection{GRB 140508A}

The optical counterpart to the \fermi-GBM detection GRB 140508A \citep{yu14} was quickly detected and identified by iPTF just 0.28 days post-burst \citep{singer14}. A late time radio detection was then made 5.2 days post-burst with the VLA at 6.1 GHz and 22 GHz, reporting a 6.1 GHz flux of $0.13 \pm 0.01$ mJy/beam \citep{horesh14}. AMI did not start observing GRB 140508A until after 8 days post-burst but there were no detections of this radio counterpart in the individual epochs. However, the concatenated image from the four AMI observations, ranging from 8.93 to 16.79 days post-burst, yielded the blind detection of a $4.9\sigma_s$ source at the UVOT position of GRB 140508A with a flux of $0.12 \pm 0.02$ mJy/beam. This flux value is comparable to the VLA  6.1 GHz detection. To investigate if this was radio afterglow of GRB 140508A, we concatenated the first two epochs and the last two epochs separately to see if there was any evidence for variability. However, the two resulting concatenated images were not sensitive enough to detect the coincident radio source seen in the full concatenated image. It is therefore not possible to determine whether the concatenated observation was deep enough to detect the radio counterpart to GRB 140508A or if we instead detected a steady background source.

\subsection{GRB 140801A}

The radio source detected in the concatenated image of GRB 140801A lies $14.6''$ SE (within $3\sigma_p$) from the XRT position with a flux of $0.16 \pm 0.03$ mJy/beam. This blindly detected source is also within $3\sigma_p$ of NVSS 025617+305552 and WENSS B0253.2+3044, which could mean that all three are the same source. However, the $3\sigma_p$ position error circles may not necessarily overlap between NVSS and WENSS \citep[the position accuracy of WENSS can range from $1.5-10''$,][]{rengelink97}. Concatenating the early and later epochs into two separate images showed the source flux to be unchanged between these two epochs. It is therefore likely that this radio source is steady and could be a backgrounds source.

\subsection{GRB 150309A}

A $4.8\sigma_s$ radio source lies $1.8'$ to the SE of the XRT position at RA (J2000.0) = 18:29:10.92 ($\pm 55.16"$) and Dec (J2000.0) = +86:24:08.04 ($\pm 4.20"$) in the concatenated image of GRB 150309A. The large error on the RA coordinate of this radio source means it is within a $1.9\sigma_p$ of the XRT position. Combining the early and late epochs into two concatenated images showed that the flux was consistent between the first concatenated epoch and the deep image, and also consistent with the poorer sensitivity of the second epoch concatenation. This radio source is therefore likely steady and not the radio afterglow of GRB 150309A. 

\section{Radio-detected GRBs not detected with AMI}
\label{appendixc}

\subsection{GRB 130603B}

The short GRB 130603B was the first claimed case for an associated ``kilonova'' \citep{tanvir13,berger13} and the third radio detected short GRB \citep{fong14}. GRB 130603B was first detected at 4.9 and 6.7\,GHz with the VLA, at fluxes of $125.1 \pm 14.4$ and $118.6 \pm 9.1\, \mu$Jy/beam, respectively, just 0.37 days post-burst, fading within 2\,days and likely representing one of the earliest radio detections of any GRB \citep{fong14}. AMI rapidly responded to the \swift\ trigger and was on-target within 5 minutes post-burst, obtaining a 1 hour observation and $4\sigma_s$ upper-limit of $0.57$\,mJy/beam. Two further observations at 0.13 and 0.95 days post-burst, both 3 hours in duration, where then manually scheduled within the next 24~hrs. Unfortunately no radio counterpart was detected with AMI within 24~hrs down to a $4\sigma_s$ upper-limit of $0.24$\,mJy/beam but this is consistent with the VLA detections.

\subsection{GRB 140419A}

The only reported radio observation of GRB 140419A is a marginal radio detection of 1.5 mJy/beam at 93 GHz with CARMA, just 77 minutes post-burst \citep{perley14}. If real, this is the earliest ever reported radio detection of a GRB. However, no other radio observations have been reported to confirm this. The earliest AMI observation of GRB 140419A, which resulted from the ALARRM trigger, took place 0.38 days post-burst but did not detect a counterpart with a $4\sigma_s$ upper limit of 0.29 mJy/beam. Another 7 follow-up AMI observations were conducted over the following month but no radio counterpart was detected with the most constraining $4\sigma_s$ upper limit of $0.14$ mJy/beam.

\subsection{GRB 140515A}

GRB 140515A was observed with the VLA at multiple frequencies 0.62 days post-burst and was detected at 21.8 GHz with a flux of $\sim0.1$ mJy/beam \citep{laskar14d}. AMI was observing GRB 140515A following the ALARRM trigger at 0.26 days post-burst with a further 5 follow-up observations occurring over the next week. Unfortunately all the AMI observations of this GRB were very poor likely due to increased terrestrial interference at this low observing elevation (Dec = $15.105^{\circ}$) and perhaps also from artefacts generated by the nearby extended source NVSS 122409+150526. As a result, no radio counterpart to GRB 140515A was detected in any of the AMI observations. However, given that the radio counterpart was reported to be $\sim0.1$ mJy/beam at 21.8 GHz it would likely have been at the same level or fainter at 15.7 GHz so no sensitive 4 hour AMI observation would have been able to detect it.

\subsection{GRB 140903A}

This short GRB was first reported as a possible burst \citep{cummings14a} and later confirmed by \citet{cummings14b}. A radio detection at 6 GHz with the VLA was reported by \citet{fong14} just 0.40 days post burst with a flux $\sim0.11$ mJy/beam. A later marginal detection ($3.1\sigma_s$) of $0.102 \pm 0.033$ mJy/beam at 1.4 GHz was then provided by \citet{nayana14} using the GMRT. An analysis of JVLA observations \citep[first reported by][]{fong15} at 6.1 and 9.8 GHz was conducted by \citet{troja16}, with the brightest detection in both bands occurring at 2.51 days post-burst with fluxes  $203 \pm 13 \mu$Jy/beam and $153 \pm 10 \mu$Jy/beam, respectively. This event was detectable for up to $\sim9$ days post-burst, the longest lived radio afterglow observed from a short GRB. Analyses performed by both \citet{troja16} and \citet{zhang17} demonstrate the radio afterglow is consistent with a standard forward-shock model involving a narrow, collimated outflow.

AMI first observed GRB 140903A at 0.93 days post-burst and detected a possible radio counterpart at the XRT position with a flux of $0.72 \pm 0.08$ mJy/beam. However, its close proximity to the bright source NVSS 155207+273501, which lies of $1.5'$ to the SE of GRB 140903A and has a 15.7 GHz flux of $11.67 \pm\ 0.58$ mJy/beam, caused uncleanable structure in the image. We are therefore unable to preclude the possibility that our detection is in fact an artefact, or partly contaminated by an artefact, generated by NVSS 155207+273501 at the XRT position of GRB 140903A. {\sc PySE} did not detect any radio sources at the GRB position in the following six AMI observations of this event. However, the significance resulting from a forced fit at the position of the GRB in these observations are extremely high ($>20 \sigma_s$). It is possibly that these high significance values could be due to structures and artefacts from NVSS 155207+273501.

\subsection{GRB 141026A}

GRB 141026A was observed with the VLA 6.0 days post-burst, resulting in a clear detection at 6.2 GHz with a flux of $91 \pm 7$ $\mu$Jy/beam \citep{corsi14b}. Marginal detections were also obtained with the VLA at 1.1 and 4.3 days post-burst at 21.8 GHz. A non-detection with the IRAM 30m antenna at 150 GHz within a day post-burst was also reported, resulting in a $3\sigma_s$ flux upper-limit of 1.2 mJy/beam \citep{castro-tirado14b}. No radio afterglow was detected in the eight AMI observations of GRB 141026A taken between 3 minutes to 18.9 days post-burst. The AMI observation taken closest in time to the VLA 6.2 GHz detection took place 6.9 days post-burst, with a $3\sigma_s$ upper limit of 0.31 mJy/beam. Given the faintness of the afterglow in the VLA observations it is likely that the radio emission was below the sensitivity of AMI for a 4 hour observation. 

\section{Coincident steady sources}
\label{appendixd}

\subsection{GRB 130216A}

When \swift-BAT detected GRB 130216A, moon constraints prevented follow-up with the XRT and UVOT until nearly $4$ days, at which point any possible counterpart had faded below detectability \citep{melandri13}. Ground based optical follow-up did not detect an optical counterpart either. As a result, the best known position was provided by BAT with an $1'$ error \citep{barthelmy13}. AMI triggered on this \swift\ event and was on-target and observing GRB 130216A for one hour within 14 minutes post-burst. A single uncatalogued radio source lying $45''$ Southwest of the BAT position, and therefore within the BAT position error, was detected at RA (J2000.0) = 04:31:34.23 ($\pm 0.50"$) and Dec (J2000.0) = +14:39:36.94 ($\pm 0.88$) (based on the concatenated image) with a flux of $0.87 \pm 0.05$ mJy/beam. However, subsequent observations demonstrated that this radio source showed little evidence for variability and was therefore unlikely to be the radio counterpart to GRB 130216A. The force fitted flux reported in Table~\ref{tab:2} for the only epoch without a detection taken on 2013-02-17 was conducted at the position of the radio source rather than at the best BAT position. 

\subsection{GRB 140320B}\label{grb140320b}

\textit{INTEGRAL} detected the long GRB 140320B \citep{mereghetti14} at 09:26:00 UT, which was quickly localised by the \swift\ XRT \citep{pagani14}. AMI obtain five $\sim4$ hr observations of GRB 140320A beginning 1.5 days up to 11.5 days post-burst. In each observation an uncatalogued radio source was detected, lying $11"$ SE of the XRT \citep[and the optical][]{guidorzi13b} position at RA (J2000.0) = 09:42:15.00 ($\pm 1.44"$) and Dec (J2000.0) = +60:15:56.04 ($\pm 1.02"$) (based on the concatenated image) and is therefore well outside the $3\sigma_p$ XRT position error. Given that the offset between the uncatalogued radio source and the GRB XRT position is much larger than the overall position error, as well as the lack of evidence for variability, this source is likely to be a steady field source rather than the radio counterpart to GRB 140320B.

\subsection{GRB 140606A}

GRB 140606A is a short hard burst that was detected by \swift\ \citep{stroh14}. Due to lack of counterpart detections, the best position of this GRB comes from the BAT instrument and has a 90\% position error of $2.4'$ \citep{cummings14d}. A blind source search of the concatenated image detected three sources within a $3\sigma_p$ position error. Of these three, the radio source that lies closest to the best BAT position is NVSS 132712+373613, the fluxes of which are reported in Table~\ref{tab:2}. The source NVSS 132720+373351, which lies within $2\sigma_p$ of the BAT position was also detected. Both NVSS sources were individually detected in the final three epochs. One other uncatalogued radio source was detected in the concatenated image on the very edge of the BAT error circle at RA (J2000.0) = 13:26:58.7 ($\pm 9.3"$) and Dec (J2000) = 37:35:04.3 ($\pm7.4"$) with a flux of $0.21 \pm 0.05$mJy/beam and a significance of 4.1. Both NVSS sources are unlikely to be associated but we cannot rule out that the uncatalogued radio source could be the GRB counterpart. 

\subsection{GRB 141015A}

AMI obtained eight observations of GRB 141015A with the majority of observations lasting between 4 and 5 hours. On both 2014 October 20 and 2014 October 23, a source was blindly detected at the XRT position of GRB 141015A with a significance of $4.5\sigma_s$ and $4.7\sigma_s$, respectively. However, the $4\sigma_s$ upper limits of the other epochs are consistent with these detections. This source is therefore near the sensitivity limit of AMI for this range of exposure times. As both the detections and the concatenated flux agree within $2\sigma_s$, it is unlikely that this source is transient. Conversely, the most sensitive late time (post-detections) observation of GRB 141015A, taken on 2014 October 25, should have been able to detect the brightest detection seen on 2014 October 20 at a $5\sigma_s$ level. Using these data it is therefore not possible to confirm if this is the radio afterglow of GRB 141015A or a steady source. 

\subsection{GRB 141020A}

During the AMI monitoring of GRB 141020A, the observation taken on 2014 October 24 detected an uncatalogued radio source with a $4.5\sigma_s$ significance just $26''$ South of the UVOT position. This same source was detected in the concatenated image with a significance of $4.9\sigma_s$ but with a flux that was a factor of $\sim2$ lower than what was detected on 2014 October 24. The faintness of this radio source resulted in large positional errors and therefore lies within $4\sigma_p$ of the UVOT position. Given that the detection on 2014 October 24 is very close to the sensitivity of the other five epochs, it is not possible to conclude any evidence for transient activity. While the concatenated flux is a factor of $\sim2$ fainter than the detection, it is still agrees within $2\sigma_s$. Given the $>3\sigma_p$ position offset between the UVOT position of GRB 141020A and the blindly detection radio source, and the general lack of statistical evidence for variability, it is unlikely that this is the radio counterpart to GRB 141020A.

\label{lastpage}


\end{document}